\documentclass[11pt]{article}
\pdfoutput=1
\usepackage{jcapmod}
\usepackage{shorthand}

\usepackage{mathtools}
\usepackage{booktabs}
\usepackage[english]{babel}
\usepackage{amsmath,amssymb,amsbsy,amstext, amsthm, simplewick, amsfonts}
\usepackage{hyperref}
\usepackage{graphicx}
\usepackage[small]{caption}
\usepackage{siunitx}
\usepackage{upgreek}
\usepackage{framed}
\usepackage{wrapfig}
\usepackage{multirow}
\usepackage{bbm}
\usepackage[svgnames,dvipsnames,x11names]{xcolor}
\usepackage[utf8x]{inputenc}
\usepackage{selinput}
\usepackage{bm}
\usepackage{float}
\usepackage{geometry}
\usepackage{hyperref}
\usepackage{yfonts}
\usepackage{caption}
\usepackage{subcaption}
\usepackage{sidecap}
\usepackage{longtable}
\usepackage{anyfontsize}
\usepackage{dsfont}
\usepackage{xcolor}
\usepackage{tikz}
\usepackage{tikz-feynman}
\usepackage[framemethod=default]{mdframed}
\newmdenv[skipabove=7pt,
skipbelow=7pt,
rightline=false,
leftline=false,
topline=false,
bottomline=false,
backgroundcolor=gray!10,
linecolor=gray,
innerleftmargin=5pt,
innerrightmargin=5pt,
innertopmargin=5pt,
innerbottommargin=5pt,
leftmargin=0cm,
rightmargin=0cm,
linewidth=4pt]{eBox}
\newmdenv[skipabove=7pt,
skipbelow=7pt,
rightline=false,
leftline=false,
topline=false,
bottomline=false,
backgroundcolor=gray!10,
linecolor=gray,
innerleftmargin=5pt,
innerrightmargin=5pt,
innertopmargin=-5pt,
innerbottommargin=5pt,
leftmargin=0cm,
rightmargin=0cm,
linewidth=4pt]{eBox2}

\definecolor{blue3}{RGB}{31, 119, 180}
\definecolor{red3}{RGB}{	214, 39, 40}
\definecolor{orange3}{RGB}{255, 127, 14}
\definecolor{green3}{RGB}{44, 160, 44}
\definecolor{repBlue}{RGB}{31, 119, 180}
\definecolor{repRed}{RGB}{	214, 39, 40}
\definecolor{repGreen}{RGB}{44, 160, 44}

\renewcommand{\(}{\left(}
\renewcommand{\)}{\right)}
\renewcommand{\[}{\left[}
\renewcommand{\]}{\right]}

\def\be{\begin{equation}}
\def\ee{\end{equation}}
\newcommand{\bea}{\begin{eqnarray}}
\newcommand{\eea}{\end{eqnarray}}
\def\fnl{f_{\rm NL}}
\def\vp{\varphi}

\usepackage{colortbl}
\definecolor{lightgreen}{cmyk}{0.2, 0, 0.2, 0.2}
\definecolor{lightgray}{cmyk}{0.1,0.2,0,0.1}
\definecolor{lightgray2}{cmyk}{0.1,0.1,0,0.1}

\makeatletter
\newlength{\apb@width}
\newcommand{\autoparbox}[2][c]{\settowidth{\apb@width}{#2}\parbox[#1]{\apb@width}{#2}}

\makeatother


\def\beq{\begin{equation}}
\def\eeq{\end{equation}}

\def\s{ \sigma}

\newcommand{\mpl}{M_{\rm Pl}}

\allowdisplaybreaks[1]
\begin{document}

\newgeometry{top=2cm, bottom=2cm, left=2.9cm, right=2.9cm}

\begin{titlepage}
\setcounter{page}{1} \baselineskip=15.5pt 
\thispagestyle{empty}

\begin{center}
{\fontsize{20}{18} \bf Bootstrapping the Cosmological Collider  \\ [12pt]
\fontsize{20}{18} with Resonant Features \\ }
\end{center}

\vskip 20pt

\begin{center}
\noindent
{\fontsize{12}{18}\selectfont Dong-Gang Wang$^{1,2}$  and  Bowei Zhang$^1$}
\end{center}

\vskip 20pt

\begin{center}
 
{$^1$ \fontsize{12}{18}\it Department of Applied Mathematics and Theoretical Physics,\\ University of Cambridge,
Wilberforce Road, Cambridge, CB3 0WA, UK}

{$^2$ \fontsize{12}{18}\it Institute for Advanced Study and Department of Physics,\\ Hong Kong University of Science and Technology, Clear Water Bay, HK, China
}
 
\end{center}

%
%

\vspace{0.4cm}
 \begin{center}{\bf Abstract} 
 \end{center}
 \noindent

Signatures of heavy particles during inflation are exponentially suppressed by the Boltzmann factor when the masses are far above the Hubble scale. 
In more realistic scenarios, however, scale-dependent features may change this conventional picture and boost the cosmological collider signals.
In this paper, we compute cosmological correlators of the primordial curvature perturbations exchanging an intermediate heavy field with periodically varying couplings.
The basic setup corresponds to inflation scenarios with globally oscillating features that enjoy a discrete shift symmetry. Adopting the bootstrap approach, we derive and solve the boundary differential equations that are satisfied by the massive-exchange three-point functions. The presence of the oscillatory couplings leads to resonance-enhanced cosmological collider signals for heavy fields when the oscillating frequency exceeds the field masses. Meanwhile, the forms of these differential equations are modified, which generates new shapes of primordial non-Gaussianity as a combination of resonant features and collider signals. 
Based on these computations, we revisit the string-inspired model of axion monodromy inflation and point out that the cosmological correlators can become sensitive to heavy moduli fields of flux compactifications. 
This finding suggests a breakdown of the conventional single-field description, as the heavy moduli may not be simply integrated out but yield detectably large signals in the primordial bispectrum.

\noindent

\end{titlepage}

\newpage

\restoregeometry
\setcounter{tocdepth}{3}
\setcounter{page}{1}
\tableofcontents

\newpage

\section{Introduction}
\label{sec:intro}

Cosmological correlators from the primordial era of cosmic inflation provide an unsurpassed opportunity for high energy physics. With the increasing amount of data from observational surveys, we may measure these correlators at high precision, and thus probe energy scales far above the ones that can be reached in any terrestrial particle accelerator \cite{Meerburg:2019qqi,Achucarro:2022qrl}. In recent years, significant progress has been made in the theoretical understanding of these inflation predictions. One particularly interesting development is the cosmological bootstrap, which brings new ideas and techniques in
scattering amplitudes, holography and conformal field theory into the study of cosmological correlators \cite{Arkani-Hamed:2018kmz,Baumann:2019oyu,Baumann:2020dch, Arkani-Hamed:2017fdk, Benincasa:2018ssx, Sleight:2019mgd, Sleight:2019hfp, Goodhew:2020hob, Cespedes:2020xqq, Pajer:2020wxk, Jazayeri:2021fvk, Bonifacio:2021azc, Melville:2021lst, Goodhew:2021oqg, Pimentel:2022fsc, Jazayeri:2022kjy,  Baumann:2022jpr,  Qin:2022fbv, Xianyu:2022jwk, Wang:2022eop,Qin:2023ejc,Stefanyszyn:2023qov,DuasoPueyo:2023viy,Cespedes:2023aal, Bzowski:2023nef, Arkani-Hamed:2023kig,Grimm:2024mbw,Donath:2024utn,   Melville:2024ove, Aoki:2024uyi, Stefanyszyn:2024msm,Liu:2024xyi, Goodhew:2024eup, Ghosh:2024aqd, Lee:2024sks, Cespedes:2025dnq, Pimentel:2025rds,Stefanyszyn:2025yhq}.
At the conceptual level, this emerging direction provides a boundary perspective towards cosmology, where we may find clues about how to reconstruct time evolution in the bulk from spatial correlations on the late-time slicing. Practically, the bootstrap itself is a powerful computational tool, and furthermore, based on minimal assumptions, it gives us a systematic way to classify all the possibilities for the inflationary observables, also known as primordial non-Gaussianities (PNG). 

\vskip4pt
In the current bootstrap analysis,
scale invariance is a key assumption, which originates from the
de Sitter (dS) dilation symmetry. However, this assumption breaks down for a large number
of inflation models, and there we expect interesting PNG signals. Therefore, for the development of the bootstrap program, one intriguing question is
whether it can be extended to these more realistic setups.
Generally speaking, once we break dS dilation symmetry, new guiding principles are needed
to perform the bootstrap analysis with concrete results. An interesting substitute to which we can
resort is the {\it discrete shift symmetry} of the inflaton field, which allows for small periodic
oscillations in the background. Remarkably, this small deviation from scale invariance leads
to the resonance effects between the Bunch-Davies vacuum and the oscillatory couplings,
which generates a brand new type of signal called resonant non-Gaussianity \cite{Chen:2008wn,Flauger:2010ja}. Recently,
a bootstrap analysis has been performed for single field inflation with this type of features \cite{DuasoPueyo:2023viy}.

\vskip4pt
This work extends the investigation of cosmological correlators with a discrete shift symmetry, particularly to scenarios with intermediate heavy fields. 
The presence of massive particles during inflation has been extensively studied in the context of cosmological collider physics \cite{Chen:2009zp, Baumann:2011nk, Noumi:2012vr, Arkani-Hamed:2015bza}. For masses of $\mathcal{O}(H)$, one can identify a characteristic signature of the new species in the squeezed limit of the primordial bispectrum. This is known as the collider signal that manifests itself as scale-invariant oscillations with frequencies determined by the masses. Thus, model-independently, one can do particle spectroscopy for these new states at extremely high energies by measuring the oscillatory pattern in the squeezed bispectrum, as has been recently attempted in both the CMB \cite{Sohn:2024xzd} and large scale structure \cite{Cabass:2024wob} data. 

\vskip4pt
However, for heavy fields with masses much above the Hubble scale, the distinctive collider signals are exponentially suppressed by a Boltzmann factor $e^{-\pi m/H}$, as the spontaneous production of particles becomes more unlikely in the dS background with heavier masses.
This factor significantly narrows the detection window of the cosmological collider proposal and thus undermines its promises.
If one wants to boost the heavy-mass signals for phenomenological interests, additional mechanisms need to be engineered, such as the chemical potential proposal \cite{Chen:2018xck,Wang:2019gbi, Bodas:2020yho, Tong:2022cdz, Jazayeri:2023kji}  and the effective field theory (EFT) with small sound speeds \cite{Lee:2016vti,Pimentel:2022fsc,Jazayeri:2022kjy,Wang:2022eop}. 
Recently, another attempt has been made by introducing scale-dependent features such as oscillatory couplings at the phenomenological level \cite{Chen:2022vzh} (also see \cite{Qin:2023ejc,Werth:2023pfl,Pinol:2023oux} for relevant studies and other types of scale dependences for cosmological colliders in \cite{Wang:2019gok,Aoki:2020wzu,McCulloch:2024hiz}). Interestingly, due to the resonance effects between the massive field and oscillating background, there may be significant particle production and enhanced collider signals in the scalar bispectrum.
This proposal, dubbed the classical cosmological collider, can be closely related to the discrete shift symmetry of the inflaton. 

\vskip4pt
Based on these recent developments, in this work we perform a bootstrap analysis
of cosmological collider with resonant features. In particular, we will apply the
boundary differential equations to compute the full analytical shapes of the primordial bispectrum and analyse the possible resonant enhancements.
There are three types of oscillations in our consideration: the Bunch-Davies vacuum $e^{ik\eta}$, the massive field $e^{imt}$ and the coupling $\cos(\omega t)$.
Because of the resonance, we show that the original boundary differential equations from the full dS isometries are revised and the frequency of the coupling oscillation $\omega$ becomes an important new parameter in the analysis.
For cosmological collider, as expected in Ref.~\cite{Chen:2022vzh},
we find that the resonance
effects can significantly boost the distinctive signal of heavy fields and overcome the Boltzmann suppression, as long as $\omega$ is larger than the massive of the heavy field, i.e. $\omega\gtrsim m \gg H$.
Furthermore, this leads to new shapes of the primordial bispectrum with more sophisticated oscillatory patterns, combining both the
resonant feature and collider signal.

\vskip4pt
One particularly interesting application of the bootstrap computation is to re-examine the UV sensitivity of axion monodromy inflation \cite{Pajer:2024ckd}. 
This class of models, commonly seen as one of the most
successful examples in string embeddings of inflation \cite{Silverstein:2008sg,McAllister:2008hb,Flauger:2009ab,Berg:2009tg,Kaloper:2011jz}, 
enjoy the discrete shift symmetry of an axion field.
Here we consider a two-field system with a light axion and a heavy modulus, which is generically expected in UV-completions.
Naively, moduli fields decouple, and thus the low-energy effective theory reduces to single field inflation. 
In contrast,
we demonstrate that, with no further changes to this minimal setup, the periodic modulation of the axion potential would continuously excite the moduli that were thought to be stabilised.\footnote{Here we note that in the multi-field language, this minimal extension of single field axion monodromy has a very small turning rate, and thus does {\it not} belong to the rapid/sharp-turn scenarios. The key mechanism here is to maintain a small turning rate but with high-frequency oscillations, such that resonance occurs and boosts the heavy particle production.}
As a result, the heavy moduli can no longer be integrated out, and instead they may lead to unsuppressed collider signals through resonance. 
We show that this simple setup just corresponds to a concrete realization of cosmological collider with resonant features.
Thus our finding has immediate implications for string inflation scenarios. On the one hand, we need to be more careful with the regime of validity of the single-field description. On the other hand, when the single-field EFT breaks down, a new type of UV-sensitive signatures emerge in cosmological correlators, which offers a remarkable avenue for probing high energy physics much above the Hubble scale during inflation.

\vskip4pt
The rest of the paper is organized as follows. 
In Section \ref{sec:res}, we introduce the basic setup for bootstrap computation and review the resonance effects during inflation through a simple example of contact interaction with an oscillatory coupling. 
In Section \ref{sec:exc}, we perform the bootstrap computation for massive-exchange bispectra with resonances. We first introduce the mixed propagator from the quadratic interaction between the inflaton and massive field, and then solve for the three-point scalar seed functions with oscillatory couplings.
Section \ref{sec:collider} investigates the phenomenological implications of the bootstrap result, especially for the corrections to the power spectrum and the leading bispectrum. In the bispectrum, we focus on the resonance enhancement and the new shapes with both scale-dependent features and scale-invariant collider signals.
In Section \ref{sec:axionM}, we turn to study the UV sensitivity of axion monodromy by looking into a string-inspired two-field extension of the original single field model. Through a detailed analysis, we elucidate the regime of validity for single field axion monodromy, and highlight the new phenomenology of resonant cosmological collider.
This part of the results has been summarized in a short companion letter \cite{Pajer:2024ckd} from the UV-completion perspective of inflation. Here we provide more details on the multi-field analysis and also elaborate on the connection with the bootstrap computation of the resonant cosmological collider.
We conclude in Section \ref{sec:concl} with an outlook for future work.
 Throughout the paper, we
 take the convention of natural units $c = \hbar = 1$, with the metric signature $(-,+,+,+)$ and the reduced Planck mass $\mpl^2=1/8\pi G$.

\section{\boldmath Physics of Resonances during Inflation}
\label{sec:res}
In this section, we start with a simple example to illustrate the resonance effects during inflation.
After introducing the basic setup,
we shall focus on the interplay among three oscillatory behaviours: the Bunch-Davies vacuum, the massive field oscillation, and the periodically varying interaction.
We take a close look at the contact three-point function $\langle\varphi\varphi\sigma\rangle$ with an oscillatory coupling. We first analyse the resonance effects by explicitly solving the bulk integral and then revisit the problem from a boundary perspective. 

\subsection{Generalities}

Let's first briefly review the Feynman rules for computing cosmological correlators and resonance behaviours in single field inflation and cosmological collider.
Readers who are familiar with these topics may skip this part and directly move on to Section \ref{sec:contact}.

\vskip4pt
We use the de Sitter (dS) space as an approximation of the inflationary spacetime 
\be
ds^2=a(\eta)^2{(-d\eta^2 + d{\bf x}^2)}~,~~~~~~a(\eta)=- \frac{1}{H\eta}
\ee
with $H$ being the Hubble parameter. We use the conformal time $\eta$ in most of the analysis, and it is related to the cosmic time $t$ through the simple expression $t=-H^{-1}\ln(-H\eta)$. On this fixed background, we consider quantum fields fluctuations. 
The free theory of canonical scalars is given by the action
\be
S_2 = \int d \eta d^3{\bf x} a^4\[ \frac{1}{2a^2} (\partial_\eta\sigma)^2-\frac{1}{2a^2}  (\partial_i\sigma)^2 - \frac{1}{2} m^2 \sigma^2 \] ~,
\ee
where $m$ is the mass parameter.  In the Fourier space,   the mode function $\sigma_k(\eta)$ satisfies
the equation of motion  
\be \label{sigmaeom}
\(\mathcal{O}_\eta +{m^2}/{H^2} \) \sigma_k ( \eta) = 0 ,
~~~~~~{\rm with}~~
\mathcal{O}_\eta \equiv  \eta^2 \partial^2_\eta - 2\eta \partial_\eta + k^2\eta^2~.
\ee
After canonical quantization,  the $\s$ mode function can be solved by imposing the vacuum choice $\s_k(\eta) =i H\eta {e^{-ik\eta}}/{\sqrt{2k}}$ in the early time limit ($\eta\rightarrow-\infty$)
\be \label{sigmak}
 \sigma_k(\eta) = -i \frac{H\sqrt{\pi}}{2} e^{i\pi/4}e^{-\pi\mu/2}
 (-\eta)^{3/2}H^{(1)}_{i\mu}(-k\eta)  
\ee
with 
$\mu=\sqrt{{m^2}/{H^2}-{9}/{4}}$.
We are mainly interested in the principal series $m>3H/2$ where $\mu$ has a real positive value.
The late-time limit ($\eta\rightarrow 0$)  of the massive scalar behaves as 
\be \label{scaling}
\lim_{\eta\rightarrow 0} \s({\bf k},\eta)= -i \frac{H\sqrt{\pi}}{2} e^{i\pi/4}e^{-\pi\mu/2} (-\eta)^{3/2} \[ \frac{1+\coth(\pi\mu)}{\Gamma(i\mu+1)} \(\frac{-k\eta}{2}\)^{i\mu} - i\frac{\Gamma(i\mu)}{\pi} \(\frac{-k\eta}{2}\)^{-i\mu} \]~.
\ee
This expression is already quite informative. The two terms in the bracket correspond to the two $e^{-imt}$ and $e^{imt}$ massive oscillations, and they have  different weight factors. In the large mass limit $\mu\gg 1$, we see the second term is suppressed by $e^{-\pi\mu}$ while the first is not. This is because the positive frequency part $e^{imt}$ can be interpreted as particle production in the expanding background, which is suppressed by a Boltzmann factor.

In the subsequent  computation, two specific fields of interest are the massless scalar $\phi$   and the conformally coupled scalar $\vp$.
Their mode functions are explicitly given by
\bea 
\phi_{k}(\eta) &=&\frac{H}{\sqrt{2k^3}}(1+i  k\eta)e^{-ik\eta}
~,~~~~~~~~m^2=0 \\ 
\vp_{k}(\eta) &=&i \frac{H\eta}{\sqrt{2k}}e^{-ik\eta}~, ~~~~~~~~~~~~~~~~~~~~m^2=2H^2 .
\eea
In this paper, they serve as external lines in Feynman diagrams, while the $\sigma$ field is taken to be internal lines of the exchanged massive particle.

\vskip4pt
For computing cosmological correlators, the standard approach is the Schwinger-Keldysh or in-in formalism~\cite{Maldacena:2002vr, Weinberg:2005vy}.  
In this computation, we use two types of propagators for the quantum field fluctuations in dS.
The {\it bulk-to-boundary propagators} are 
associated with the external lines.
For the massless scalar $\phi$, they are given by
\be
K_+(k,\eta) =  \phi_{k}(\eta_0) \phi^*_{k}(\eta)~, ~~~~~~~~
K_-(k,\eta) =  \phi^*_{k}(\eta_0) \phi_{k}(\eta)~,
\ee
where we use $+$ and $-$ for the time-ordered and anti-time-ordered pieces in the integration contour.
For the conformally coupled scalar, $K_\pm^\vp(k,\eta) $ are simply given by the same form but with the $\vp_k(\eta)$ mode function. 
Another type of propagator is the {\it bulk-to-bulk} one. For the massive field $\sigma$ that appears in the internal lines, these propagators are given by
\bea
G_{++}^\sigma(k,\eta, \eta') &=& \sigma_k(\eta) \sigma^*_k(\eta') \Theta(\eta - \eta') +
\sigma^*_k(\eta) \sigma_k(\eta') \Theta(\eta' - \eta) \nn\\ 
G_{+-}^\sigma(k,\eta, \eta') &=&  
\sigma^*_k(\eta) \sigma_k(\eta')  \nn\\ 
G_{-+}^\sigma(k,\eta, \eta') &=&  
\sigma_k(\eta) \sigma^*_k(\eta') \nn\\ 
G_{--}^\sigma(k,\eta, \eta') &=& \sigma_k(\eta) \sigma^*_k(\eta') \Theta(\eta' - \eta) +
\sigma^*_k(\eta) \sigma_k(\eta') \Theta(\eta - \eta')~,
\eea
where $\Theta$ is the Heaviside function. 
From the equation of motion \eqref{sigmaeom}, the $++$ and $--$ propagators satisfy the following inhomogeneous equation
\be \label{G-diff-eq}
\(\mathcal{O}_\eta +{m^2}/{H^2}\)G_{\pm\pm}^\sigma(k,\eta, \eta') = \mp i H^2 \eta^2\eta'^2 \delta(\eta-\eta') ~.
\ee
With this set of propagators, now it is straightforward to derive the Feynman rules for the computation of cosmological correlators. 
This is also called the bulk perspective.
For contact interactions,  we just need the bulk-to-boundary propagators for each external line, and there is one time integral capturing interactions from the deep sub-horizon regime to the end of inflation.
For exchange diagrams, we associate each internal line with bulk-to-bulk propagators, and then perform multiple nested time integrals, which in general are very difficult.

\vskip4pt

As a warmup, we follow  this bulk perspective and briefly summarize the resonance effects in two typical scenarios of cosmic inflation:

\begin{itemize}
    \item {\it Resonance in Single Field Inflation}. The simplest example is given by oscillatory couplings in single field inflation, which generate the so-called resonant non-Gaussianity. This can be achieved through small periodic modulations on the slow-roll potential Ref.\cite{Chen:2008wn,Flauger:2010ja,Chen:2010bka}. The coupling is supposed to take the following form
    \begin{equation}
    \lambda(t) =\lambda_{\alpha} \cos(\omega t+\theta_f) =\frac{\lambda_{\alpha}}{2}\[\left(\frac{\eta}{\eta_1}\right)^{i\alpha}+ \left(\frac{\eta}{\eta_1}\right)^{-i\alpha}\]
    \label{2.1E2},
\end{equation}
where we change from the cosmic time to conformal time and define $\eta_1 \equiv - \mathrm{exp}\left(-{\theta_f}/{\alpha}\right)$.
Then one master integral that we encounter in the computation of resonant non-Gaussianity has the following form \cite{DuasoPueyo:2023viy}
\be
\int_{-\infty}^0 d\eta \eta^{m-1} \[(-\eta)^{i\alpha}+ (-\eta)^{-i\alpha}\] e^{ik_T\eta} \simeq -e^{\pi\alpha/2} \frac{i^m}{k_T^{m+i\alpha}} \Gamma(m+i\alpha) \[ 1+ \mathcal{O}(e^{-\pi\alpha}) \]~,
\ee
where $m$ is an integer, $k_T=k_1+k_2+k_3$ for the bispectrum and we have used $\alpha\gg 1$ in the second step.
This corresponds to the resonance between the Bunch-Davies vacuum of inflaton fluctuations $e^{ik\eta}$ and the positive frequency part $(-\eta)^{i\alpha}$ of the oscillatory coupling.
Using the large-$\alpha$ limit of the Gamma function $\Gamma(m\pm i\alpha) \sim \alpha^{m-1/2} e^{-\pi\alpha/2} $, we find there can be enhancement from the resonance effects.  
A full bootstrap analysis from the boundary perspective for this type of correlators is performed in \cite{DuasoPueyo:2023viy}.

\item {\it Resonance in Cosmological Collider}. The oscillations of massive fields during inflation $e^{\pm i m t}$ can also resonate with the inflaton fluctuations, which leads to the cosmological collider signals. This perspective has been elaborated in \cite{Chen:2015lza,Tong:2021wai}. It is informative to look at the three-point function from the contact $\vp^2\s$ interaction. Then one typical integral is given by
\bea \label{collider}
&&\int_{-\infty}^0  {d\eta}  a(\eta)^4 K_+^\vp(k_1,\eta)K_+^\vp(k_2,\eta) K_+^\s(k_3,\eta) \sim
\int_{-\infty}^0 \frac{d\eta}{\eta^2} e^{ik_{12}\eta} \s_{k_3}^*(\eta) \\
&& ~~ \xrightarrow{{k_3}/{k_1}\ll  {1}/{\sqrt{\mu}}\ll 1} 
 \frac{ iH}{\sqrt{2\mu}}   \int_{-\infty}^0d\eta \frac{ e^{ik_{12}\eta}}{(-\eta)^{1/2}}  \[  \frac{i}{2} \(\frac{-k_3\eta}{2}\)^{-i\mu} -e^{-\pi\mu} \(\frac{-k_3\eta}{2}\)^{i\mu} \]\nn\\
 &&  ~~ =  \frac{ iH}{\sqrt{2\mu}} e^{ -\pi\mu/2  -i\pi/4} \[\frac{i}{2} \Gamma\(\frac{1}{2}-i\mu\) \(\frac{k_3}{2k_{12}}\)^{-i\mu}- \Gamma\(\frac{1}{2}+i\mu\) \(\frac{k_3}{2k_{12}}\)^{i\mu} \]~, \nn
\eea
with $k_{12}=k_1+k_2$.
In the second line, we have taken the heavy mass and squeezed limits, and used the late-time asymptotic behaviour of $\s$ field in \eqref{scaling}.
From this computation, we see resonance happens between the Bunch-Davies mode function $e^{ik_{12}\eta}$ and the positive frequency massive oscillation $(-\eta)^{i\mu}$, which leads to a $e^{\pi\mu/2}$ factor. Meanwhile the negative frequency term  $(-\eta)^{-i\mu}$ does not resonate and generates a $e^{-\pi\mu/2}$ suppression. But since the  $(-\eta)^{i\mu}$ component of the $\sigma$ field has an extra suppression $e^{-\pi\mu}$,  the two contributions are comparable in the final result, both suppressed by the Boltzmann factor $e^{-\pi\mu}$. This is basically the origin of the generic exponential suppression of cosmological collider  with heavy masses: while the resonance does happen, its effect is to partially alleviate the suppressed signals. 
The full analytical form of the correlator from this simple example has been derived by using boundary conformal Ward identities in \cite{Arkani-Hamed:2015bza}.

\end{itemize}

\subsection{A Contact Example: $\langle\varphi^2\sigma\rangle$}
\label{sec:contact}

Now we move to the main focus of this paper --  cosmological collider physics with oscillatory couplings. For simplicity, let's first illustrate the physics of resonance using the contact interaction $\varphi^2\s$ but with a time-dependent coupling  given by \eqref{2.1E2}. Then in this system, there are three types of oscillations during inflation: Bunch-Davies vacuum, heavy field mass, and the coupling 
\be \label{3osci}
\vp\sim e^{\pm ik\eta}~,~~~~~~~~~~\s\sim e^{\pm imt}~,~~~~~~~~~~\lambda\sim (-\eta)^{\pm i \alpha}~.
\ee
In the following, we shall see how the oscillations in $\lambda$ will change the resonance behaviour in \eqref{collider} and bypass the usual Boltzmann suppression.


The contact three-point function $\langle\varphi\varphi\sigma \rangle$ is not observable as it vanishes at the future boundary of dS spacetime, but it provides helpful insight into the resonance effect among three types of oscillations in \eqref{3osci}.  Explicitly it is given by
\begin{align}
\langle\varphi_{\bold{k_1}}\varphi_{\bold{k_2}}\sigma_{\bold{k_3}}\rangle' &= i\int_{-\infty}^{0} d\eta a(\eta)^4\lambda(\eta)\left[K_+^{\varphi}(k_1, \eta)K_+^{\varphi}(k_2, \eta)K_+^{\sigma}(k_3, \eta)-c.c.\right]+\mathrm{perms} \nonumber
    \\&=\frac{iH\eta_0^2\sigma_{k_3}(\eta_0)}{8k_1 k_2 k_3^{\frac{3}{2}}} \left[k_3^{1-i\alpha}\hat{{I}}_a(k_3/k_{12})+k_3^{1+i\alpha}\hat{{I}}_b(k_3/k_{12})\right]+c.c.+\mathrm{perms}.
    \label{2.2E3}
\end{align}
where the two integrals are  functions  of the momentum ratio $u\equiv k_3/k_{12}$  only. We define $\hat{I}_a$ as
\be
\hat{I}_a (u)   \equiv \frac{1}{k_3^{1 - i\alpha}}\int_{-\infty}^{0}\frac{d\eta}{\eta^2} \left(\frac{\eta}{\eta_1}\right)^{ i\alpha} \hat{\sigma}^*(k_3\eta)e^{ik_{12}\eta}
\label{Ia}
\ee
with $\hat{\sigma}^*(k\eta)\equiv {k^{3/2}}\sigma^*_k(\eta)/H$. Then $\hat{I}_b$ is given by replacing $\alpha \rightarrow -\alpha$. 
They are  simple extensions of \eqref{collider} with  oscillatory couplings, which
 can be solved analytically \cite{Arkani-Hamed:2018kmz} (also see Appendix B in \cite{Pimentel:2022fsc} for more details). The result for $ \hat{I}_a$ is given by
\bea \label{Isol}
&&\hat{I}_a = \frac{ie^{\frac{\pi\alpha}{2}}}{\sqrt{\pi}}\left(-\frac{1}{\eta_1}\right)^{i\alpha}\sum_{\pm}\frac{\Gamma\left(\frac{1}{2}\pm i\mu+i\alpha\right)\Gamma(\mp i\mu)}{2^{\pm i\mu}}u^{\frac{1}{2}\pm i\mu+i\alpha}\nonumber\\
&&~~~~~~~~~~~~~~~~~~\times{}_2F_1\left[\frac{1}{4}\pm\frac{i\mu}{2}+\frac{i\alpha}{2},\frac{3}{4}\pm\frac{i\mu}{2}+\frac{i\alpha}{2};1\pm i\mu; u^2\right]~,
\eea
where we have used the two identities of hypergeometric functions in \eqref{hyperg1} and \eqref{hyperg2} to reduce the expression.
This integral contains the resonance between the positive frequency oscillation $(-\eta)^{i\alpha}$ of the coupling and the Bunch-Davies mode function. The resonance with the heavy field is more easily seen
in the squeezed limit $u\rightarrow0$, where the hypergeometric functions go to unity. Then we can estimate the size of the prefactor in the heavy field limit $\mu\gg 1$:  
\begin{align}
e^{{\pi\alpha}/{2}}\Gamma\left(\frac{1}{2}- i\mu+i\alpha\right)\Gamma(+ i\mu) \rightarrow  
\begin{cases}
   & {\rm const.}~~~~~~{\rm for}~~ \alpha \gtrsim \mu \\
   & e^{-\pi(\mu-\alpha)}~~~~~~{\rm for} ~~\alpha < \mu 
\end{cases}~.
\end{align}
For  $\alpha=0$, this prefactor gives us the usual Boltzmann suppression; while it is no longer suppressed when the oscillating frequency becomes larger than the heavy field mass.
A similar analysis can be applied to another integral $\hat{I}_b$, where the negative frequency part of $\lambda(\eta)$ does not resonate with the inflaton fluctuations, and thus no enhancement to overcome the Boltzmann suppression.

\subsection{A Boundary Approach}
\label{sec:bdry}
Although the above contact example can be directly computed by solving the bulk integrals, for the later convenience of the bootstrap analysis, it is helpful to re-investigate the problem from a different perspective based on boundary differential equations.

The starting point is to notice that from \eqref{sigmaeom} the massive field satisfies the following homogeneous  equations
\be
    \left(\mathcal{O}_{k}+\frac{m^2}{H^2}\right)\hat\sigma^*_k(\eta) = 0,~~~~{\rm with}~ \mathcal{O}_{k} \equiv k^2\partial^2_{k}-2k\partial_{k}+k^2\eta^2~,
    \label{2.3E5}
\ee
where we have traded time derivatives with $k$-derivatives.
Again, we take $\hat{I}_a$ for demonstration, while the analysis for  $\hat{I}_b$ can be performed by changing $\alpha\rightarrow-\alpha$.
Using the above equation, we derive the boundary differential equation satisfied by $\hat{I}_a$
\be
\(\Delta_u^\alpha + \mu^2+ \frac{1}{4} \pm i \alpha -\alpha^2 \)\hat{I}_a(u) = 0~,~~~~ {\rm with} ~~ \Delta_u^\alpha =u^2(1-u^2)\partial_u^2-2u(u^2+ i\alpha)\partial_u~.
\ee
Thus for computing correlators, this transforms the bulk time integration into a boundary description using differential equations in terms of the momentum ratio $u\equiv k_3/k_{12}$.
One can simply check that by setting $\alpha=0$, $\Delta_u^\alpha$ becomes the boundary differential operator $\Delta_u$ in the conventional cosmological bootstrap and the above equations reduce to the one from conformal Ward identities in \cite{Arkani-Hamed:2015bza}. 
As the oscillatory coupling explicitly breaks the dS dilation and boost symmetries, we are left with additional terms in the boundary  equations. 
This second order homogeneous differential equation has an analytical solution in terms of hypergeometric functions 
\begin{equation}
    \hat{I}_a (u) = \sum_{\pm}C_{\pm}u^{\frac{1}{2}\pm i\mu+i\alpha}{}_2F_1\left[\frac{1}{4}(1\pm 2i\mu+2i\alpha), \frac{1}{4}(3\pm 2i\mu +2i\alpha), 1\pm i\mu; u^2\right]
    \label{2.3E8}~,
\end{equation}
where the two free coefficients $C_{\pm}$ can be fixed by imposing boundary conditions.
The first condition from the choice of the Bunch-Davies vacuum requires that $\hat{I}_a$ at the folded limit $u\rightarrow1$ must be regular. The second condition corresponds to the normalization at the total-energy singularity $k_t=k_1+k_2+k_3\rightarrow 0$, or $u\rightarrow -1$. In this limit, the integral \eqref{Ia} picks up contributions from $\eta = -\infty$, which means that the mode function of the massive scalar appearing in the integrand can be approximated by its early-time asymptotic expression 
\begin{equation}
    \lim_{\eta\rightarrow -\infty}\hat{\sigma}^*(k\eta) = -\frac{ik\eta}{\sqrt{2}}e^{ik\eta}.
\end{equation}
Then we find the leading $k_t$-pole of the integral $\hat{I}_a$
\begin{align}
    \lim_{u\rightarrow-1}\hat{I}_a&=\frac{i}{\sqrt{2}}\left(-\frac{k_3}{\eta_1}\right)^{i\alpha}\int_{-\infty}^{0}d\eta \left(-\frac{1}{\eta}\right)^{1-i\alpha}e^{ik_t\eta} 
    \rightarrow  
    \frac{ie^{\frac{\pi\alpha}{2}}\Gamma(i\alpha)}{\sqrt{2}}\left(-\frac{1}{\eta_1}\right)^{i\alpha}\(\frac{k_3}{k_t}\)^{i\alpha}
    .
    \label{2.3E18}
\end{align}
By adding up the complex conjugate, the correlator with an oscillatory coupling has a singularity of the form $\langle \vp^2\sigma \rangle \propto \cos(\alpha\log(k_t))$. For a constant coupling ($\alpha=0$), the total-energy singularity takes a logarithmic form $\log(k_t)$. The oscillating $k_t$-pole is similar with the complex deformation observed in the bootstrap analysis of single-field resonant non-Gaussianity \cite{DuasoPueyo:2023viy}.
 Meanwhile, the residue of the pole carries information of the enhancement, as a consequence of the resonance effect at the subhorizon limit.
Imposing these two conditions at $u\rightarrow\pm1$ fixes the free coefficients to be
\be
C_\pm = \frac{ie^{\frac{\pi\alpha}{2}}}{\sqrt{\pi}}\left(-\frac{1}{\eta_1}\right)^{i\alpha} \frac{\Gamma\left(\frac{1}{2}\pm i\mu+i\alpha\right)\Gamma(\mp i\mu)}{2^{\pm i\mu}}~,
\ee
which matches the result \eqref{Isol} from the explicit bulk integration.

As a simple exercise, this boundary computation shows a way about how to perform the bootstrap analysis for resonance effects with massive field. 
Comparing with the standard approach of using conformal Ward identities for fully dS-invariant theories, we see that the main differences caused by the time-dependent couplings come at two places: i) the differential operator $\Delta_u^a$ is changed for nonzero $\alpha$; ii) the total-energy singularity takes an oscillatory form and provides an enhancement factor for the overall normalization. These lessons will help us with the following bootstrap computation for exchange diagrams with resonances.

\section{Massive Exchanges with Oscillatory Couplings}
\label{sec:exc}
In this section, we perform the bootstrap analysis for the three-point correlator with an intermediate massive state and oscillatory couplings. 
Following the methodology in \cite{Pimentel:2022fsc}, we start with introducing the mixed propagator from the $\dot\phi\s$ two-point vertex, and then construct the three-point scalar seed with two conformally coupled scalars, which can be fully solved by using the boundary differential equations. With this seed function as a building block, we derive the inflaton bispectra using  weight-shifting operators.

We note that the resonance enhancement for cosmological collider was discussed in \cite{Chen:2022vzh}, and some of the results have already been computed in \cite{Qin:2023ejc}. Here we make more explicit connections with phenomenology with a detailed bootstrap analysis.

\subsection{The Mixed Propagator}
\label{sec:mix}

Let's consider the simplest quadratic interaction $\dot{\phi}\sigma$ with a general time-dependent coupling $\lambda(\eta)$. Then the mixed bulk-to-boundary propagator can be defined as $  \mathcal{K}_{\pm}(k, \eta)\equiv \langle \s_\pm(\eta)\phi(\eta_0) \rangle$, which physically describes the conversion from one massive field $\sigma$ at bulk time $t$ to the massless inflaton at the boundary. The explicit expressions can be written as
\begin{equation}
    \mathcal{K}_{\pm}(k, \eta) = \pm i \int_{- \infty}^{0} d\eta'a(\eta')^3\lambda(\eta')[G_{\pm\pm}^{\sigma}(k, \eta, \eta')\partial_{\eta'}K_{\pm}(c_sk,\eta')-G_{\pm\mp}^{\sigma}(k, \eta, \eta')\partial_{\eta'}K_{\mp}(c_sk,\eta')] .
    \label{3.1E1}
\end{equation}
This integral has a complicated solution. For the purpose of deriving the late-time correlator, performing the explicit integration is an intermediate step that may not be necessary.  Instead, here we take a different approach 
by acting the differential operator $\mathcal{O}_{\eta}+m^2/H^2$ onto the mixed propagator. Then applying equation (\ref{G-diff-eq}), we find that the mixed propagator satisfies the following inhomogeneous differential equation:
\begin{equation}
    \left(\mathcal{O}_{\eta}+\frac{m^2}{H^2}\right)\mathcal{K}_{\pm}(k, \eta) = -\lambda(\eta)\frac{H\eta^2}{2k}e^{\pm ik\eta}.
    \label{3.1E2}
\end{equation}
As illustrated in Section \ref{sec:bdry}, we would like to achieve a boundary description for the bootstrap analysis.
When the coupling $\lambda$ is constant, we can introduce a dimensionless mixed propagator, which then allows us to change the $\eta$ derivatives to the $k$ derivatives in the above equation. 
For a time-dependent $\lambda$, in general we cannot rescale $\mathcal{K}_{\pm}$ to become a combination of $k\eta$, which then forbids the trading of $\eta$ derivatives with $k$ derivatives. 

This technical challenge can be addressed in different ways.
For the special case of an oscillatory $\lambda(\eta)$, we could follow the strategy in Section \ref{sec:bdry}: we write the coupling as a sum of two power-law terms (like \eqref{2.1E2}), and then separate the mixed propagator into two parts such that we can derive their bootstrap equations respectively. 
Here we propose another method which can be generalized to deal with arbitrary time-dependent couplings. 
The trick is to introduce a new ``constants" $x_1\equiv k\eta_1$, and then we can rewrite the coupling as $\lambda(\eta) = \lambda(k\eta / x_1)$. 
In this way the ``constant"  $x_1$  absorbs the remnant $k$-dependence. Then we keep it as a constant when deriving and solving the boundary differential equations. Once obtaining the full solution, we restore the $k$-dependence by changing it back using $x_1 = k\eta_1$. This method was first introduced by \cite{Wang:2022eop} to deal with the extra scale-dependence introduced by a late-time cutoff. 

In the case of oscillatory interactions, by defining $x_1 = k\eta_1$ (and setting the  constant piece to unity), we rewrite the coupling as 
\begin{equation}
    \lambda(k\eta, x_1) = \cos\left[\alpha \log\left(\frac{k\eta}{x_1}\right)\right]~.
    \label{3.1E3}
\end{equation}
Now we are allowed to introduce a dimensionless mixed propagator as
\begin{equation}
    \hat{\mathcal{K}}_{\pm}(k\eta)\equiv (k^3/H)\mathcal{K}_{\pm}(k, \eta, x_1)~,
    \label{dml_mppg}
\end{equation}
which satisfies the following differential equation
\begin{equation}
    \left(\mathcal{O}_{\eta}+\frac{m^2}{H^2}\right)\hat{\mathcal{K}}_{\pm}(k\eta, x_1) =\left(\mathcal{O}_{k}+\frac{m^2}{H^2}\right)\hat{\mathcal{K}}_{\pm}(k\eta, x_1) = -\frac{1}{2}\lambda(k\eta, x_1)k^2\eta^2e^{\pm ik\eta}.
    \label{3.1E4}
\end{equation}
Taking the soft-limit ($k \rightarrow 0$), or equivalently, the late-time limit ($\eta \rightarrow 0$), the mixed propagator can be explicitly computed as 
\begin{equation}
\lim_{k\rightarrow0}\hat{\mathcal{K}}_+(k\eta, x_1) = \sum_{\pm}A_{\pm}(x_1)\left(-\frac{k\eta}{2}\right)^{\frac{3}{2}\pm i\mu},
\label{3.1E7}
\end{equation}
where
\begin{align}
    A_{\pm}(x_1) = &\left(\frac{1}{2}\right)^{2+i\alpha}\frac{\sqrt{\pi}}{\sinh{\pi\mu}}\frac{\Gamma(\frac{1}{2}+i\alpha-i\mu)\Gamma(\frac{1}{2}+i\alpha+i\mu)}{\Gamma(1+i\alpha)\Gamma(1\pm i\mu)}\left(-\frac{1}{x_1}\right)^{i\alpha}\\[5pt]
    & \times
    \left[e^{\mp i\frac{\pi}{4}+\frac{\pi\mu}{2}\pm\frac{\pi\alpha}{2}}+e^{\mp i\frac{3\pi}{4}-\frac{\pi\mu}{2}\mp\frac{\pi\alpha}{2}}\right]\nonumber+\alpha\rightarrow -\alpha.
    \label{3.1E8}
\end{align}
We notice that $A_- = A_+^*$, so in the soft-limit, $\hat{\mathcal{K}}_+(k\eta, x_1) =\hat{\mathcal{K}}_-(k\eta, x_1) \equiv \hat{\mathcal{K}}(k\eta, x_1)$ is a real function. 
More importantly, in the $\eta\rightarrow0$ limit, the $\eta'$ integral becomes similar with the one in \eqref{Ia} and then the enhancement is manifested in these two coefficients. 
For instance, the leading contribution to $A_+$ comes from the resonance among the massless oscillation $\phi^*\sim e^{ik\eta}$, the positive frequency oscillatory coupling $\lambda\sim(-\eta)^{i\alpha}$ and the massive mode $\sigma^*$. However, depending on the specific values of $\mu$ and $\alpha$, these resonances occur in different regimes and exhibit distinct behaviors \cite{Chen:2022vzh}. When $\alpha\geq\mu\gg1$, the integral (\ref{3.1E1}) is dominated by stationary phase contributions, which are suppressed by $\alpha^{-\frac{1}{2}}\mu^{-\frac{1}{2}}$, instead of a Boltzmann factor. 
In contrast, for $\mu > \alpha \gg1$, stationary phases are absent, and the leading contribution comes from the late-time part of the integral. In this regime, the massive mode can be approximated by its asymptotic expression $\sigma^* \sim (-\eta)^{-i\mu}+e^{-\pi\mu}(-\eta)^{i\mu}$. Consequently,
neglecting any power-law dependence, $A_+$ reduces to a softened Boltzmann factor $e^{-\pi(\mu-\alpha)}$, similar to what we have observed in the $\vp^2\sigma$ contact example.

\subsection{The Scalar Seed Function $\langle\varphi^2\phi\rangle$}

Using the mixed propagator with an oscillatory coupling, now we come to compute the three-point function of two conformally coupled scalars and one massless scalar, exchanging a massive particle.
Our analysis incorporate two different scenarios for the cubic vertex: one has a constant coupling; the other is oscillatory.
We shall derive the boundary differential equations for the corresponding scalar seed functions, and solve for their solutions for any kinematics.

\subsubsection{One Oscillatory Coupling}
When the cubic interaction $\varphi^2\sigma$ has a constant coupling (which we set to 1 for simplicity), the exchange bispectrum with the mixed propagator \eqref{3.1E1} can be expressed into the following form:
\begin{align}
\langle\varphi_{\bf{k_1}}\varphi_{\bf{k_2}}\phi_{\bf{k_3}}\rangle' &= i\int_{-\infty}^{0} d\eta a(\eta)^4\left[K_+^{\varphi}(k_1, \eta)K_+^{\varphi}(k_2, \eta)\mathcal{K}_+(k_3 \eta, x_1)-c.c\right]+\mathrm{perms.} \nonumber
    \\&=\frac{i
    H\eta_0^2}{4 k_1k_2k_3^2}\hat{\mathcal{I}}(k_{12}, k_3, x_1)+\mathrm{perms.},
\end{align}
where we define the three-point scalar seed as
\begin{equation}
    \hat{\mathcal{I}}(k_{12}, k_{3}, x_1) = \frac{1}{k_3}\int_{-\infty}^0\frac{d\eta}{\eta^2}\left[e^{ik_{12}\eta}\hat{\mathcal{K}}_+(k_3\eta,x_1)- e^{-ik_{12}\eta}\hat{\mathcal{K}}_-(k_3\eta,x_1)\right].
    \label{3.2E2}
\end{equation}
Here the dependency of the oscillation frequency $\alpha$ is absorbed into the dimensionless mixed propagator $\hat{\mathcal{K}}$.
By rescaling the integral variable to $k_3\eta$, it is easily seen that $\hat{\mathcal{I}}$ is a dimensionless function of $u \equiv k_3/k_{12}$ and $x_1$ only. Acting the differential operator $\mathcal{O}_{k_3}+{m^2}/{H^2}$ on both sides of equation (\ref{3.2E2}), 
we find the following boundary differential equation after some algebra
\begin{align}
\left[\Delta_u+\left(\mu^2+\frac{1}{4}\right)\right]\hat{\mathcal{I}}(u,x_1) = \frac{i}{4}(e^{-\frac{\pi\alpha}{2}}+e^{\frac{\pi\alpha}{2}})\sum_{\pm}\Gamma(1\pm i\alpha)\left(-\frac{1}{x_1}\right)^{\pm i\alpha}\left(\frac{u}{1+u}\right)^{1\pm i\alpha} ,
    \label{3.2E3}
\end{align}
with $\Delta_u \equiv (1-u^2)u^2\partial_u^2 - 2u^3\partial_u $.
This is a second-order inhomogeneous ordinary differential equation. The  source terms contain the positive- and negative-frequency parts, originating from the oscillating quadratic coupling. One can simply check that by setting $\alpha=0$, equation \eqref{3.2E3} turns to the featureless version for the primary scalar seed in \cite{Pimentel:2022fsc}.

To find its analytical solution, let's first look at the particular part, where we decompose it into two pieces, $\hat{\mathcal{S}}^+(u,x_1)$ and $\hat{\mathcal{S}}^-(u,x_1)$, corresponding to the positive- and negative-frequency components in the source term. Defined over the range $u\in [0,1]$, they can be expressed as series expansions around $u=0$ with the following ansatz
\begin{equation}
   \hat{\mathcal{S}}^{\pm}(u, x_1) = g^{\pm}(x_1)\sum_{n = 0}^{\infty}c_{n}^{\pm}u^{n+1\pm i\alpha},
   \label{3.2E6}
\end{equation}
where
\begin{equation}
    g^{\pm}(x_1)\equiv\frac{i}{4}(e^{-\frac{\pi\alpha}{2}}+e^{\frac{\pi\alpha}{2}})\Gamma(1\pm i\alpha)\left(-\frac{1}{x_1}\right)^{\pm i\alpha}.
    \label{3.2E7}
\end{equation}
Substituting (\ref{3.2E6}) into equation (\ref{3.2E3}), we find the recursive relation  
\begin{align}
    &c_0^{\pm} =\frac{\binom{\mp i\alpha-1}{0}}{\mu^2+(\frac{1}{2}\pm i\alpha)^2},\nonumber\\
    &c_1^{\pm} =\frac{\binom{\mp i\alpha-1}{1}}{\mu^2+(\frac{1}{2}+1 \pm i\alpha)^2},\nonumber\\
    &c_n^{\pm} = \frac{\binom{\mp i\alpha-1}{n}+(n \pm i\alpha)(n-1 \pm i\alpha)c_{n-2}^{\pm}}{(\frac{1}{2}+n\pm i\alpha)^2+\mu^2},
    \label{3.2E8}
\end{align}
where $\binom{m}{n}\equiv {m(m-1)(m-2)...(m-n+1)}/{n!}$ is the generalised binomial coefficient. 
The general formulae can be solved as
\begin{equation}
    c_n^{\pm} = \sum_{m = 0}^{\lfloor n/2 \rfloor}\frac{(n\pm i\alpha)!/(n-2m \pm i\alpha)!\binom{\mp i\alpha-1}{n-2m}}{\left[(n+\frac{1}{2}\pm i\alpha)^2+\mu^2\right]\left[(n-\frac{3}{2}\pm i\alpha)^2+\mu^2\right]...\left[(n+\frac{1}{2}-2m\pm i\alpha)^2+\mu^2\right]}.
    \label{3.2E9}
\end{equation}
Next, we derive the homogeneous solution $\hat{\mathcal{H}}$, which takes the form
\begin{equation}
    \hat{\mathcal{H}}(u, x_1) = \sum_{\pm}C_{\pm}(x_1)u^{\frac{1}{2}\pm i\mu}{}_2F_1\left[\frac{1}{4}\pm\frac{i\mu}{2},\frac{3}{4}\pm\frac{i\mu}{2};1\pm i\mu, u^2\right].
    \label{3.2E4}
\end{equation}
To fix the two free coefficients $C_{\pm}$, we impose the non-analytic behaviour of the primary scalar seed at $u\rightarrow 0$. More explicitly, we require that in the squeezed limit the oscillations of the homogeneous solution match the result from the bulk integral, which can be computed by substituting (\ref{3.1E7}) into (\ref{3.2E2}). This simply leads to
\begin{align}
    C_{\pm}(x_1) = & A_{\pm}(x_1)\left(\frac{1}{2}\right)^{\frac{3}{2}\pm i\mu}\Gamma\left(\frac{1}{2}\pm i\mu\right)\left(e^{-i\frac{\pi}{4}\pm\frac{\pi \mu}{2}}-e^{i\frac{\pi}{4}\mp\frac{\pi\mu}{2}}\right) 
    \label{3.2E5}
\end{align}
These two coefficients determine the size of the cosmological collider signals.
In cases without oscillatory couplings ($\alpha=0$), we find the conventional Boltzmann suppression factors for $\mu\gg 1$.
For $\mu \geq\alpha \gg 1$, $C_+$ yields a factor $e^{-\pi(\mu-\alpha)}$. As analysed in \cite{Chen:2022vzh}, this can be seen by tracing the bulk time evolution: at the linear mixing vertex, the massless oscillation $e^{ik_{3}\eta}$ resonates with the positive frequency oscillatory coupling $(-\eta)^{i\alpha}$, leading to the above softened Boltzmann suppression; 
at the cubic interaction vertex, the resonance between the conformally coupled scalar $\varphi^*\sim e^{ik_{12}\eta}$ and the massive oscillation $(-\eta)^{i\mu}$ gives a factor of $e^{\frac{\pi\mu}{2}}\Gamma(1/2+i \mu)$, which is of $\mathcal{O}(1)$. As a result, the overall factor, neglecting any power-law dependence on $\alpha$ and $\mu$, is $e^{-\pi(\mu-\alpha)}$.
{For $\alpha>\mu$, the power-law dependence on $\mu$ and $\alpha$ from the asymptotic expressions of $C_{\pm}$ becomes important, which can be explained by the stationary phase resonance \cite{Chen:2022vzh}.}
Moreover, the $(-1/x_1)^{\pm i\alpha}$ factors, which appear both in homogeneous and particular solutions, exhibit explicit scale dependence, distinguishing them from the scale-invariant cosmological collider signals from exchange diagrams without oscillatory coupling. Both the softened Boltzmann factor and the breakdown of the scale invariance from the bootstrap computation here confirm the result of the general analysis and the approximate computations performed in \cite{Chen:2022vzh}.

\subsubsection{Two Oscillatory Couplings}
\label{sec:2osc}

When both the quadratic and cubic interactions oscillate,
the exchange three-point functions present a more significant resonance enhancement and richer structures in the analytical expressions.
To keep the analysis general, we use the following parameterisation for two couplings $\lambda_{\mathrm{quad}} \dot{\phi}\sigma$ and $\lambda_{\mathrm{cub}}(\eta)\varphi^2\sigma$ with different oscillating frequencies and phases\footnote{In realistic models, it is more natural to have oscillations of these two couplings with the same frequency, as we shall see in a specific example in Section \ref{sec:axionM}. Here we allow $\alpha_1$ and $\alpha_2$ to be different with the goal of deriving the general result.}
\begin{equation}
    \lambda_{\mathrm{quad}}(\eta) = \cos\left[\alpha_1\log\left(\frac{\eta}{\eta_1}\right)\right] \equiv \cos\left[\alpha_1\log\left(\frac{k_3\eta}{x_1}\right)\right];
    \label{3.2E10}    
\end{equation}
\begin{equation}
    \lambda_{\mathrm{cub}}(\eta) = \cos\left[\alpha_2\log\left(\frac{\eta}{\eta_2}\right)\right] \equiv \cos\left[\alpha_2\log\left(\frac{k_3\eta}{x_2}\right)\right].
    \label{3.2E11}    
\end{equation} 
where we assumed the constant factors to be unity. 
For the cubic coupling, we have introduced another ``constant" $x_2$ to absorb the extra $k_3$-dependence.
Again,  the three-point correlator is given by
\begin{align}
\langle\varphi_{\bold{k_1}}\varphi_{\bold{k_2}}\phi_{\bold{k_3}}\rangle' &= i\int_{-\infty}^{0} d\eta a(\eta)^4\lambda_{\mathrm{cub}}(\eta)\left[K_+^{\varphi}(k_1, \eta)K_+^{\varphi}(k_2, \eta)\mathcal{K}_+(k_3 \eta, x_1)-c.c\right]+\mathrm{perms} \nonumber
    \\&=\frac{iH\eta_0^2}{4k_1k_2k_3^2}[\hat{\mathcal{I}}_a(k_{12}, k_3, x_1, x_2)+\hat{\mathcal{I}}_b(k_{12}, k_3, x_1, x_2)]+\mathrm{perms}.
    \label{3.2E12}
\end{align}
In this case, we have introduced two scalar seed functions $\hat{\mathcal{I}}_a$ and $\hat{\mathcal{I}}_b$ to reflect the positive- and negative-frequency components of $\lambda_{\rm cub}$ respectively. 
The one associated to the positive frequency part $\hat{\mathcal{I}}_a$ is defined as 
\begin{equation}
    \hat{\mathcal{I}}_a (u,x_1,x_2) \equiv \frac{k_3^{i\alpha_2-1}}{2(-x_2)^{i\alpha_2}}\int_{-\infty}^{0}d\eta (-\eta)^{i\alpha_2-2}\left[e^{ik_{12}\eta}\hat{\mathcal{K}}_+(k_3\eta, x_1)-c.c.\right] .
    \label{3.2E13}
\end{equation}
The boundary differential equation follows
\begin{align}
    \left[u^2(1-u^2)\partial^2_u-2u(u^2+i\alpha_2)\partial_u+\mu^2+\frac{1}{4}+i\alpha_2-\alpha_2^2\right]\hat{\mathcal{I}}_a
    = \sum_{\pm} g_{a}^{\pm}(x_1, x_2) \left(\frac{u}{1+u}\right)^{1\pm i\alpha_1+i\alpha_2} ,
    \label{3.2E15}
\end{align}
where
\begin{align}
    g_{a}^{\pm}(x_1, x_2)= \frac{i}{4}\Gamma(1\pm i\alpha_1+i\alpha_2)\left(-\frac{1}{x_1}\right)^{\pm i\alpha_1}\left(-\frac{1}{x_2}\right)^{i\alpha_2}\cosh\left[\frac{\pi}{2}(\pm \alpha_1+\alpha_2)\right] . 
    \label{3.2E20}
\end{align}
A derivation with full technical details is presented in Appendix \ref{A:Derive}. 
Compared with other boundary differential equations in the bootstrap computation, \eqref{3.2E15} has a more complicated structure. On the left-hand side, the differential operator takes the form of $\Delta_u^\alpha$ in Section \ref{sec:bdry}, which is determined by the cubic interaction. On the right-hand side, the source term contains oscillations from both vertices. 
The other scalar seed $\hat{\mathcal{I}}_b$, which corresponds to the negative frequency part of the cubic interaction, can be easily obtained by replacing $\alpha_2 \rightarrow -\alpha_2$.

\begin{figure}[t]
     \centering
    \begin{subfigure}{0.45\textwidth}
        \includegraphics[width=\columnwidth]{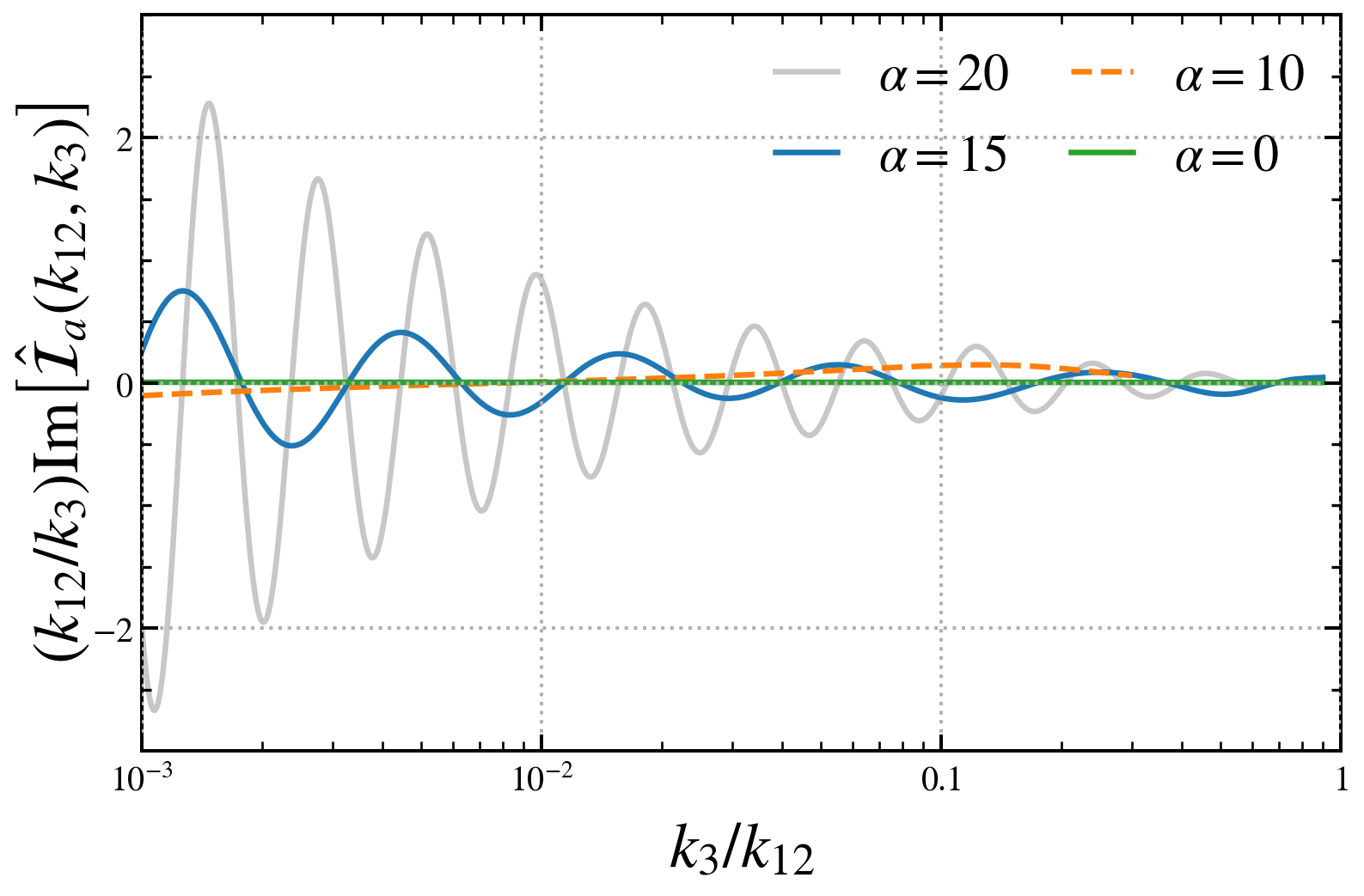}
        \caption{$k_1 = k_2 = 0.5~ \mathrm{Mpc}^{-1}$}
        \label{3.2F1.1}
    \end{subfigure}
    \begin{subfigure}{0.45\textwidth}
        \includegraphics[width=\columnwidth]{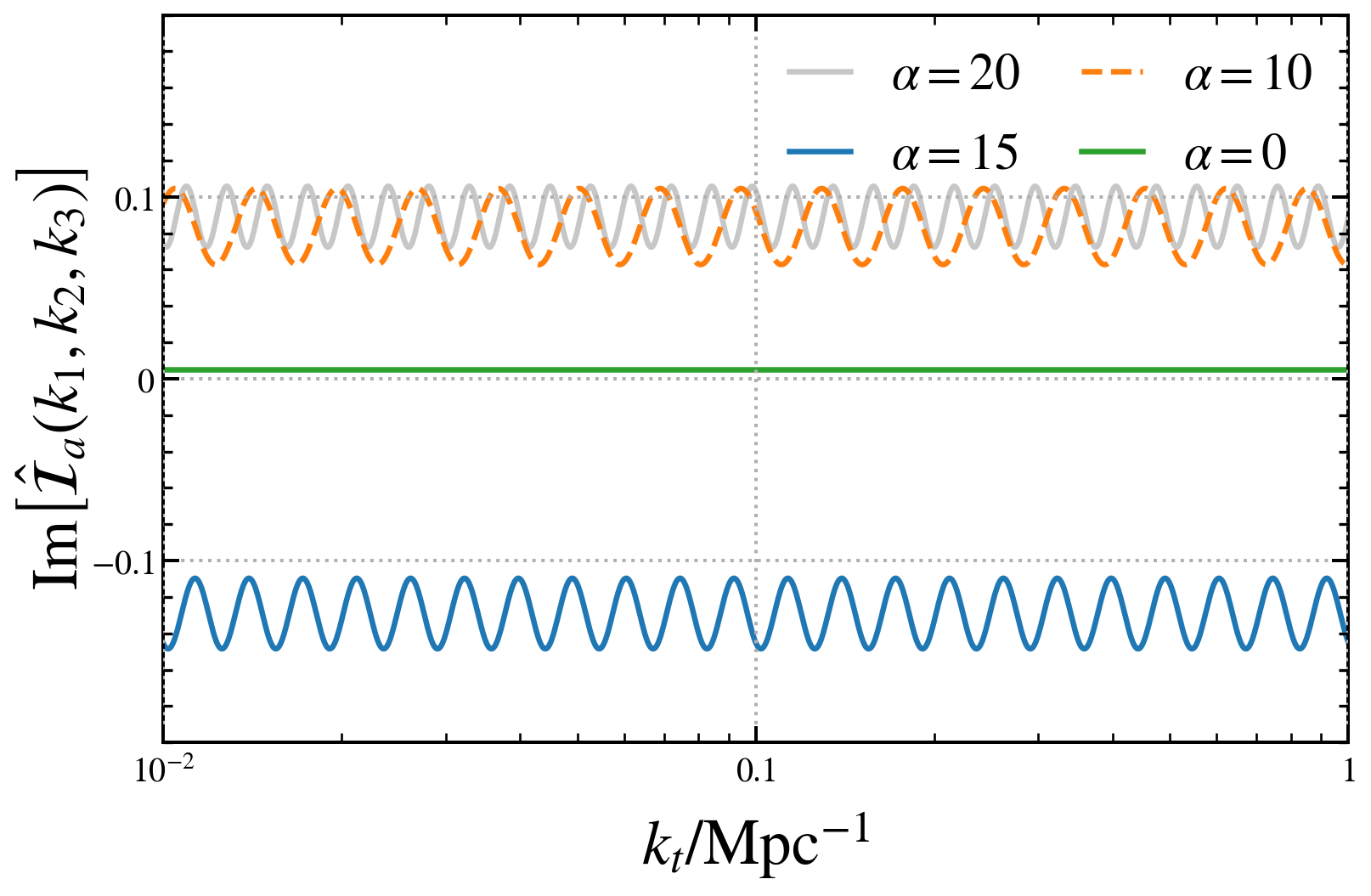}
        \caption{$k_t = k_1+k_2+k_3, ~ k_1=k_2=k_3$}
        \label{3.2F1.2}
    \end{subfigure}   
    \caption{The analytical solution of the positive-frequency primary scalar seed $\hat{\mathcal{I}}_{a}$, with $\mu = 10$ and $\alpha \equiv \alpha_1 = \alpha_2= 0, 15, 20$. The left plot illustrates the cosmological collider signals at the squeezed limit, at $k_1 = k_2 = 0.5 \mathrm{Mpc}^{-1}$, while the right plot highlights the scale dependence.}
    \label{3.2F1}
\end{figure}

\begin{figure}[t]
     \centering
    \begin{subfigure}{0.45\textwidth}
        \includegraphics[width=\columnwidth]{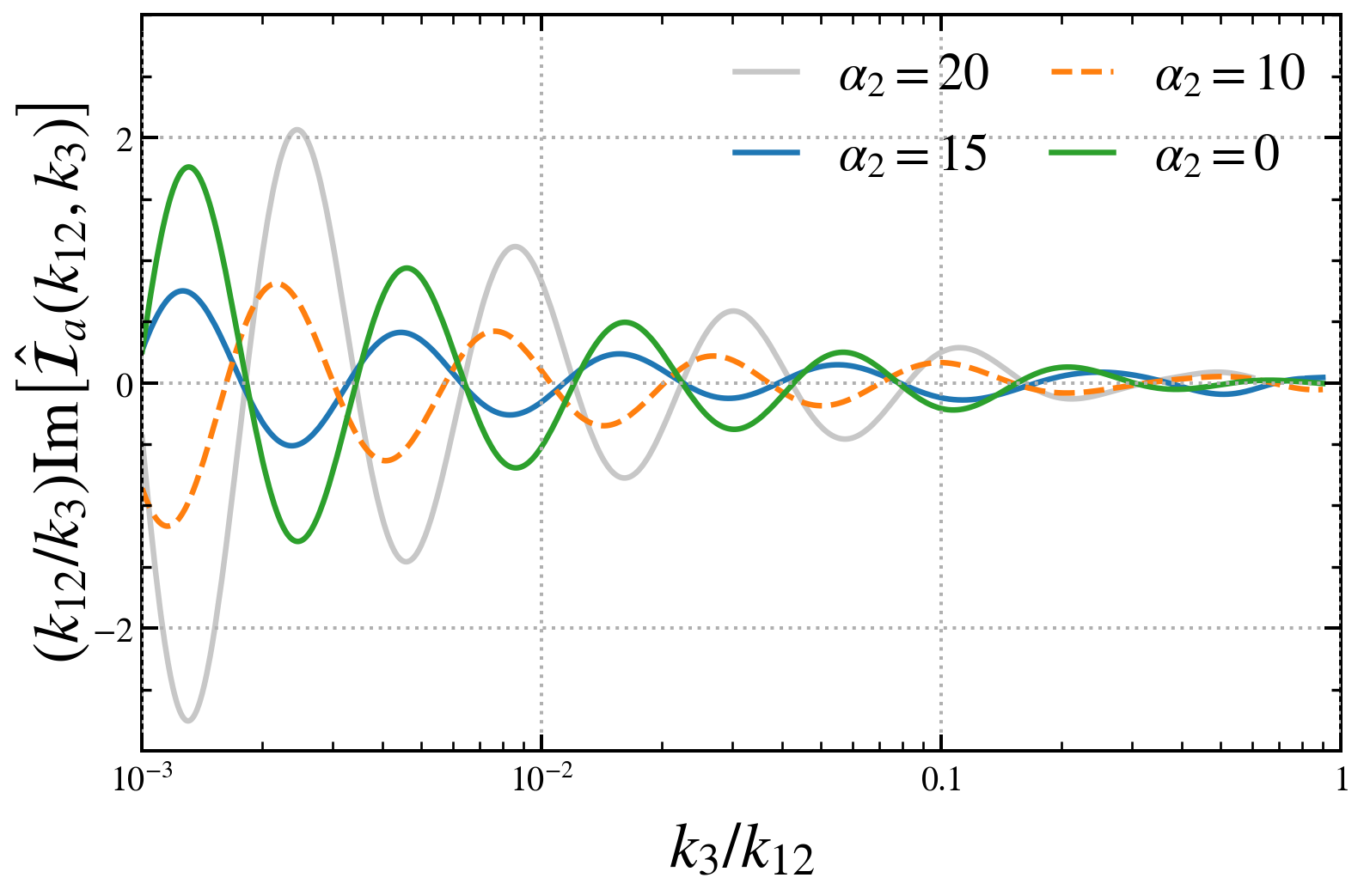}
        \caption{$k_1 = k_2 = 0.5~ \mathrm{Mpc}^{-1}$}
        \label{3.2F2.1}
    \end{subfigure}
    \begin{subfigure}{0.45\textwidth}
        \includegraphics[width=\columnwidth]{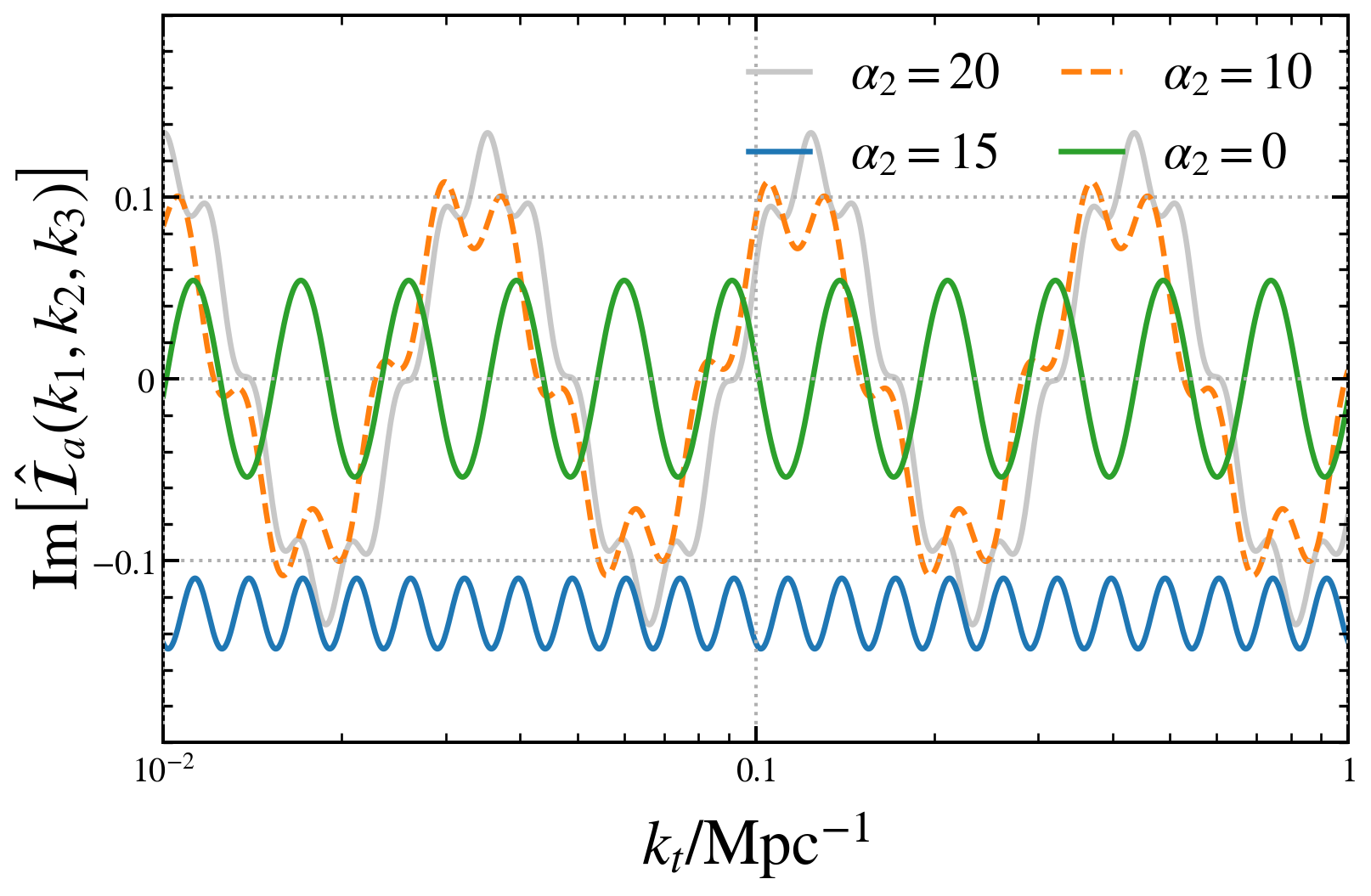}
        \caption{$k_t = k_1+k_2+k_3, ~ k_1=k_2=k_3$}
        \label{3.2F2.2}
    \end{subfigure}   
    \caption{(a) Shapes and (b) runnings of the scalar seed correlator, with $\mu = 10$, $\alpha_1 = 15$, $\alpha_2 = 0,10,15,20$. From the shapes plot, we observe that $\alpha_2$ does not enter the frequency of the cosmological collider signal, and only leads to the variation in phase. However, the oscillatory running of the bispectrum has an explicit dependence on $\alpha_2$, and a superposition-like behavior appears when $\alpha_1\neq \alpha_2$.}
    \label{3.2F2}
\end{figure}

Now let's solve equation \eqref{3.2E15}.
The particular solution is given by the ansatz
\begin{equation}
    \hat{\mathcal{S}}_a(u, x_1, x_2)=\sum_{\pm}g_{a}^{\pm}(x_1, x_2)\sum_{n=0}^\infty d_n^{\pm} u^{n+1\pm i\alpha_1+i\alpha_2},
    \label{3.2E19}
\end{equation}
Substituting (\ref{3.2E19}) into equation (\ref{3.2E15}), we find the general formula of the coefficients 
\begin{align}
    &d_n^{\pm} = \sum_{m = 0}^{\lfloor n/2 \rfloor}\frac{(n\pm i\alpha_1+i\alpha_2)!/(n-2m\pm i\alpha_1+i\alpha_2)!\binom{-1 \mp i\alpha_1-i\alpha_2}{n-2m}}{\left[(n+\frac{1}{2}\pm i\alpha_1)^2+\mu^2\right]\left[(n-\frac{3}{2}\pm i\alpha_1)^2+\mu^2\right]...\left[(n+\frac{1}{2}-2m\pm i\alpha_1)^2+\mu^2\right]}.\nonumber\\
    \label{3.2E21}
\end{align}
The homogeneous solution takes the form
\begin{equation}
    \hat{\mathcal{H}}_a(u, x_1, x_2) = \sum_{\pm}C_{a\pm}(x_1, x_2)u^{\frac{1}{2}\pm i\mu+i\alpha_2}{}_2F_1\left[\frac{1}{4}(1\pm 2i\mu+2i\alpha_2), \frac{1}{4}(3 \pm 2i\mu+2i\alpha_2), 1\pm i\mu; u^2\right],
    \label{3.2E17}
\end{equation}
where coefficients can be determined by taking the squeezed limit of \eqref{3.2E13}
\begin{align}
    C_{a\pm}(x_1,x_2) = & \left(\frac{1}{2}\right)^{\frac{9}{2}\pm i\mu+i\alpha_1}\frac{\sqrt{\pi}}{\sinh (\pi\mu)}\frac{\Gamma(\frac{1}{2}\pm i\mu+i\alpha_2)}{\Gamma(1\pm i\mu)}\nonumber\frac{\Gamma(\frac{1}{2}-i\mu+i\alpha_1)\Gamma(\frac{1}{2}+i\mu+i\alpha_1)}{\Gamma(1+i\alpha_1)}\nonumber\\[5pt]
    &\left(-\frac{1}{x_1}\right)^{i\alpha_1}\left(-\frac{1}{x_2}\right)^{i\alpha_2}\left(ie^{-\pi\mu\mp\frac{\pi\alpha_1}{2}\mp\frac{\pi\alpha_2}{2}}-ie^{\pi\mu\pm\frac{\pi\alpha_1}{2}\pm\frac{\pi\alpha_2}{2}} \mp e^{-\frac{\pi\alpha_1}{2}+\frac{\pi\alpha_2}{2}}\mp e^{\frac{\pi\alpha_1}{2}-\frac{\pi\alpha_2}{2}}\right)\nonumber\\[5pt]
    &+\alpha_1 \rightarrow -\alpha_1.
    \label{3.2E18}
\end{align}
Combining the two solutions above and restoring the $k_3$-dependence of $x_1, x_2$, we obtain the full result for the positive-frequency scalar seed $\hat{\mathcal{I}}_a=\hat{\mathcal{H}}_a+\hat{\mathcal{S}}_a$.  Figure \ref{3.2F1} and Figure \ref{3.2F2} demonstrate the shape and scale dependences of this scalar seed function. 

There are two types of oscillations in this scalar seed function.
The typical cosmological collider signals are the shape-dependent oscillations in terms of the momentum ratio $k_3/k_{12}$, which manifest in the squeezed limit $k_3 \ll k_{12}$. 
This corresponds to the $u\rightarrow0$ limit of the homogeneous solution, where the hypergeometric functions ${}_2F_1$ in (\ref{3.2E17}) approach unity.
Another type of oscillations is the scale-dependent ones from restoring the $k_3$ dependence of $x_1$ and $x_2$, which are similar with the resonant non-Gaussianity of single field feature models.

In Figure \ref{3.2F1}  we take $\mu=10$ and the two oscillating frequencies to be the same $\alpha_1=\alpha_2=\alpha$: the left panel corresponds to the shape dependence on the momentum ratio $k_3/k_{12}$; the right panel is the scale dependence on $k_t$. For $\alpha=0$, this goes back to the conventional cosmological collider with no resonance enhancement: we see that the oscillatory collider signal in Figure \ref{3.2F1.1} is highly suppressed and the seed function in Figure \ref{3.2F1.1} is featureless. For $\alpha\geq 10$, the shape-dependent collider signals are amplified in the squeezed limit; while the plots in the right panel also demonstrate global oscillatory features. 
Figure \ref{3.2F2} shows the situation with two different oscillating frequencies. In addition to the resonance enhancement of collider signals for $\alpha_1,\alpha_2 \geq \mu$, we see novel patterns in Figure \ref{3.2F2.2}: for $\alpha_1\neq\alpha_2$, we find superpositions of two  scale-dependent oscillations with $k_t$ as shown by the grey and orange lines.

Let's take a closer look at the squeezed limit. If we fix $k_{12}$, we observe that the oscillations over $k_3$ have two distinct frequencies: $|\mu-\alpha_1|$ and $|\mu+\alpha_1|$. Examining their prefactors reveals that the latter component is always more suppressed. 
As a result, there is no superposition of wavefronts with different frequencies in Figure \ref{3.2F1.1} and Figure \ref{3.2F2.1}. 
Another observation is that the factors $\left(- {1}/{k_3\eta_2}\right)^{i\alpha_2}$ in the coefficients (\ref{3.2E18}) exactly cancel the $k_3^{i\alpha_2}$ term in the solution \eqref{3.2E17}. Thus, the frequency of the cubic coupling  $\alpha_2$ does not contribute to the oscillatory frequency of the bispectrum in $k_3$, as shown in Figure \ref{3.2F2.1}. However, after cancellation, the remaining factor of $(1/k_{12})^{i\alpha_2}$ in the solution breaks the scale invariance and induces a $\cos(\log k)$-type of running in the bispectrum like resonance non-Gaussianity. For $\alpha_1 \neq \alpha_2$, this scale-dependent feature leads to the superposition of oscillations that we observed in Figure \ref{3.2F2.2}. 

Likewise, the resonance enhancement of cosmological collider signals manifest in the prefactors of the two homogeneous solutions.
From a bulk evolution perspective, the leading contribution in $C_{a+}$ can be seen as follows: at the linear mixing vertex, the massless scalar mode function $\phi^*\sim e^{ik_3 \eta}$ first resonates with the positive frequency coupling $(-\eta)^{i \alpha_1}$, and then at the cubic vertex, the resonance occurs among the conformally coupled scalar $\varphi^* \sim e^{ik_{12}\eta}$, positive frequency coupling $(-\eta)^{i\alpha_2}$ and positive frequency massive mode $\sigma \sim (-\eta)^{i \mu}$. Taking the limit $\mu \geq \alpha_1\sim\alpha_2 \gg 1$, we find $C_{a+}\sim e^{-\pi(\mu-\alpha_1)}$, which corresponds to a softened Boltzmann suppression. 
The leading term in $C_{a-}$   is not negligible either
when $\alpha_1$ and $\alpha_2$ get close to or exceed $\mu$. There the resonance first happens between the massless mode $\phi\sim e^{-ik_3\eta}$ and the negative frequency coupling $(-\eta)^{-i\alpha_1}$ at the linear mixing vertex, and then there is another resonance at the cubic vertex between the conformally coupled scalar $\varphi^* \sim e^{ik_{12}\eta}$ and the positive frequency coupling $(-\eta)^{i\alpha_2}$, leading to an overall factor of $e^{-\pi(2\mu-\alpha_1-\alpha_2)}$.
We shall present a more detailed analysis about resonance enhancement factors with numerical plots in the phenomenological studies in Section \ref{sec:collider}.


\subsection{Weight-Shifting}
With the scalar seed functions, the next step of the bootstrap computation is to apply the weight-shifting operators which map the two external conformally coupled scalars to massless ones.   These differential operators depend on the form of cubic interactions. Here we present the weight-shifting results of two explicit examples with oscillatory couplings: $(\partial_{\mu}\phi)^2\sigma$ and $\dot{\phi}\phi\sigma$, which will be needed later for the phenomenological studies in Section \ref{sec:collider} and the model analysis in Section \ref{sec:axionM}. 
See Refs. \cite{Arkani-Hamed:2018kmz,Baumann:2019oyu, Pimentel:2022fsc} for the systematic constructions of the weight-shifting procedure. Here we mainly follow the approach introduced in \cite{Pimentel:2022fsc}.

For the cubic vertex $\cos\left[\alpha_2\log \left({\eta}/{\eta_2}\right)\right](\partial_{\mu}\phi)^2\sigma$, as the effects of the oscillatory coupling have already been incorporated in the scalar seed functions, we can simply apply the dS-invariant weight-shifting operator  \cite{Arkani-Hamed:2015bza} 
\begin{equation}
\mathcal{W}_{12}^{\mathrm{dS}}\equiv \frac{1}{2}(k_{12}^2-k_3^2)\partial_{k_{12}}^2-\frac{1}{2k_1 k_2}(k_3^2-k_1^2-k_2^2)(1-k_{12}\partial_{k_{12}})~.
    \label{3.3E1}
\end{equation}
Then we consider the three-point correlator of the massless scalar field $\langle \phi\phi\phi \rangle$ with the above cubic vertex and another oscillatory quadratic interaction $\cos\left[\alpha_1\log \left( {\eta}/{\eta_1}\right)\right]\dot{\phi}\sigma$. The full result can be derived by using the solution of the scalar seed $\hat{\mathcal{I}}_a$ in \eqref{3.2E19} and \eqref{3.2E17}, and the one for $\hat{\mathcal{I}}_b$ 
\begin{align}
\langle\phi_{\bold{k_1}}\phi_{\bold{k_2}}\phi_{\bold{k_3}}\rangle' &= \frac{iH^3}{4 k_1^2 k_2^2 k_3^2}\mathcal{W}_{12}^{\mathrm{dS}}(\hat{\mathcal{I}}_a+\hat{\mathcal{I}}_b)+\mathrm{perms.} .
    \label{3.3E2}
\end{align}

For inflation theories with oscillatory features, another important cubic interaction between the inflaton and massive field is given by $\cos\left[\alpha_2\log\left( {\eta}/{\eta_2}\right)\right]\dot{\phi}\phi\sigma$. This can be seen as an operator in the EFT of inflation that breaks both boost and time translation symmetries. In Section \ref{sec:axionM} we shall see that in a concrete example this interaction leads to the dominant contribution to the final exchange bispectrum of curvature perturbations. For the corresponding weight-shifting operator, we simply notice that its bulk integral can be written into the following form
\be
\int d\eta a(\eta)^3 \cos\left[\alpha_2\log\left( {\eta}/{\eta_2}\right)\right] \partial_\eta K^\phi(k_1,\eta) K^\phi(k_2,\eta) \mathcal{K}(k_3,\eta) \sim (1-k_2\partial_{k_2})(\hat{\mathcal{I}}_a+\hat{\mathcal{I}}_b)~.
\ee
As a result, the final bispectrum of the massless scalar is given by
\begin{align}
\langle\phi_{\bold{k_1}}\phi_{\bold{k_2}}\phi_{\bold{k_3}}\rangle' &= \frac{iH^2}{4 k_1^2 k_2^2 k_3^2}\mathcal{W}_{12}^{\dot\phi\phi\s}(\hat{\mathcal{I}}_a+\hat{\mathcal{I}}_b)+\mathrm{perms.}.
     \label{3.3E5}
\end{align}
with the differential operator 
\begin{equation}
    \mathcal{W}_{12}^{\dot\phi\phi\s} \equiv -\frac{k_2}{k_1}(1-k_1\partial_{k_1})-\frac{k_1}{k_2}(1-k_2\partial_{k_2})~.
    \label{3.3E4}
\end{equation}

\section{When Cosmological Collider Meets Resonant Non-Gaussianity}
\label{sec:collider}

With the bootstrap computation, now we move to investigate the phenomenological implications of cosmological collider physics with resonant features. 
In this section, we build a concrete connection with cosmological observables, and highlight the distinctive signatures of the resonance effect on the inflationary power spectrum and bispectrum. 


\subsection{Power Spectrum}

While the main observational signals of our interest lie in the primordial bispectrum, the oscillatory couplings introduce nontrivial corrections to the scalar power spectrum as well. 
Now its amplitude and scale dependence are tightly constrained by CMB experiments such as Planck \cite{Planck:2018jri}.
Thus, before studying the phenomenology of primordial non-Gaussianities, we first examine the predictions on the power spectrum from massive exchange with oscillatory couplings, and ensure that they remain compatible with the latest observational constraints.

\begin{figure}
    \centering
    \includegraphics[width=0.8\linewidth]{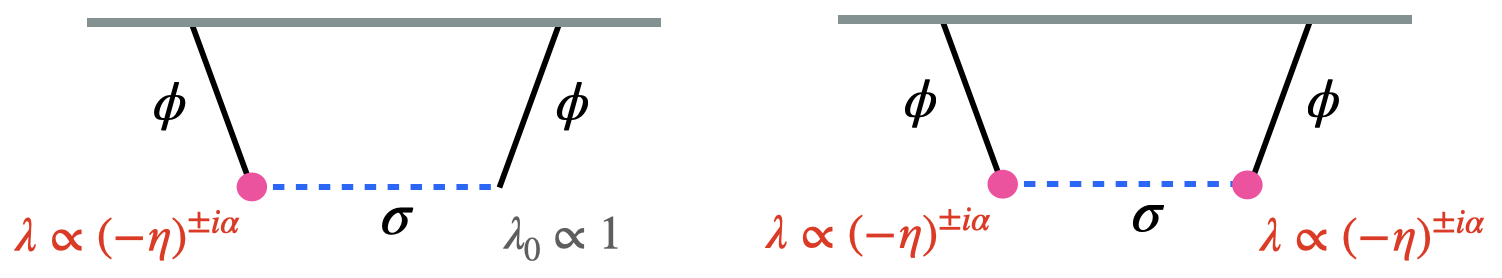}
    \caption{Massive single-exchange corrections to the power spectrum}
    \label{4.1F1}
\end{figure}

Up to the leading order correction from oscillatory couplings with massive fields, the power spectrum of canonically normalised massless scalar $\phi$ can be expressed as
\begin{align}
    P(k) &= P_0(k)+\delta P(k)\nonumber\\
    &=\frac{H^2}{2k^3}[1+\Delta+\delta n^{\alpha}\cos(\alpha\log k+\theta_1)+\delta n^{2\alpha}\cos(2\alpha \log k+\theta_2)],
    \label{4.1E1}
\end{align}
where $ P_0=H^2/2k^3$ is the free theory power spectrum; $\Delta$ denotes the constant correction; $\delta n^{\alpha}$ and $\delta n^{2\alpha}$ denote the amplitude of the oscillatory correction with frequencies $\alpha$ and $2\alpha$ respectively.
Within perturbation theory, the leading corrections to the power spectrum arise from the single-exchange diagrams involving a massive scalar with two linear mixing vertices $\lambda \dot{\phi} \sigma$. Here, we consider the scenario where at least one vertex has an oscillatory coupling: we use $\lambda_0$ to represent the constant quadratic coupling, and use $ \lambda(\eta, \eta_1)$ in \eqref{2.1E2} for the oscillatory one. The two massive exchange diagrams are shown in Figure \ref{4.1F1}.

Let's first consider the case where the coupling of one linear mixing vertex oscillates with frequency $\alpha$ and the coupling of the other is constant.
The correction to the two-point  function is given by 
\begin{align}
    \langle\phi_{\bold{k}}\phi_{-\bold{k}}\rangle_{\alpha}'&=i\int d\eta a(\eta)^3\lambda_0[\partial_{\eta}K_+(k, \eta)\mathcal{K}_+(k,\eta,\eta_1)-c.c]\nonumber\\
    &=\frac{i\lambda_{0}\lambda_{\alpha}}{2k^3}\hat{\mathcal{I}}(u=1, x_1),
    \label{4.1E2}
\end{align}
where $\hat{\mathcal{I}}(u=1,x_1)$ is the scalar seed defined by (\ref{3.2E2}), evaluated at the folded limit $u \rightarrow 1$. 
In the first line of \ref{4.1E2}, the oscillatory coupling $\lambda(\eta, \eta_1)$ has been absorbed into the mixed propagator, while we plug the magnitude factor $\lambda_{\alpha}$ outside $\hat{\mathcal{I}}$ in the second line to make it consistent with the previous definition.
Similarly, the contribution from the Feynman diagram with two oscillatory vertices is given by 
\begin{align}
    \langle\phi_{\bold{k}}\phi_{-\bold{k}}\rangle_{2\alpha}'&=i\int d\eta a(\eta)^3\lambda(\eta, \eta_1)[\partial_{\eta}K_+(k, \eta)\mathcal{K}_+(k,\eta,\eta_1)-c.c]\nonumber\\
    &=\frac{i\lambda_{\alpha}^2}{2k^3}[\hat{\mathcal{I}}_a(u=1, x_1, x_1)+\hat{\mathcal{I}}_b(u=1, x_1,x_1)],
    \label{4.1E3}
\end{align}
where the scalar seeds $\hat{\mathcal{I}}_a$ and $\hat{\mathcal{I}}_b$ are defined by (\ref{3.2E12}) and (\ref{3.2E13}). 

One technical challenge remains in evaluating the scalar seed functions in the $u\rightarrow1$ limit. 
In our solutions, there are unphysical poles in both the homogeneous and particular parts of each scalar seed solutions, which cancel out each other, ensuring that the correlators remain regular in the folded limit. 
However, accurately determining the finite pieces of the scalar seed in this limit, in principle, requires a resummation of infinite series of the particular solutions, which is a difficult task in general. As shown in \cite{Qin:2023ejc}, this issue can be bypassed by changing the variable of the boundary differential equation, and then the solution of the scalar seed takes a well-behaved closed-form. Here we apply the results derived in \cite{Qin:2023ejc} and find that the unsuppressed corrections to the power spectrum given by
\begin{small}
\begin{align}\label{4.1E21}
    \frac{\delta P(k)}{P_0(k)}= &2k^3\langle\phi_{\bold{k}}\phi_{-\bold{k}}\rangle_{\alpha}'+2k^3\langle\phi_{\bold{k}}\phi_{-\bold{k}}\rangle_{2\alpha}' \\
    =&-\lambda_0\lambda_{\alpha}E_1^P(\mu,\alpha)\left(\frac{-1}{2k\eta_1}\right)^{i\alpha}-\lambda_{\alpha}^2 E_2^{P}(\mu, \alpha)\left(\frac{-1}{2k\eta_1}\right)^{2i\alpha} 
    -\lambda_{\alpha}^2 E_3^P(\mu,\alpha)+\lambda_{\alpha}^2 E_4^P(\mu,\alpha)  +\mathrm{c.c.}, \nonumber
\end{align}
\end{small}where the four $E^P_i(\mu,\alpha)$ factors reflect the enhancement/suppression of the resonance effect. Their analytical expressions in terms of hypergeometric and Gamma functions are explicitly shown in \eqref{4.1E18}--\eqref{4.1E20.1} in Appendix.
The first and second terms in (\ref{4.1E21}) introduce oscillatory features in the power spectrum, and the third and fourth terms are constant corrections. Comparing (\ref{4.1E21}) with (\ref{4.1E1}), we find
\begin{equation}
    \delta n^\alpha=-\lambda_0 \lambda_{\alpha}|E_1^P(\mu, \alpha)|~~~~\&~~~~ \delta n^{2\alpha}=-\lambda_{\alpha}^2|E_2^P(\mu,\alpha)|.
    \label{4.1E22}
\end{equation}

\begin{figure}[t]
     \centering
    \begin{subfigure}{0.45\textwidth}
        \includegraphics[width=\columnwidth]{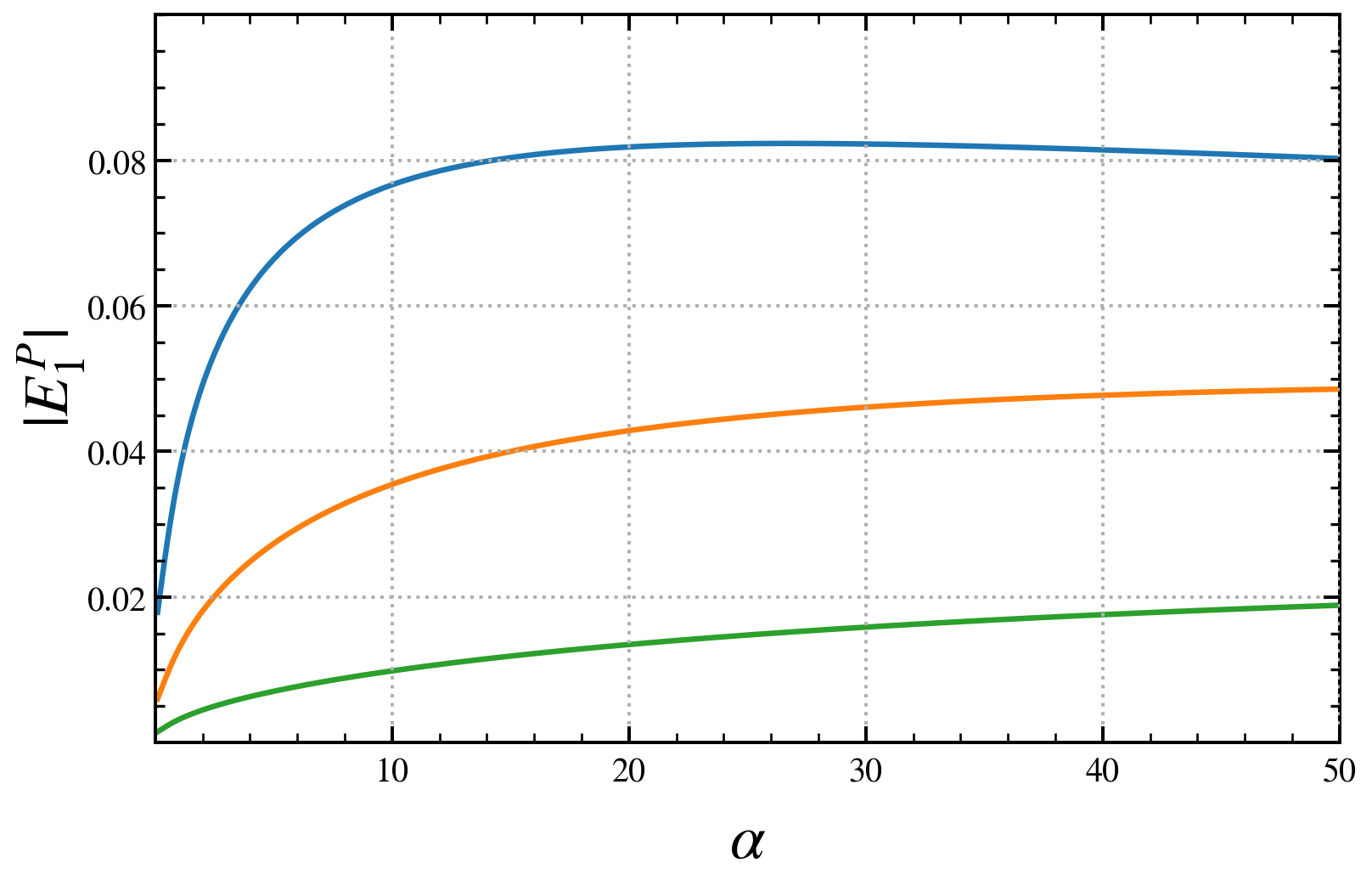}
        \caption{}
        \label{4.1F1.1}
    \end{subfigure}
    \begin{subfigure}{0.45\textwidth}
        \includegraphics[width=\columnwidth]{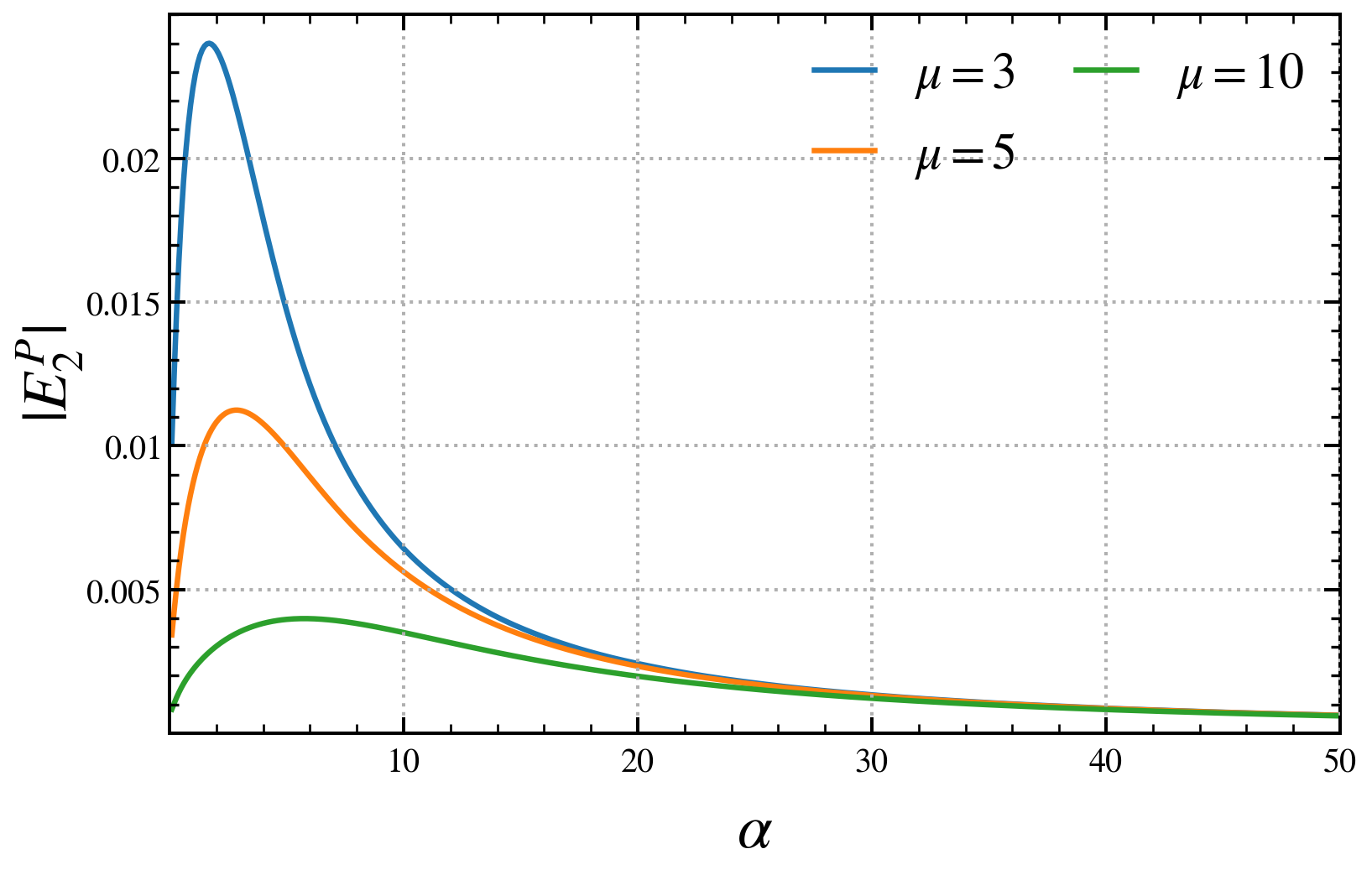}
        \caption{}
        \label{4.1F1.2}
    \end{subfigure}
    \begin{subfigure}{0.45\textwidth}
        \includegraphics[width=\columnwidth]{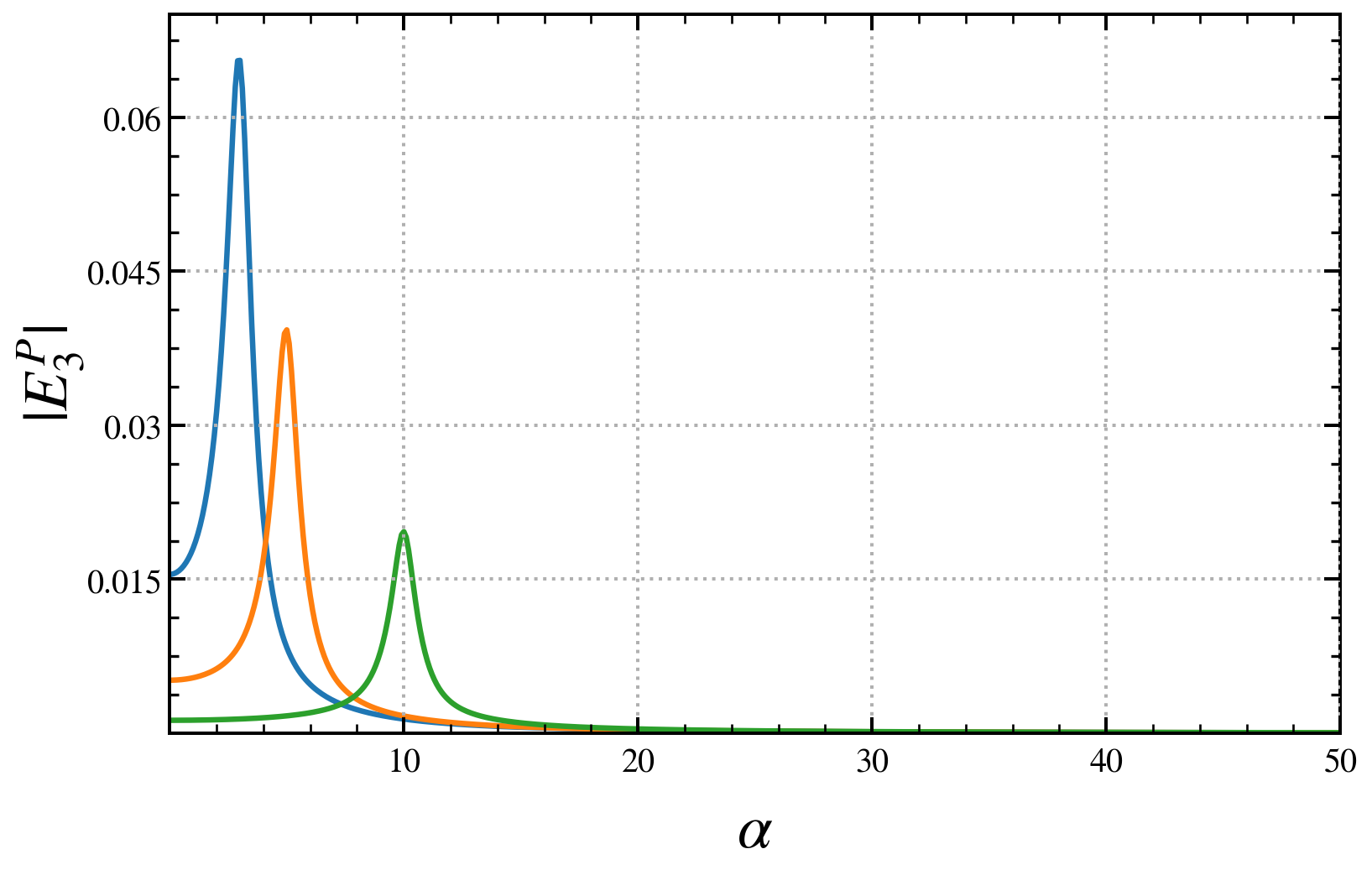}
        \caption{}
        \label{4.1F1.3}
    \end{subfigure}
    \begin{subfigure}{0.45\textwidth}
        \includegraphics[width=\columnwidth]{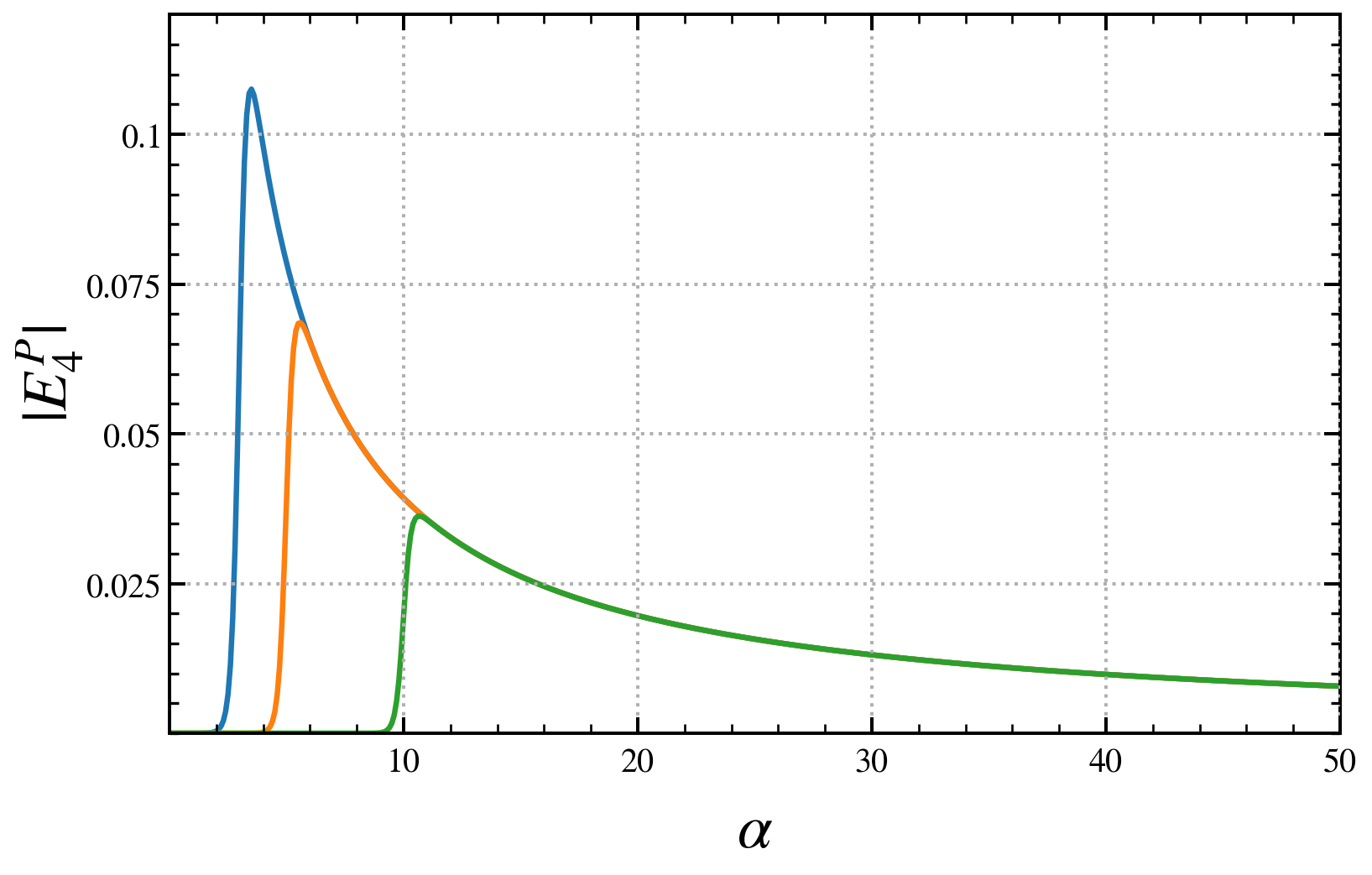}
        \caption{}
        \label{4.1F1.4}
    \end{subfigure}
    \caption{$|E_1^P|$, $|E_2^P|$, $|E_3^P|$ and $|E_4^P|$ vs $\alpha$, with $\mu=3$, $5$ and $10$.}
    \label{4.1F2}
\end{figure}

Let's take a close look at the enhancement factors. While their full analytical expressions are less informative, the size of these corrections can be seen from the numerical plots in Figure \ref{4.1F2}.
In the conventional scenario with no oscillatory couplings, the massive exchange corrections to the power spectrum are not exponentially suppressed by the mass of the field, as they can be seen as local EFT contributions instead of the non-local particle production effects. 
Introducing oscillations in the quadratic coupling makes the analysis more complicated.
By varying the frequency $\alpha$, we first see
a modest power-law enhancement on the correction of the power spectrum up to several times of the non-oscillating case, and the maximum usually happens around $\alpha \sim \mu$. While the frequency exceeds the mass, the enhancement factors will be suppressed, as the bulk time integral becomes gradually dominated by the contribution around the stationary phase. Phenomenologically, this means that corrections to the power spectrum can be easily under control even for highly oscillatory scenarios.


\begin{figure}[t]
     \centering
    \begin{subfigure}{0.45\textwidth}
        \includegraphics[width=\columnwidth]{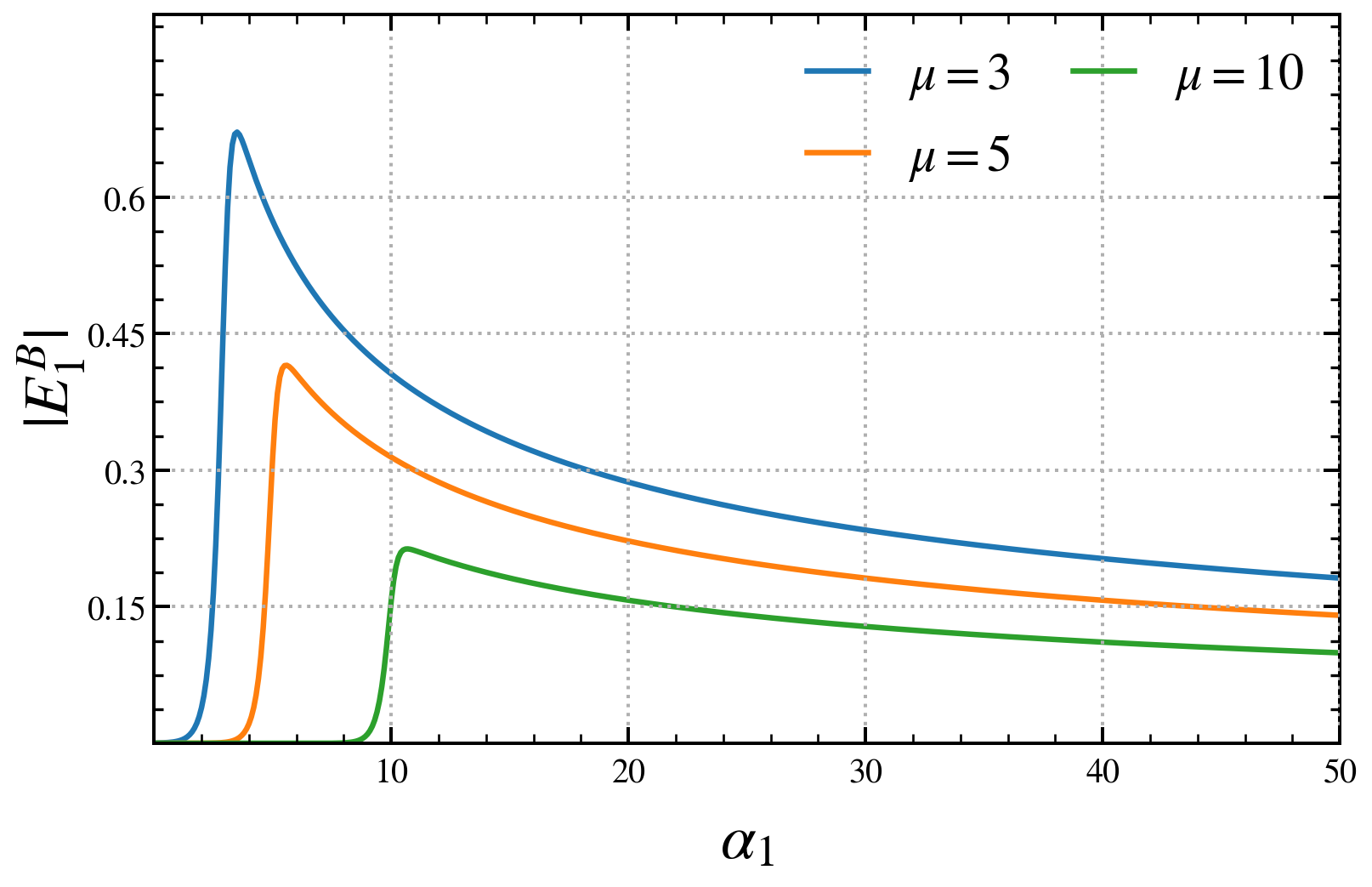}
        \caption{$\alpha_2=5$; $\mu = 3, 5, 10$}
        \label{4.2F1.1}
    \end{subfigure}
    \begin{subfigure}{0.45\textwidth}
        \includegraphics[width=\columnwidth]{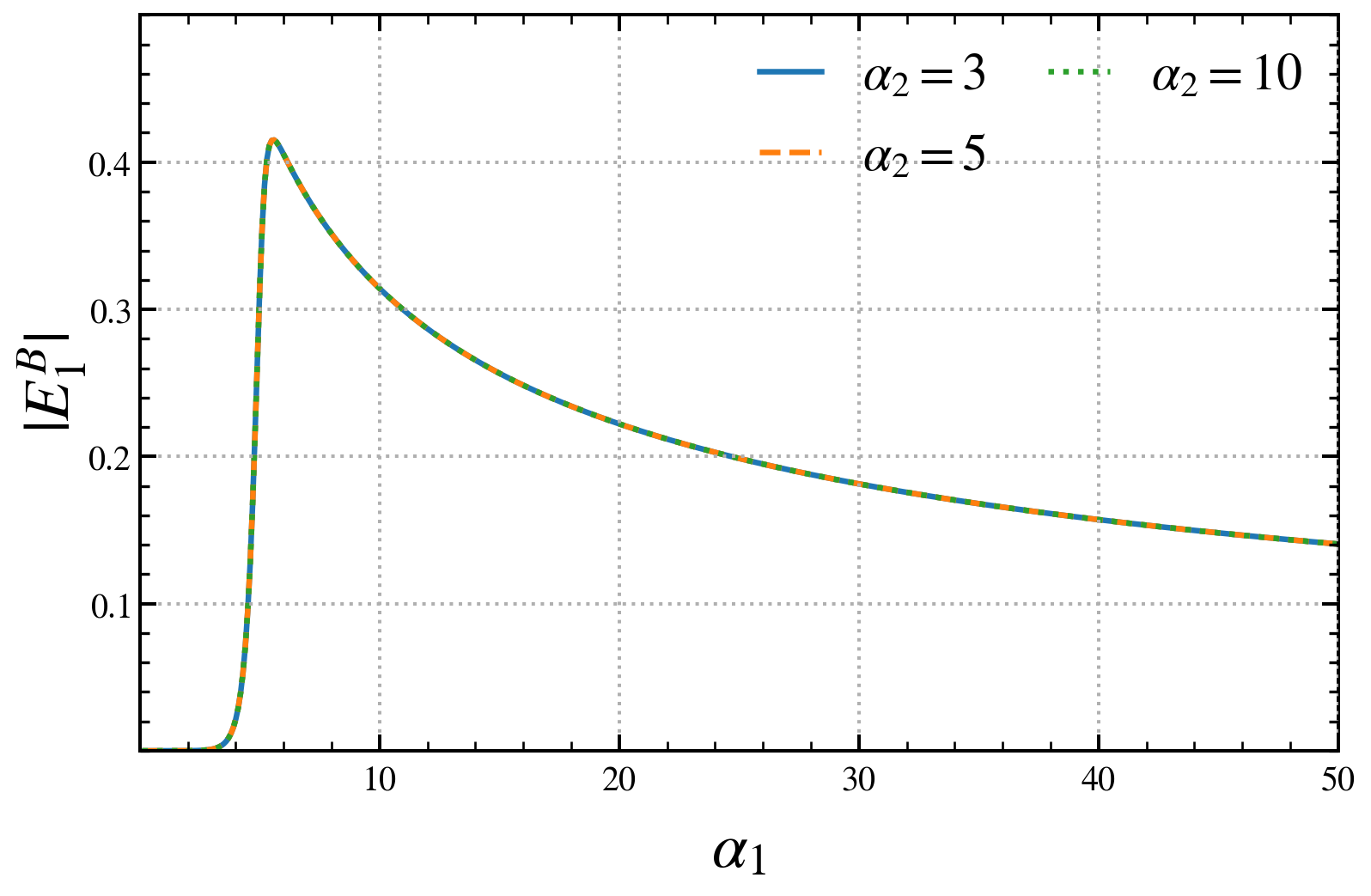}
        \caption{\centering $\mu=5$; $\alpha_2 = 3, 5, 10$}
        \label{4.2F1.2}
    \end{subfigure}
    \begin{subfigure}{0.45\textwidth}
        \includegraphics[width=\columnwidth]{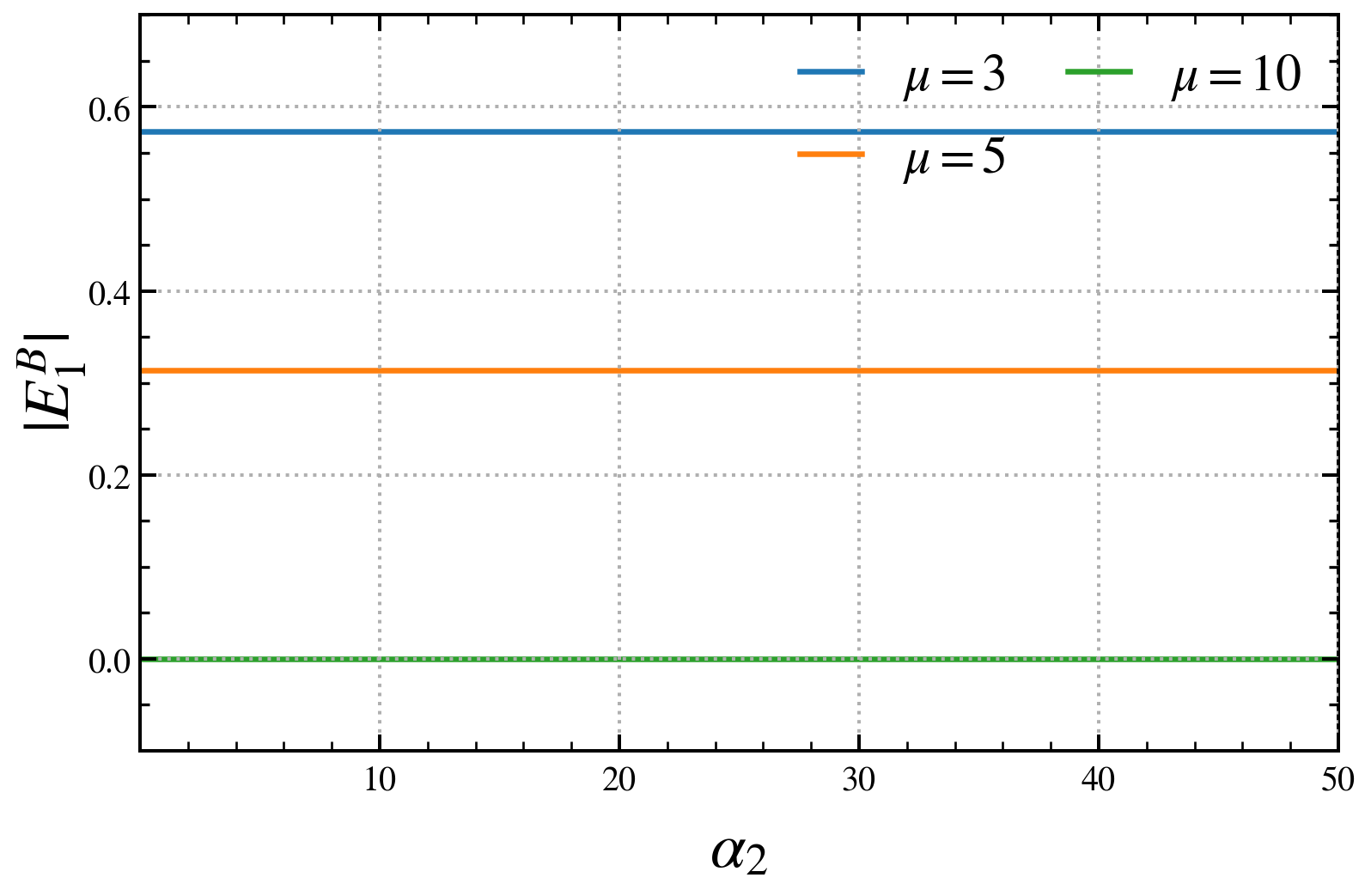}
        \caption{$\alpha_1=5$; $\mu = 3, 5, 10$}
        \label{4.2F1.3}
    \end{subfigure}
    \begin{subfigure}{0.45\textwidth}
        \includegraphics[width=\columnwidth]{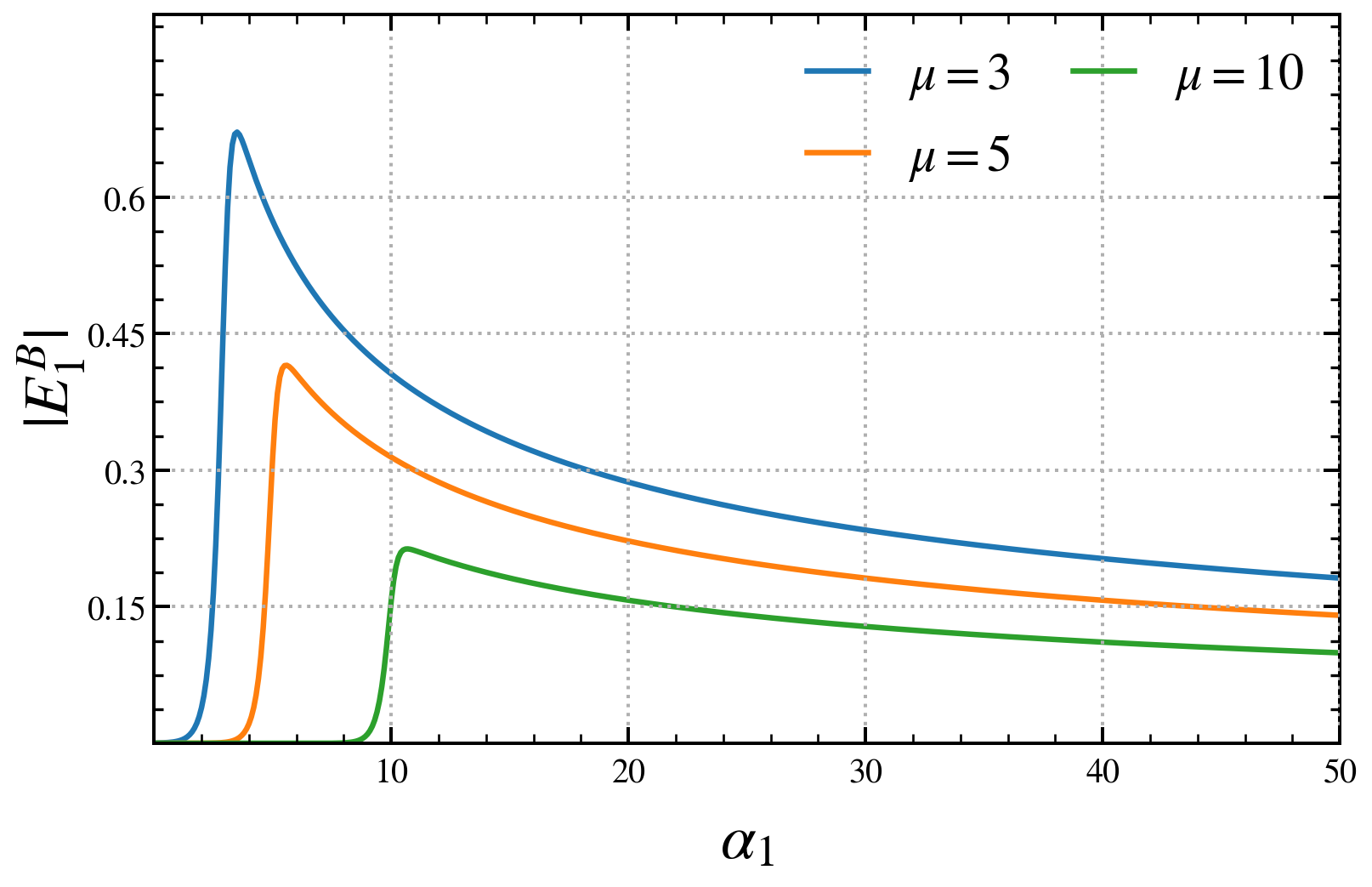}
        \caption{$\alpha_1=\alpha_2$; $\mu = 3, 5, 10$}
        \label{4.2F1.4}
    \end{subfigure}    
    \caption{$|E_1^B|$ vs $\alpha_1$ and $\alpha_2$.}
    \label{4.2F1}
\end{figure}  
\begin{figure}[t]
     \centering
    \begin{subfigure}{0.45\textwidth}
        \includegraphics[width=\columnwidth]{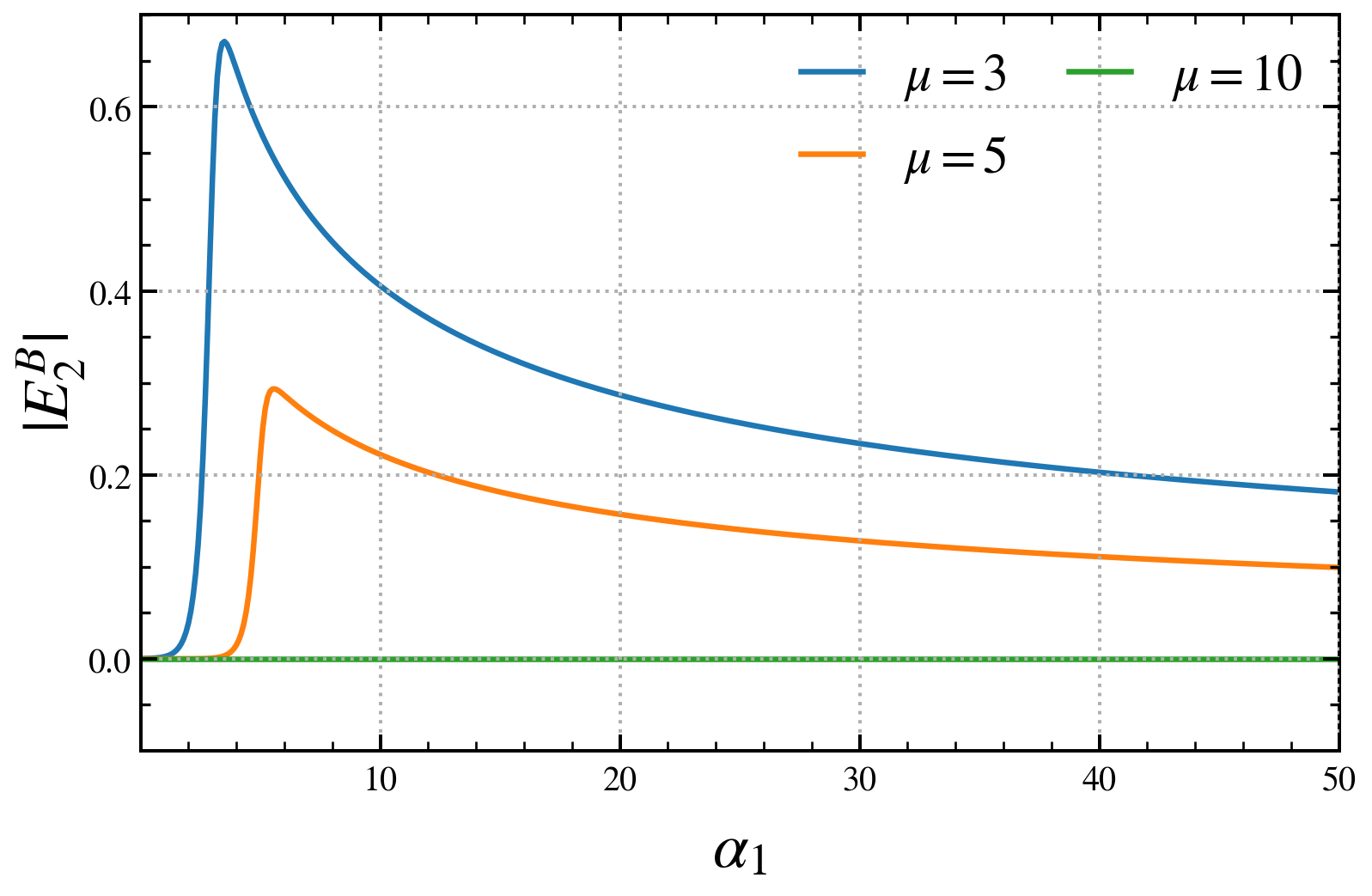}
        \caption{$\alpha_2=5$, $\mu = 3,5,10$}
        \label{4.2F2.1}
    \end{subfigure}
    \begin{subfigure}{0.45\textwidth}
        \includegraphics[width=\columnwidth]{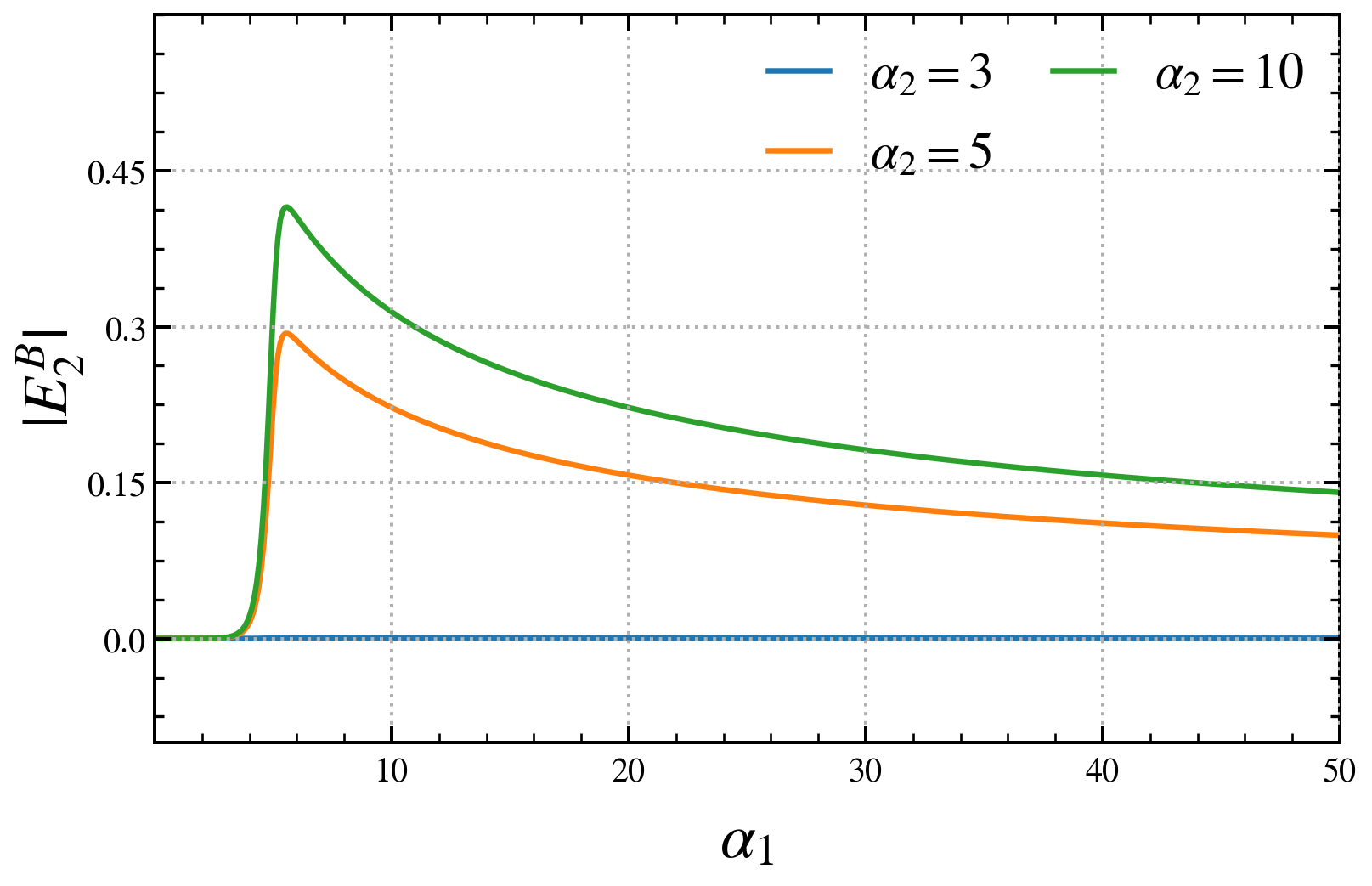}
        \caption{$\mu=5$, $\alpha_2 = 3,5,10$}
        \label{4.2F2.2}
    \end{subfigure}
    \begin{subfigure}{0.45\textwidth}
        \includegraphics[width=\columnwidth]{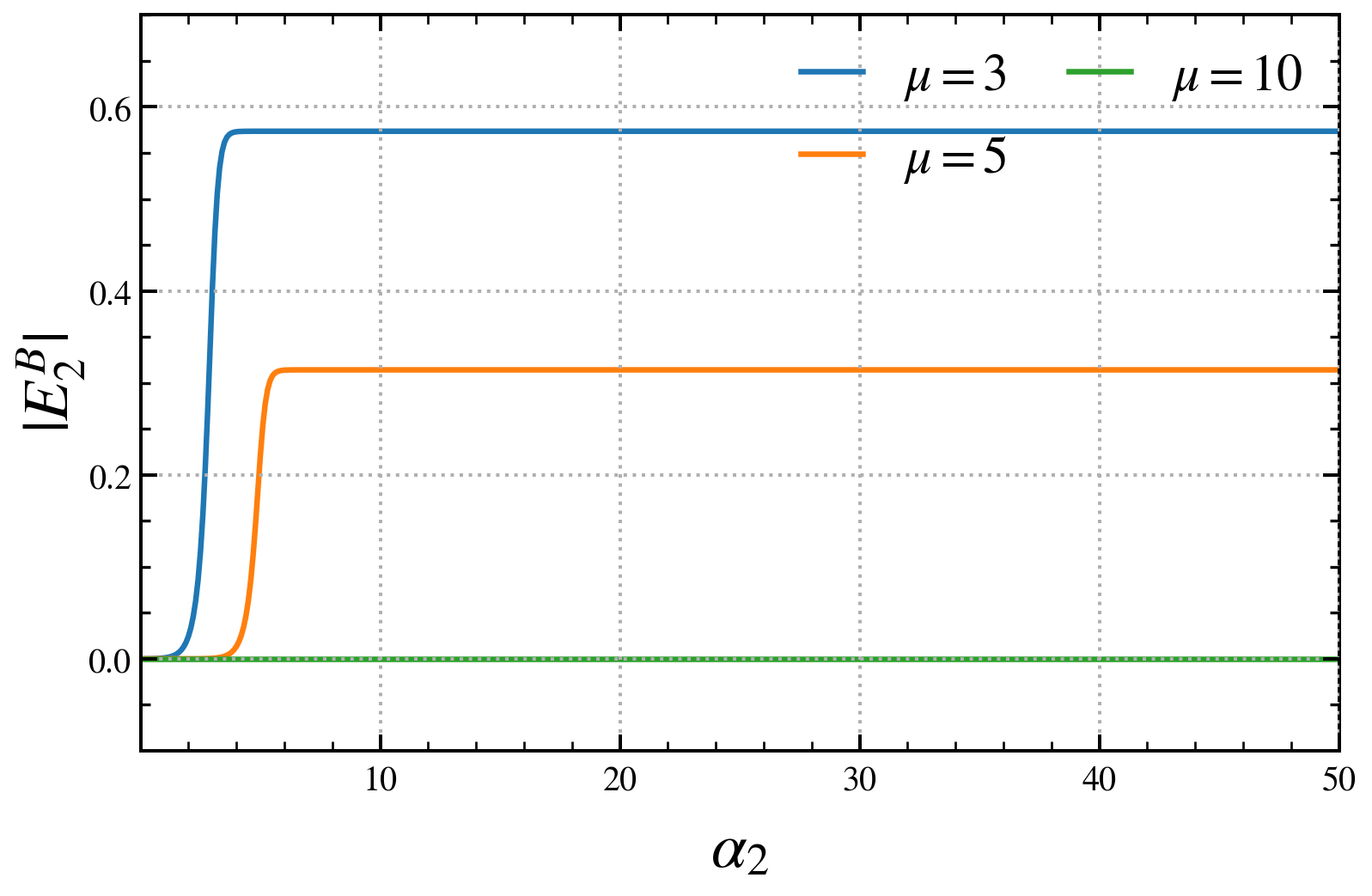}
        \caption{$\alpha_1=5$, $\mu = 3,5,10$}
        \label{4.2F2.3}
    \end{subfigure}
    \begin{subfigure}{0.45\textwidth}
        \includegraphics[width=\columnwidth]{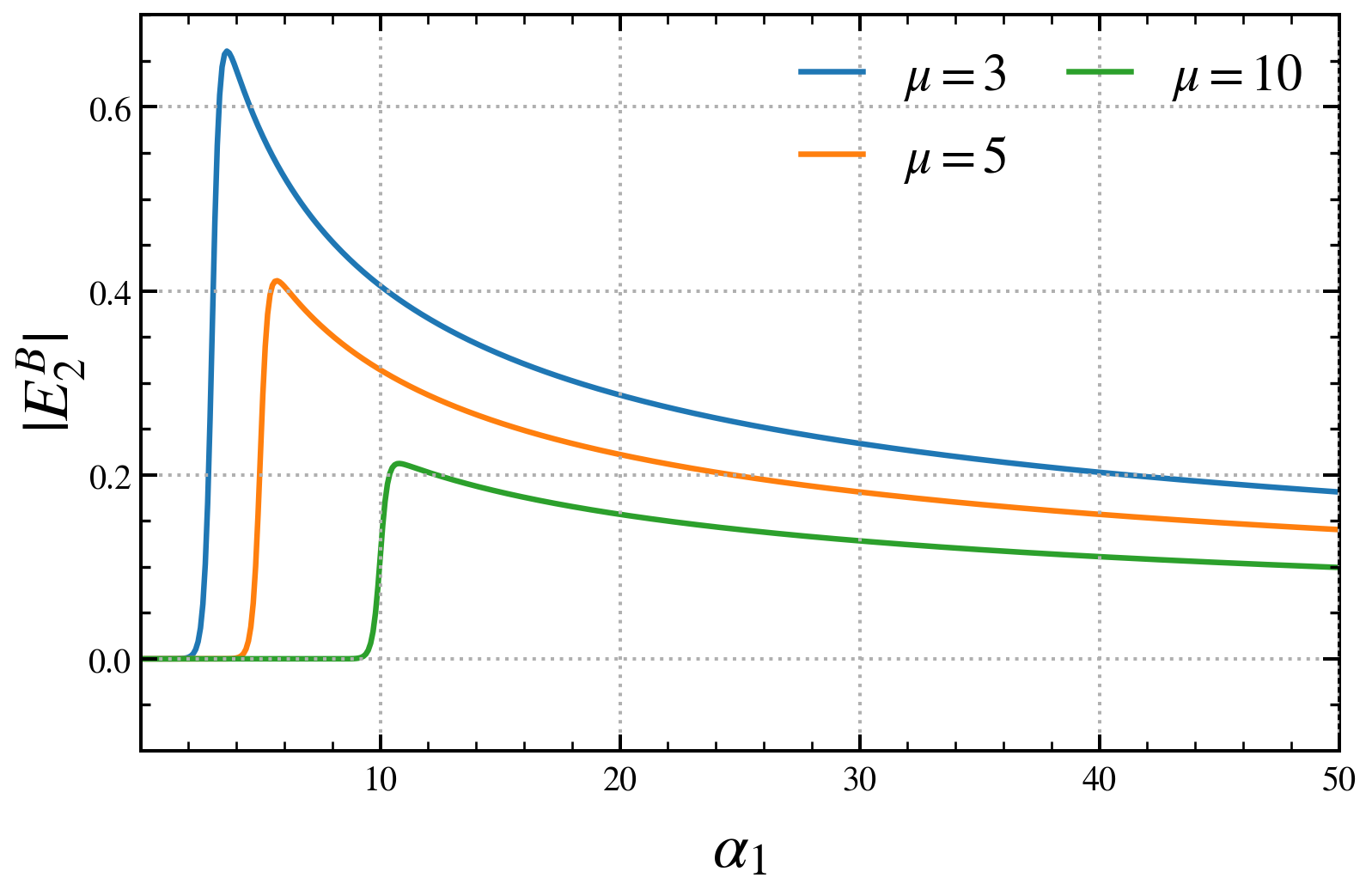}
        \caption{$\alpha_1=\alpha_2$, $\mu = 3,5,10$}
        \label{4.2F2.4}
    \end{subfigure}    
    \caption{$|E_2^B|$ vs $\alpha_1$ and $\alpha_2$.}
    \label{4.2F2}
\end{figure}  

\subsection{Non-Gaussianity}
\label{pheno_ng}

Now we look into the phenomenology of the primordial bispectrum.
The shapes and amplitudes of non-Gaussianity are associated with
the form of the interaction and the size of the couplings, which are model-dependent. 
Here we mainly focus on the model-independent properties of the resulting inflaton three-point function $\langle \phi\phi\phi \rangle$ after weight-shifting process.
For cosmological colliders with resonant couplings, there are two major novelties in the scalar bispectrum: the resonance enhancement of its size; and the new shapes combining both collider signals and oscillatory features.

The enhancement of the non-Gaussian signals can already be seen from the scalar seed functions before performing the weight-shifting operators. 
Let's take $\hat{\mathcal{I}}_a$ in \eqref{3.2E13} with two oscillatory couplings as a concrete example. As we have already discussed in Section \ref{sec:2osc}, the two coefficients $C_{a\pm}$ of the homogeneous solutions contain information about the size of the collider signals. 
From these analytical results of the bootstrap computation, we introduce the bispectrum
enhancement factors as
\begin{equation}
    E_1^B \equiv\frac{\sqrt{\pi}}{4\sinh(\pi\mu)}\frac{\Gamma\left(\frac{1}{2}+i\mu+i\alpha_2\right)}{\Gamma(1+i\mu)}\frac{\Gamma\left(\frac{1}{2}-i\mu+i\alpha_1\right)\Gamma\left(\frac{1}{2}+i\mu+i\alpha_1\right)}{\Gamma(1+i\alpha_1)}e^{\pi\mu+\frac{\pi\alpha_1}{2}+\frac{\pi\alpha_2}{2}},
    \label{4.2E1}
\end{equation}
\begin{equation}
    E_2^B \equiv\frac{\sqrt{\pi}}{4\sinh(\pi\mu)}\frac{\Gamma\left(\frac{1}{2}-i\mu+i\alpha_2\right)}{\Gamma(1-i\mu)}\frac{\Gamma\left(\frac{1}{2}-i\mu-i\alpha_1\right)\Gamma\left(\frac{1}{2}+i\mu-i\alpha_1\right)}{\Gamma(1-i\alpha_1)}e^{\frac{\pi\alpha_1}{2}+\frac{\pi\alpha_2}{2}}~.
    \label{4.2E1.1}
\end{equation}
Again, it is more illuminating to look at their numerical plots, as shown in Figure \ref{4.2F1} and Figure \ref{4.2F2}. Let's carefully break down the parameter regions and analyse the corresponding behaviours:
\begin{itemize}
    \item For $\alpha_1,\alpha_2\leq \mu$, these factors are exponentially suppressed at large $\mu$, just like the conventional cosmological collider with no resonant features. They simply reproduce the Boltzmann factors, reflecting that the non-local effects of heavy particle production are negligible. As a result, we can safely integrate out these heavy fields to obtain a low-energy EFT of single field inflation.
    \item When we increase the frequency of the linear mixing $\alpha_1$, we see that both factors get enhanced and reach their maxima at $\alpha_1 \simeq \mu$. The Boltzmann suppression is overcome due to the resonance between the coupling and the massless field at the two-point vertex. Then the heavy particles get continuously excited by the resonance. Although they have a mass $m\gg H$, we can no longer integrate out these heavy fields to achieve a single-field EFT.\footnote{As we shall see in Section \ref{sec:moduli}, in multi-field inflation models with heavy scalars, this breakdown of the single-field EFT just corresponds to the violation of the adiabaticity condition proposed in Refs. \cite{Cespedes:2012hu, Achucarro:2012yr}.}
    For $\alpha_1>\mu$, we observe a power-law suppression as $\alpha_1^{-\frac{1}{2}}$, arising from the asymptotic expression of the corresponding $\Gamma$ functions.
    \item For the frequency of the cubic coupling $\alpha_2$, its effects are more subtle, which lead to the main difference between these two factors $E_1^B$ and $E_2^B$. As we can see in Figure \ref{4.2F1.3} and Figure \ref{4.2F2.3}, $E_1^B$ is almost independent of $\alpha_2$, while $E_2^B$ is exponentially suppressed when $\alpha_2<\mu$. 
\end{itemize}    
Specifically, in two different parameter regimes, the two factors can be approximated as 
    \begin{align}
    &E_1^B(\mu,\alpha_1,\alpha_2) \simeq \frac{\pi}{\sqrt{2}}\frac{e^{-\pi(\mu-\alpha_1)}}{\sqrt{\mu\alpha_1}}; ~~  E_2^B(\mu,\alpha_1,\alpha_2) \simeq \frac{\pi}{\sqrt{2}}\frac{e^{-\pi(2\mu-\alpha_1-\alpha_2)}}{\sqrt{\mu\alpha_1}} ~~{\rm for} ~ \mu> \alpha_1,\alpha_2 \gg 0 \nonumber\\ 
    &E_1^B(\mu,\alpha_1,\alpha_2) \simeq \frac{\pi}{\sqrt{2}}\frac{1}{\sqrt{\mu\alpha_1}};
        ~~~~~~~  E_2^B(\mu,\alpha_1,\alpha_2) \simeq\frac{\pi}{\sqrt{2}}\frac{1}{\sqrt{\mu\alpha_1}}~~~~~~~~{\rm for} ~ \alpha_1,\alpha_2 \geq \mu \gg 0.
        \label{asymp}
    \end{align}
As the two enhancement factors come from the scalar seed function, they are unaffected by the weight-shifting process and thus remain independent of the choice of cubic interactions.
 

Next, we investigate the new shapes of the scalar bispectrum with both collider signals and resonant features.
Unlike the enhancement factors, the shape functions are model-dependent. 
For illustration we consider two examples of the cubic interactions $(\partial_\mu\phi)^2\s$ and $\phi\dot\phi\s$, which will give us the leading contributions to the exchange bispectrum in the concrete model in Section \ref{sec:axionM}.
Their full bispectra $\langle\phi\phi\phi\rangle$ have already been derived in \eqref{3.3E2} and \eqref{3.3E5} using the corresponding weight-shifting operators.
From the analytical results, we can see that outside of the squeezed corner the shape function resembles the resonant non-Gaussianity. Thus it is more interesting to look at their squeezed-limit behaviour and highlight the new features in the cosmological collider signal.

\begin{figure}
    \centering
    \includegraphics[width=0.6\linewidth]{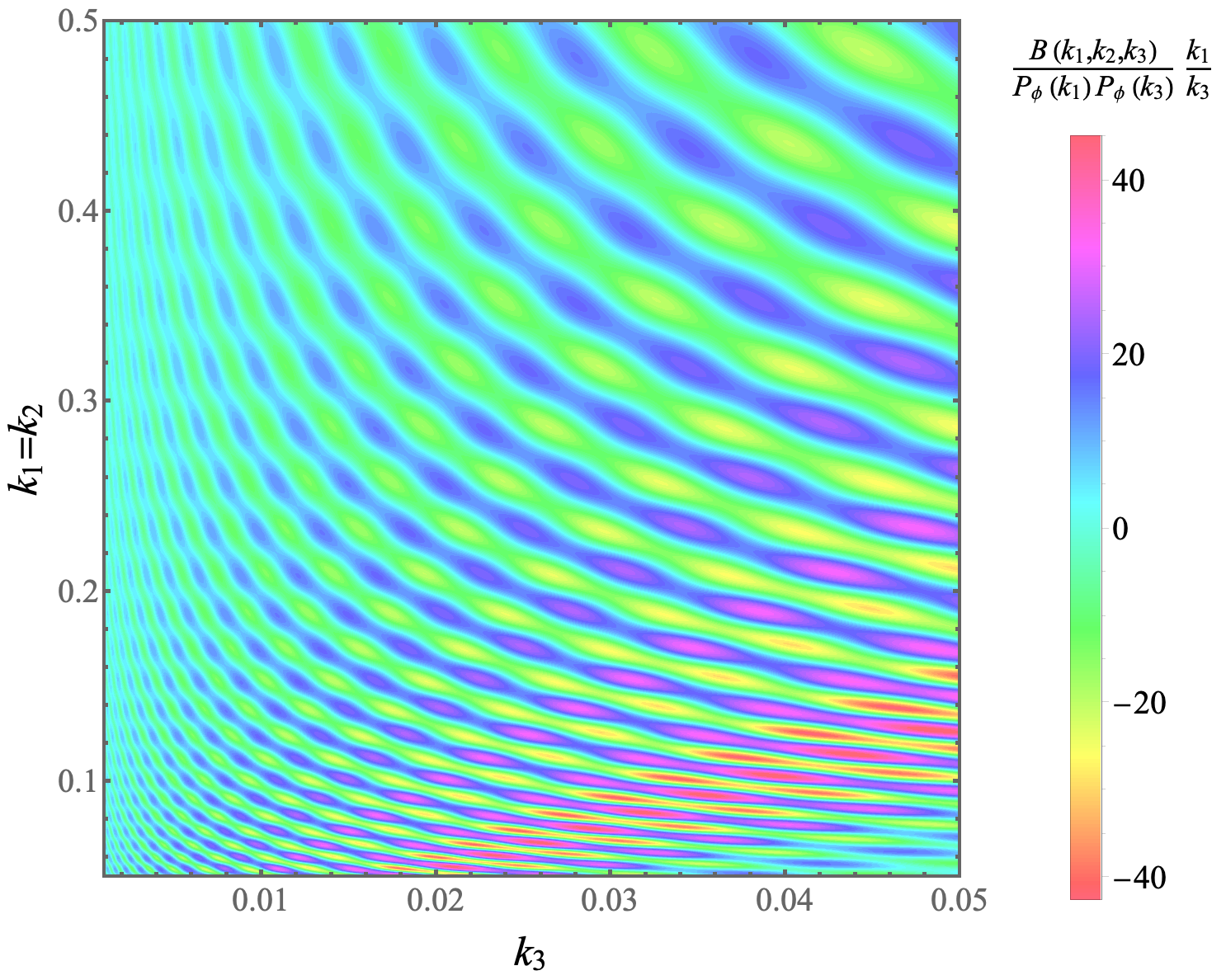}
    \caption{Oscillatory pattern in the squeezed corner of the bispectrum from the dS-invariant interaction $(\partial_{\mu}\phi)^2\sigma$ for $\alpha_1 = \alpha_2 = 30$ and $\mu=10$ (which has $|E_1^P|=0.016$ and $|E_1^B|=0.128$). The two coupling constants are set to be unity. A similar plot but for $\phi\dot\phi\s$ is presented in Ref.\cite{Pajer:2024ckd}.}
    \label{fig:dS_shape}
\end{figure}

For the dS-invariant cubic interaction $(\partial_{\mu}\phi)^2\sigma$, the squeezed bispectrum becomes
\begin{align}
    \lim_{k_3\ll k_1=k_2}\langle\phi\phi\phi\rangle'_{\mathrm{dS}}
    &\equiv \frac{\lambda_{\mathrm{quad}}^{\dot{\phi}\sigma}\lambda_{\mathrm{cub}}^{(\partial_{\mu}\phi)^2\sigma}}{32 H}|E_1^B||F^{(\partial_{\mu}\phi)^2\sigma}(\mu, \alpha_2)|P_{\phi}(k_1)P_{\phi}(k_3)\tilde{S}^{\mathrm{dS}}(k_1=k_2,k_3)
    \label{bisp_dS}
\end{align}
where $\lambda_{\mathrm{quad}}^{\dot{\phi}\sigma}$ and $\lambda_{\mathrm{cub}}^{(\partial_{\mu}\phi)^2\sigma}$ denote the strengths of the quadratic and cubic couplings respectively. In addition to the enhancement factor $E_1^B$, this bispectrum also contains another dimensionless coefficient
\begin{equation}
F^{(\partial_{\mu}\phi)^2\sigma}(\mu, \alpha_2)\equiv 15+16i\mu-4\mu^2+16i\alpha_2-8\mu\alpha_2-4\alpha_2^2,
    \label{F_factor}
\end{equation}
which comes from applying the weight-shifting operator \eqref{3.3E1} on the primary scalar seed. 
It characterises the dependence of the bispectrum size on the form of the cubic interaction.
For $(\partial_\mu\phi)^2\s$, this factor goes as $-4\alpha_2^2$ with large frequencies of the cubic coupling.
The dimensionless shape function in the squeezed limit is given by
\begin{align}
    \tilde{S}^{\mathrm{dS}}(k_1,k_3)=&\left(\frac{k_3}{k_1}\right)^{\frac{3}{2}}\left\{\cos\left[\mu\log\left(\frac{k_3}{k_1}\right)+\alpha_1\log\left(\frac{-1}{ k_3\eta_1}\right)+\alpha_2\log\left(\frac{-1}{k_1     \eta_2}\right)+\theta_1^{\mathrm{dS}}(\mu,\alpha_1,\alpha_2)\right]\right.\nonumber\\[5pt]     
    &\left.+\Delta f^{\mathrm{dS}}\cos\left[\mu\log\left(\frac{k_3}{k_1}\right)+\alpha_1\log\left(\frac{-1}{k_3\eta_1}\right)-\alpha_2\log\left(\frac{-1}{k_1\eta_2}\right)+\theta_2^{\mathrm{dS}}(\mu,\alpha_1,\alpha_2)\right]\right\},
    \label{sqshape_dS}
\end{align}
where we have introduced the phase parameters
\begin{align}
    &\theta_1^{\mathrm{dS}}=-(2\mu+\alpha_1+\alpha_2)\log 2+\mathrm{Arg}[E_1^B(\mu, \alpha_1, \alpha_2)F^{(\partial_{\mu}\phi)^2\sigma}(\mu, \alpha_2)];\nonumber\\[10pt]
    &\theta_2^{\mathrm{dS}}=-(2\mu+\alpha_1-\alpha_2)\log 2-\mathrm{Arg}[iE_2^B(\mu, \alpha_1, \alpha_2)F^{(\partial_{\mu}\phi)^2\sigma}(-\mu, \alpha_2)].
    \label{phase_dS}
\end{align}
There are two cosine oscillations in \eqref{sqshape_dS}, and their relative size is determined by the $ \Delta f^{\mathrm{dS}}$ factor 
\begin{equation}
    \Delta f^{\mathrm{dS}} = \frac{|E_2^B(\mu, \alpha_1, \alpha_2)||F^{(\partial_{\mu}\phi)^2\sigma}(-\mu, \alpha_2)|}{|E_1^B(\mu, \alpha_1, \alpha_2)||F^{(\partial_{\mu}\phi)^2\sigma}(\mu, \alpha_2)|}~.
    \label{del_f}
\end{equation}
When $\alpha_1,\alpha_2\geq\mu\gg1$, this factor is generically $\mathcal{O}(1)$. In contrast, for $\alpha_2<\mu$,  $\Delta f^{\mathrm{dS}}$ goes towards $0$. When $\alpha_2=0$, the squeezed-limit bispectrum reproduces the result in \cite{Chen:2022vzh}.
In physically motivated scenarios of our interest, such as the model in Section \ref{sec:axionM}, it is more natural for the two couplings to oscillate at the same frequency $\alpha_1 = \alpha_2 \equiv \alpha$.
Then the first cosine in \eqref{sqshape_dS} has both the scale-dependent and shape-dependent oscillations,
while only the shape-dependent collider signals remain in the second cosine. 
The overall oscillatory pattern is shown in Figure \ref{fig:dS_shape}. We observe that scale-invariant collider signals with frequency $|\mu-\alpha|$ emerge perpendicular to the $k_3/k_1={\rm const.}$ lines. On top of that, there are scale-dependent oscillations aligned along those $k_3/k_1={\rm const.}$ lines, introduced by the first cosine in \eqref{sqshape_dS}. 

\begin{figure}[t]
     \centering
    \begin{subfigure}{0.45\textwidth}
        \includegraphics[width=\columnwidth]{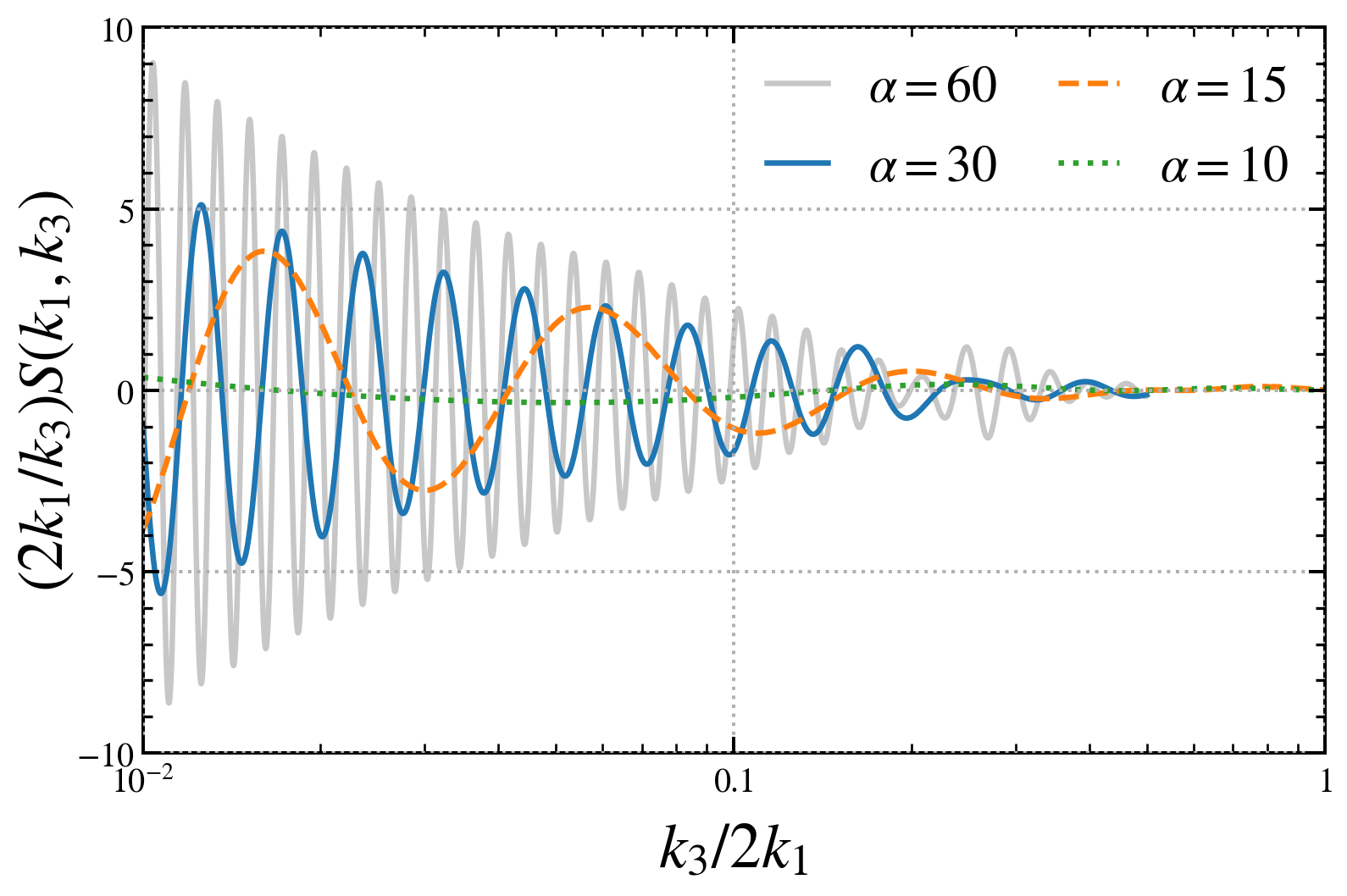}
        \caption{$k_1 = k_2 = 0.5 \mathrm{Mpc}^{-1}$}
        \label{4.2F3.1}
    \end{subfigure}
    \begin{subfigure}{0.45\textwidth}
        \includegraphics[width=\columnwidth]{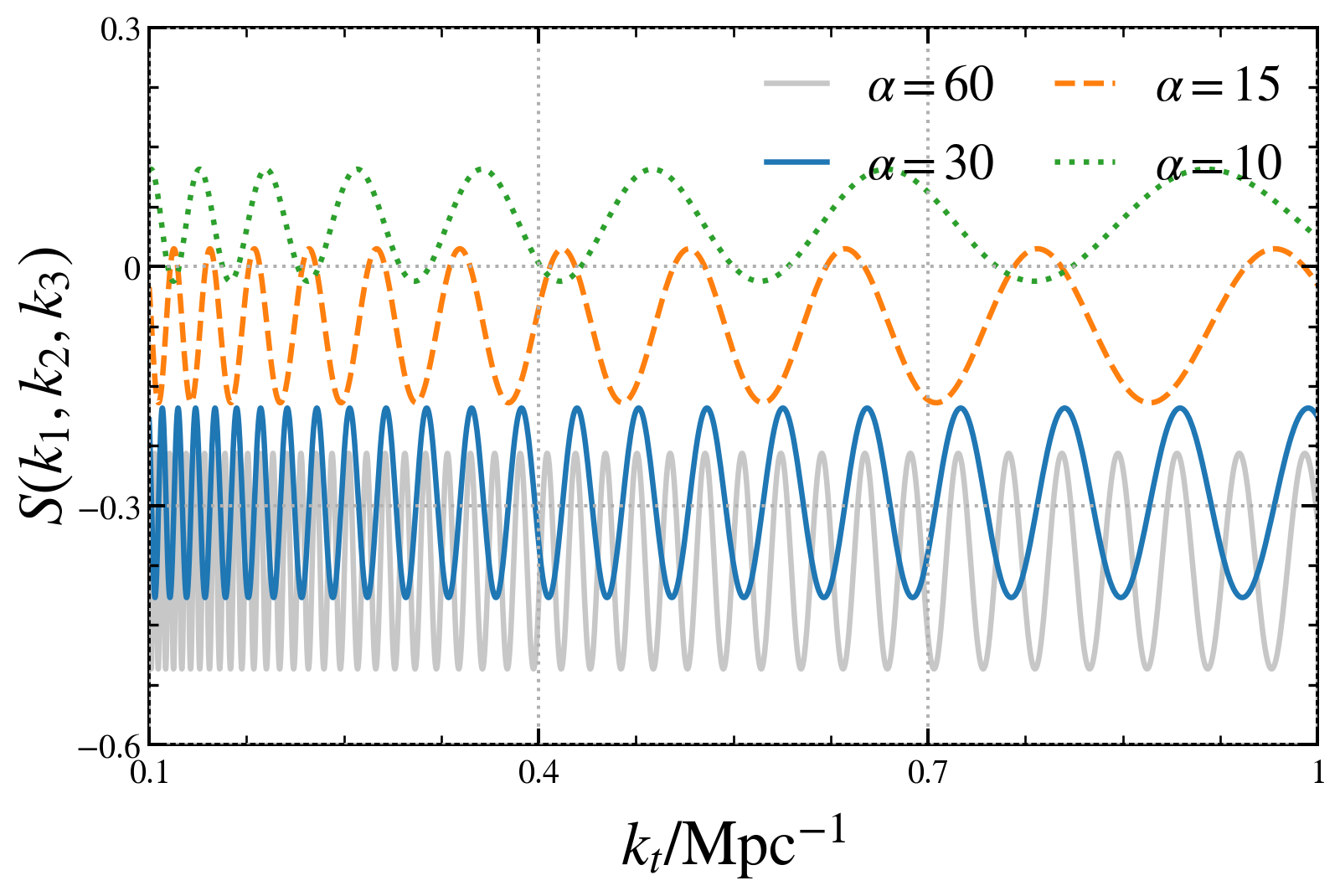}
        \caption{$k_t = k_1+k_2+k_3,~ k_1=k_2=k_3$}
        \label{4.2F3.2}
    \end{subfigure}   
    \caption{Oscillations in the bispectrum (\ref{3.3E5}) from the $\phi\dot\phi\s$ cubic interaction: (a) collider signal in the squeezed limit and (b) scale-dependent resonant features, with $\mu = 10$, $\alpha_1 = \alpha_2 = 10,15,30,60$. We set the sizes of the quadratic and cubic couplings to unity.}
    \label{4.2F3}
\end{figure}

A similar analysis applies to the $\dot\phi\phi\s$ interaction, which as we shall see in the next section, is the main contribution to the exchange bispectrum in a concrete model of axion monodromy.
In the squeezed limit the three-point function (\ref{3.3E5}) becomes
\begin{align}
    \lim_{k_3\ll k_1=k_2}\langle\phi\phi\phi\rangle'_{\dot{\phi}\phi\sigma}
    &\equiv -\frac{\lambda_{\mathrm{quad}}^{\dot{\phi}\sigma}\lambda_{\mathrm{cub}}^{\dot{\phi}\phi\sigma}}{8 H^2}|E_1^B||F^{\dot{\phi}\phi\sigma}(\mu, \alpha_2)|P_{\phi}(k_1)P_{\phi}(k_3)\tilde{S}^{\dot{\phi}\phi\sigma}(k_1=k_2,k_3),
    \label{4.2E2}
\end{align}
where we have $F^{\dot{\phi}\phi\sigma}(\mu, \alpha_2)\equiv 5+i\mu+i\alpha_2$. The shape function $\tilde{S}^{\dot{\phi}\phi\sigma}(k_1, k_3)$ takes the same form as \eqref{sqshape_dS} while with different phase angles
\begin{align}
    &\theta_1^{\dot{\phi}\phi\sigma}=-(2\mu+\alpha_1+\alpha_2)\log 2+\mathrm{Arg}[E_1^BF^{\dot{\phi}\phi\sigma}(\mu, \alpha_2)];\nonumber\\[10pt]
    &\theta_2^{\dot{\phi}\phi\sigma}=-(2\mu+\alpha_1-\alpha_2)\log 2-\mathrm{Arg}[iE_2^BF^{\dot{\phi}\phi\sigma}(-\mu, \alpha_2)],
    \label{phase_singledot}
\end{align}
and a different relative factor $\Delta f^{\dot{\phi}\phi\sigma}\equiv {|E_2^B(\mu, \alpha_1, \alpha_2)F^{\dot{\phi}\phi\sigma}(-\mu, \alpha_2)|}/{|E_1^B(\mu, \alpha_1, \alpha_2)F^{\dot{\phi}\phi\sigma}(\mu, \alpha_2)|}$, which remains of order $\mathcal{O}(1)$ for $\alpha_1 \sim \alpha_2 \sim \mu$. 
The shapes and runnings of this type of bispectra are plotted in Figure \ref{4.2F3}. The 2D plot for oscillatory patters similar to Figure \ref{fig:dS_shape} was presented by Figure 3 in the companion paper \cite{Pajer:2024ckd}. 

\section{On the UV Sensitivity of Axion Monodromy}
\label{sec:axionM}

The bootstrap computation above has important UV implications on inflation theories. In this section, we focus on the axion monodromy models from a UV-inspired construction of string cosmology, and discuss the validity of the single field effective description.
The key result has been presented in a short companion paper \cite{Pajer:2024ckd}, while in this section we provide a more systematic treatment with technical details.

\vskip5pt
Originated from string theory, the 4D effective theory of axion monodromy normally corresponds to {\it single field} slow-roll inflation with small oscillatory terms in the potential. As a result, one typical signature of this class of models is the resonant features in cosmological correlators \cite{Chen:2008wn,Flauger:2009ab,Flauger:2010ja,Chen:2010bka,Behbahani:2011it,Leblond:2010yq,Cabass:2018roz} (also see \cite{DuasoPueyo:2023viy,Creminelli:2024cge} for recent discussions). 
If we take a more careful look at the UV embedding, in principle
its full effective description should also include couplings to heavy states, such as
the stabilised moduli fields. Naively, one may expect these UV physics can be integrated out and will not affect the low-energy EFT. 
However, the oscillating background may completely change this oversimplified picture.
For instance, the effects of heavy physics in axion monodromy have been studied previously in \cite{Dong:2010in,Flauger:2016idt,Pedro:2019klo,Chen:2022vzh,Bhattacharya:2022fze,Chakraborty:2019dfh}.
As we have shown, resonances may excite the heavy fields continuously, and leave enhanced signals in the cosmological correlators. Thus for this type of models, one particularly interesting question is whether there are UV-sensitive signatures that cannot be captured by the single-field EFT. In the following we shall tackle this problem by using our bootstrap computation of resonant cosmological collider.

\subsection{A String Inflation Model}

Let's first give a brief review about some generic features of compactifications in string cosmology. 
One can find a more detailed and systematic introduction in \cite{Baumann:2014nda,Kaloper:2011jz}. 
Normally we start with a 10D supergravity theory, and then perform the dimensional reduction $\mathcal{M}_{10}\rightarrow \mathcal{M}_4\times X_6$ to get back to our 4D spacetime $\mathcal{M}_4$. The extra dimensions $X_6$ are compactified in microscopic scales.  
Thus at low energies, we are left with an effective field theory in 4D, where the compactification of $X_6$ in general leads to a large number of fields. For the purpose of this work, we shall focus on two of them -- axions and moduli. 

\begin{itemize}
    \item {\it Axion}. In string compactification, an axion field $\theta$ arises from the integration of a gauge potential over nontrivial cycles in the compact dimensions $X_6$. The kinetic term is given by $\frac{1}{2}f_\theta^2 (\partial_\mu \theta)^2$, where $f_\theta$ is the axion decay constant that carries information about the extra dimensions.
    Meanwhile, due to non-perturbative instanton effects, the continuous shift symmetry $\theta\rightarrow \theta+\mathrm{const}.$ is broken to a discrete one $\theta\rightarrow \theta+2\pi$. As a result, we find a periodic correction to the potential
    \be \label{modulation}
    \Delta V(\theta) = \Lambda_{uv}^4\sum_n c_n\cos \(n\theta\) ,
    \ee
    where $c_n=e^{-n S}$ and $S$ is the action for one instanton. Thus this correction is dominated by the $n=1$ term, and its size is determined by a dynamically generated scale $A^4=\Lambda_{uv}^4e^{-S}$.
    \item {\it Radion moduli}. This type of moduli fields control the volume of the extra dimensions $\mathcal{V}$. 
    As a general consequence of dimensional reduction, constants in the 4D theory, such as the Planck mass, are derived quantities depending on the radion modulus. 
    For the 4D Lagrangian of the axion field above, both the axion decay constant and the instanton action in $c_n$ are functions of the moduli fields.
    The simplest example is to consider an isotropic compactification of $X_6$ with characteristic length $L$. Then its volume becomes $\mathcal{V} \propto L^6 \propto \exp{(\sqrt{3}\rho / \mpl)}$, where $\rho$ is a canonically normalized modulus field describing volume fluctuations.  
    In this case, one find an axion decay constant $ f_\theta^2 \propto \mathcal{V}/L^4= f^2 \exp{({\rho}/{\sqrt{3}\mpl})}~\cite{Silverstein:2008sg,McAllister:2008hb} $, where $f$ is the stabilised value at $\rho=0$. Thus the kinetic term from the UV provides a Planck-suppressed coupling between axion and moduli fields. We may go beyond this simplest isotropic example and consider compactified geometries with hierarchically different volumes \cite{Balasubramanian:2005zx}, which generates stronger axion-modulus couplings in the kinetic term.\footnote{We acknowledge stimulating discussions on the string embeddings with Gonzalo Villa.}

\end{itemize}

\begin{figure}
    \centering
    \includegraphics[width=0.6\linewidth]{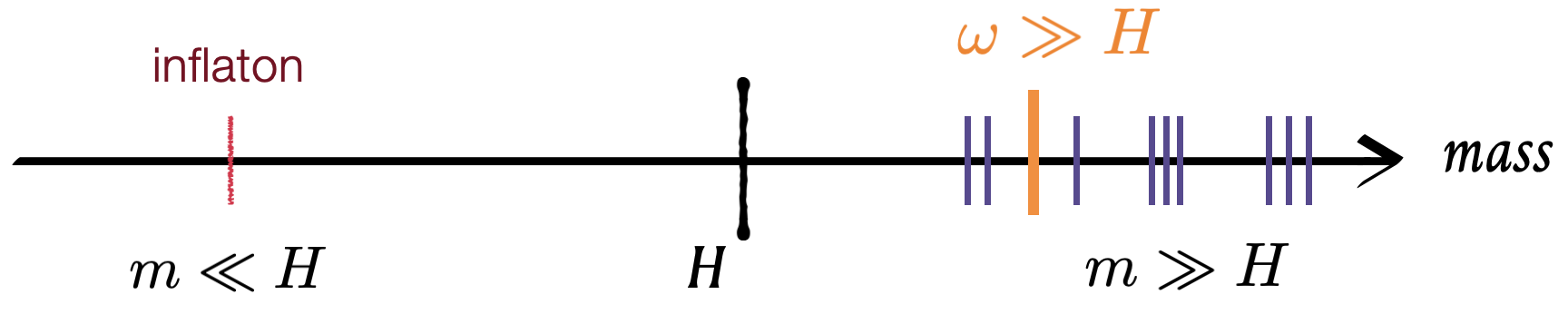}
    \caption{Mass scales in axion monodromy. Conventionally, like many string inflaton models, we treat the heavy states as decoupled and get a single-field EFT at low energies. However, the axion oscillating frequency $\omega$ appears as a new heavy scale, which may change this naive picture.}
    \label{fig:mass}
\end{figure}

For all fields in the 4D description, normally we assume there is a large hierarchy between one light scalar and additional heavy states, as shown in Figure \ref{fig:mass}.
This leads to the conventional single field inflation scenario, where the light scalar plays the role of the inflaton, and the extra fields are stabilized with $m\gg H$.
As one particular realization of string inflation, the basic idea of axion monodromy is to take one axion field as the inflaton candidate and generate a slow-roll potential by using a monodromy, which means the axion discrete shift symmetry is further broken and its potential becomes multi-valued. 
In its UV embeddings, we may simply expect that the heavy states are also present, but normally they can be integrated out and then become decoupled from the low-energy single-field description. But this naive expectation may be a bit too quick, as a {\it new scale} is introduced due to the periodic modulation of the axion potential in \eqref{modulation}. 
This new scale corresponds to the frequency of the background oscillations of the inflaton, which is likely to be higher than the masses of certain heavy states. One natural question then is whether we can still integrate out these heavy states to achieve a single-field effective theory.

\vskip5pt
In the following we shall try to tackle the above question by carefully re-examining a two-field toy model of axion monodromy.
Our consideration includes an axion field $\Phi\equiv f\theta$ as the inflaton and a canonically normalized radial modulus field $\rho$. The two-field Lagrangian is given by
\be \label{model}
\mathcal{L} = -\frac{1}{2} L(\rho)(\partial \Phi)^2 - \frac{1}{2} (\partial \rho)^2 - V(\Phi, \rho) ,
\ee
where  $L(\rho)$ parametrizes the geometry of the compactification manifold. Here we assume $\rho$ stabilized around $0$ is the lightest modulus field that controls of the volume of the compactification. The coupling in the axion kinetic term takes the following form
\be
L(\rho) = e^{\rho/\Lambda} \simeq 1+\frac{\rho}{\Lambda} + \frac{\rho^2}{2\Lambda^2} + ...~,
\ee
where $\Lambda$ is a scale that can be as high as the Planck mass.
For $\Lambda\sim \mpl$, we have a weak coupling and return to the simple case of isotropic compactification.
In general, we may lower $\Lambda$ to parametrize the nontrival geometries of the extra dimension and achieve sizable couplings between the axion and modulus.
The potential is given by 
\be
 V(\Phi, \rho) = V_{\rm sr}(\Phi) + \tilde{c}_1(\rho) A^4 \cos\(\frac{\Phi}{f}\) + W(\rho) .
\ee
Here $V_{\rm sr}(\Phi)$ from the monodromy effect is assumed to admit  slow-roll dynamics for the inflaton: one simple example in axion monodromy is the linear potential $V_{sr} \propto \Phi $. The second term is the small periodic modulation from \eqref{modulation}, where the axion decay constant is sub-Planckian with $f\ll \mpl$. 
We introduce $\tilde{c}_1$ to parametrize the $\rho$-dependence of this correction\footnote{In Ref.\cite{Pajer:2024ckd}, we neglect its $\rho$-dependence and set $\tilde{c}_1=1$ to highlight the effects of the kinetic coupling.}
\be
\tilde{c}_1 = 1+ \frac{\rho}{\tilde\Lambda}~,
\ee
where $\tilde\Lambda$ controlling the coupling strength here is another scale that can be as high as the Planck mass.
The modulus potential $W(\rho)=b \rho+\frac{1}{2}m^2\rho^2$ is supposed to stablize the field around $\rho=0$ with $m\gg H$. Here we also introduce the linear term in order to cancel the tadpole caused by the background motion of the axion field, which will be shown very soon. 

\vskip5pt
Several scales are involved in this system: $f$, $\Lambda$, $\tilde\Lambda$, $A$, $m$ and the slow-roll field velocity $\dot\Phi^{1/2}\simeq 60H$. 
For the parameter regime of interest, there are certain requirements from UV completion and phenomenology. Meanwhile, two dimensionless parameters are needed for the later convenience of model analysis.  We leave the details in the next subsection and simply collect the results here 
\be \label{conditions}
\alpha \equiv \frac{\dot\Phi}{Hf} \gg 1~,~~~~~~ b_* \equiv \frac{A^4}{V_{\rm sr}' f} \ll 1~,~~~~~~ \Lambda,\tilde\Lambda \gg \frac{\dot\Phi}{\sqrt{\alpha}H}~,~~~~~~ \frac{\dot\Phi}{\Lambda}\ll H.
\ee
The first condition simply means that the frequency of the oscillation is much higher than the Hubble scale $\omega\equiv \dot\Phi/ f \gg H$, or equivalently the axion decay constant is much smaller than the Planck mass $f\ll 3600H \simeq \sqrt{2\epsilon} \mpl$. The second condition ensures the monotonicity of the potential, which basically means that the periodic modulation must be very small. 
These two conditions are the standard ones used in single field axion monodromy models.
The third condition is a special one for our two-field extension, which is needed to exclude large deviations from the inflaton trajectory from the axion direction.
The last one is the stability condition of the modulus field such that the mass correction from the kinetic mixing is small.

\subsection{Background Dynamics}

In FRLW spacetime, the background equations for this two-field system are given by
\be
L(\rho) \( \ddot\Phi +3H \dot \Phi \) + L'(\rho) \dot\rho \dot\Phi + V_{\rm sr}'(\Phi) - \tilde{c}_1 \frac{A^4}{f} \sin\(\frac{\Phi}{f}\) =0 ~,
\ee
\be
\ddot\rho + 3 H\dot\rho + W'(\rho) +\tilde{c}_1'(\rho) A^4 \cos\(\frac{\Phi}{f}\) -\frac{1}{2}L'(\rho) \dot\Phi^2=0~,
\ee
\be
3H^2\mpl^2 = \frac{1}{2}L(\rho) \dot\Phi^2 + \frac{1}{2} \dot\rho^2 +V_{\rm sr}(\Phi) + \tilde{c}_1 A^4 \cos\(\frac{\Phi}{f}\) + W(\rho) ~.
\ee
Since the oscillatory modulation is small at the background level, we can treat this part as a perturbation. Then the background solutions can be expanded as $\Phi=\Phi_0+\Phi_1$,  $\rho=\rho_0+\rho_1$ and $H=H_0+H_1$. 
Then the $0$th order solution with no modulation is given by the slow-roll result
\be \label{0th}
\dot\Phi_0 = -\frac{V_{\rm sr}'(\Phi_0)}{3H_0}~, ~~~~~~ \frac{1}{2}L'(\rho_0) \dot\Phi_0^2 =  W'(\rho_0)~,~~~~~~ 3H_0^2\mpl^2 = V_{\rm sr}(\Phi_0)~. 
\ee
The first and third are the standard single field slow-roll equation.
For simplicity, we assume that $\dot\Phi_0$ is constant in our analysis.
The second equation is the centrifugal force equation in the $\rho$ direction. Without losing generality, we choose a linear term in the modulus potential to stabilize the modulus at  $\rho_0 = 0$, and we have used $L(\rho_0)=1$.

\vskip5pt
Equations for the first-order background quantities ($\Phi_1$, $\rho_1$,  $H_1$) are given by
\begin{small}
\begin{gather} \label{Phi1}
\ddot\Phi_1 +3H_0 \dot \Phi_1+3H_1 \dot \Phi_0 +L'(\rho_0)\rho_1 \( \ddot\Phi_0 +3H_0 \dot \Phi_0 \) + L'(\rho_0) \dot\rho_1 \dot\Phi_0 + V_{\rm sr}''(\Phi_0)\Phi_1 = b_*  V_{\rm sr}' \sin\(\frac{\Phi_0}{f}\) ~,
\\ \label{rho1}
\ddot\rho_1 + 3 H_0\dot\rho_1 + W''(\rho_0)\rho_1 +\tilde{c}_1'(\rho_0) A^4 \cos\(\frac{\Phi_0}{f}\) = L'(\rho_0) \dot\Phi_0\dot\Phi_1+\frac{1}{2}L''(\rho_0)\rho_1 \dot\Phi_0^2~,
\\ \label{Hubble1}
 6H_0 H_1\mpl^2 =  V_{\rm sr}'(\Phi_0)\Phi_1 +  \tilde{c}_1 A^4 \cos\(\frac{\Phi_0}{f}\) + W'(\rho_0)\rho_1 ~.
\end{gather}
\end{small}Approximate solutions can be achieved by using the assumption that we have oscillations with very high frequency, i.e. $\alpha\gg1$. Because of this, first-order background quantities with more time derivatives are more dominant. In addition, the third condition in \eqref{conditions} ensures that the $L'(\rho)$ terms are subleading in \eqref{Phi1}. 
Thus, for $\Phi_1$, only the first term on the left hand side of \eqref{Phi1} will contribute, and this becomes the equation for an oscillator driven by an oscillatory force. Its solution is given by
\be
\Phi_1 (t) = - \frac{A^4 f}{\dot\Phi_0^2} \sin\(\frac{\Phi_0(t)}{f}\)~,
\ee
which is the same as the one in single field axion monodromy models.
For equation \eqref{rho1}, the second term on the right hand side is subleading once we take into account the last condition in \eqref{conditions}. 
We can use the solution $\dot\Phi_1$ for the first term there, which then becomes similar with the last term on the left hand side. Then the equation has the following approximate solution 
\be
\rho_1 (t) = \sum_\pm C_\pm e^{-\(\frac{3}{2}\pm i\mu\)H_0 t }  - {A^4}\(\tilde{c}_1'(\rho_0)+L'(\rho_0)\) \frac{ 3\omega H_0 \sin \(\frac{\Phi_0(t)}{f}\) - (\omega^2-m^2) \cos\(\frac{\Phi_0(t)}{f}\)   }{9\omega^2 H_0^2 + (\omega^2-m^2)^2}~,
\ee
where $m^2=W''(\rho_0)$, $\omega = \dot\Phi_0/f$ and $\mu = \sqrt{m^2/H^2 - 9/4}$.
The two homogeneous solutions correspond to the oscillatory decaying behaviour of a heavy field in dS, with a frequency $\mu$ determined by the mass of moduli. The last term is the inhomogeneous solution driven by the cosine modulation of the potential.  In the following, we set $C_\pm=0$ and study the effects of the term oscillating with $\alpha$, which is a unique feature of axion monodromy.
From \eqref{Hubble1}, we can check that the oscillatory component of the Hubble parameter is further suppressed.

\vskip5pt
In summary, the above background analysis shows that, first, due to the periodic modulation of the axion potential, $\Phi$ acquires a small oscillatory part on top of its slow-roll constant motion; then, the `stabilized' modulus also becomes oscillating due to the centrifugal force.
It is informative to rewrite the leading oscillatory part of the background solution in terms of fewer parameters, like $\alpha$ and $b_*$, such that we can compare the sizes.
Two quantities are given by
\be
\dot\Phi_1 = - \frac{A^4}{\dot\Phi_0} \cos \(\frac{\Phi_0(t)}{f}\) = \frac{3b_*}{\alpha} \dot\Phi_0 \cos\(\frac{\Phi_0(t)}{f}\) ~,~~~~
\rho_1 = B \cos \(\frac{\Phi_0(t)}{f}+\delta \)~,
\ee
with 
\be
B= -{A^4}\(\frac{1}{\Lambda}+\frac{1}{\tilde{\Lambda}}\) \frac{1}{\sqrt{9\omega^2 H_0^2 + (\omega^2-m^2)^2}}~,~~~~\delta = \arcsin{\frac{m^2-\omega^2}{\sqrt{9\omega^2 H_0^2 + (\omega^2-m^2)^2}}} ~.
\ee
We can check that indeed the approximate solution is valid  as long as the conditions in \eqref{conditions} are satisfied. 
We also see that $\dot\Phi_1 \ll \dot\Phi_0$, which confirms that the oscillatory solutions are small perturbations of the slow-roll background.
To estimate the amplitude of $\rho_1$, let's consider a parameter regime $\Lambda\simeq\tilde{\Lambda}$ and $\omega\gtrsim m\gg H$, and then take $B \simeq  b_* f^2/\Lambda$  as a benchmark example. The relative ratio of the two oscillatory field velocities here is given by
\be \label{ratio}
\frac{\dot\rho_1}{\dot\Phi_1} \sim \alpha \frac{f}{\Lambda}~.
\ee
Thus the oscillation in the radial direction can become more significant when we have large frequency and small $\Lambda$. One cartoonish illustration of this wiggly trajectory is shown in Fig. 1 in \cite{Pajer:2024ckd}. 
At the background level, one important observation in this two-field example is that the heavy moduli cannot be seen as stabilized like before, which is expected in general for axion monodromy inflation with cosine potentials.
We should notice that this type of trajectories with constant wiggles essentially differs from those with damped oscillations. In multi-field literature, they are normally consequences of sharp turns and heavy masses \cite{Chen:2011zf,Shiu:2011qw,Gao:2012uq,Chen:2014cwa}, instead of the periodic modulation of the axion potential.

\vskip5pt
For the convenience of the perturbation analysis, here we also introduce the covariant formalism of multi-field inflation. In this description, the inflaton trajectory in field space is decomposed into tangent and normal directions  at each point. Our toy model is in a 2D hyperbolic field space with metric $G_{ab}={\rm diag}\{1,L(\rho)\}$, where the 0th order solution corresponds to a non-geodesic motion along $\Phi$ direction only, while the 1st order solution also incorporates $\rho$-direction oscillations. The tangent and normal unit vectors are defined as 
\be
T^a\equiv \frac{\dot\Phi^a}{\dot\Phi_{\rm t}} =  \frac{1}{\dot\Phi_{\rm t}}\(\dot\rho_1 ,\dot\Phi_0 + \dot\Phi_1\)~,~~~~ 
N^a\equiv \sqrt{\det G} \epsilon^{ab} T_b = \frac{L(\rho_1)}{\dot\Phi_{\rm t}}\(\dot\Phi_0 + \dot\Phi_1, - \frac{\dot\rho_1}{L(\rho_1)} \)~,
\ee
with the total field velocity $\dot\Phi_{\rm t}^2 \equiv\dot\rho_1^2 + L (\dot\Phi_0+\dot\Phi_1)^2 $. The first slow-roll parameter is given by
\bea
\epsilon &\equiv & \frac{\dot\Phi_{\rm t}^2}{2H^2\mpl^2} = \epsilon_0+\epsilon_1+... = \frac{\dot\Phi_{\rm 0}^2}{2H^2\mpl^2} + \frac{\dot\Phi_{0}\dot\Phi_{1}}{H^2\mpl^2}+...~,
\eea
where we expand it into the slow-roll and leading oscillatory components.
In this approach, we have the covariant definition of the turning rate of the inflaton trajectory 
\bea \label{turn}
\Omega &\equiv& -N_a D_t T^a   \\
&=& \frac{1}{2\sqrt{L}\dot\Phi^2_{\rm t}} \[ \(\dot\Phi_0+\dot\Phi_1\) \(2L'(\rho)\dot\rho_1^2 + L(\rho) L'(\rho)(\dot\Phi_0+\dot\Phi_1)^2- 2L(\rho)\ddot\rho_1 \) +2L(\rho) \ddot\Phi_1\dot\rho_1 \]  \nn\\
&\simeq & \frac{L'(\rho_0)}{2\sqrt{L} }\dot\Phi_0 + \frac{L'(\rho_0)}{2\sqrt{L} }\dot\Phi_1 - \frac{\ddot\rho_1}{2\sqrt{L}\dot\Phi_0}  + ...\nn\\
&=&  \frac{\dot\Phi_0}{2\Lambda} \left[ 1 + 3\frac{b_*}{\alpha} \cos
\left(\frac{\Phi_0(t)}{f}  \right) + b_*  \cos\left(\frac{\Phi_0(t)}{f} +\delta  \right) +... \right]
\eea
where in the third step we have applied the conditions in \eqref{conditions} to identify the leading oscillatory contributions, and we have used $B \simeq  b_* f^2/\Lambda$ in the last step.
When $\Omega\neq 0$, the inflaton trajectory is not a geodesic of the field space.
In the final expression, the first term from the 0th order slow-roll background simply corresponds to a constant turn in the hyperbolic field space. The other two are oscillatory contributions of the wiggly trajectory which are subleading. It is important to emphasize that this turning rate is always much smaller than the Hubble scale during inflation, i.e. $\Omega\ll H$, as required by the last condition in \eqref{conditions}. 
Nevertheless, due to the oscillating trajectory, its time dependence can become significant, i.e. $\dot\Omega/(\Omega H) > 1$.
This distinguishes our consideration from other multi-field inflation models with rapid/sharp-turn trajectories \cite{Bhattacharya:2022fze,Chakraborty:2019dfh}.

\subsection{When Heavy Moduli Matter}
\label{sec:moduli}

Now we come to investigate the consequences of the oscillating background on inflationary fluctuations. 
The goal is twofold here: one on hand, we would like to re-examine the validity of the single field effective theory of axion monodromy; on the other hand, we aim to identify the most important interactions between two scalar fields to prepare for the phenomenological studies.

\vskip5pt
Let's first recapitulate the conventional computation where we integrate out heavy fields at the perturbation level to achieve a single field inflation \cite{Tolley:2009fg, Achucarro:2010da, Baumann:2011su,Achucarro:2012sm}.
In the flat gauge, the two field perturbations are given by
\be
\Phi(t,{\bf x}) = \Phi_0(t)+\Phi_1(t)+\phi(t,{\bf x})~,~~~~\rho(t,{\bf x})=\rho_1(t) + \delta\rho(t,{\bf x})~.
\ee
If we neglect the oscillating components and just look at the 0th order solution \eqref{0th}, the background evolution corresponds to a constant-turn trajectory along the flat $\Phi$ direction with a steep valley given by the heavy $\rho$ mass. 
The turning rate $\Omega_0$, which is the first term in \eqref{turn}, provides a linear mixing between the two field fluctuations $2\Omega_0 \dot\phi\delta\rho$.
Then the $\rho$ field fluctuation satisfies the following equation of motion
\be \label{integrateout}
\Box\delta\rho -m^2\delta\rho = -2 \Omega_0 \dot\phi ~~~~~
\Rightarrow ~~~~~ \delta\rho\simeq \frac{2\Omega_0}{m^2}\dot\phi~,
\ee
where $\Box \equiv- {\partial_t^2}-3H{\partial_t}+a^{-2}\partial_i^2$ can be neglected in the heavy field limit $m\gg H$.
Next, to integrate out $\delta\rho$, we substitute the approximate solution above into the quadratic action, which leads to the standard result of single-field EFT with a reduced inflaton sound speed
\be
c_s^{-2} -1 = \frac{\dot\Phi_0^2}{\Lambda^2 m^2} = 2\epsilon \frac{H^2\mpl^2}{m^2\Lambda^2}~.
\ee
The reduction of the sound speed becomes significant if we consider strong couplings from small $\Lambda$, and then we expect enhanced self-interaction of the inflaton and equilateral non-Gaussianity.
But for sufficiently large $\Lambda$, $c_s\rightarrow 1$ and we find a canonical single field model where the heavy moduli are decoupled.

\vskip5pt
However, the oscillating part of the background solution may completely change this conventional picture of single field EFT.
The first thing to notice here is that once $\Phi_1$ and $\rho_1$ are included, the perturbative analysis becomes more complicated.
In the perturbed Lagrangian, we find the following quadratic and cubic interactions of the axion and modulus fields
\be \label{int23}
\mathcal{L} = L'(\rho_1) {(\dot\Phi_0+\dot\Phi_1)} \dot\phi\delta\rho - \frac{1}{2} L'(\rho_1) \(\partial\phi\)^2\delta\rho 
+ \tilde{c}_1'(\rho_1) \frac{A^4}{f} \sin\(\frac{\Phi}{f}\) \phi\delta\rho
+ \tilde{c}_1'(\rho_1) \frac{A^4}{f^2} \cos \(\frac{\Phi}{f}\) \phi^2\delta\rho~,
\ee
where $L' \simeq ({1} + {\rho_1}/{\Lambda})/{\Lambda}$.
So all the couplings become oscillatory and we find non-derivative mixings from the potential between the two field fluctuations. 
But this result is less informative and helpful. One reason is that field fluctuations are not observable, and in the end we need to look at the primordial curvature perturbation $\zeta$. 
While normally we can use $\zeta = (H/\dot\Phi)\phi$,
in cases with an oscillating background this simple relation may miss certain effects of oscillatory couplings. 
Another subtlety comes from the wiggly trajectory, because of which the adiabatic and isocurvature modes are no longer well approximated by $\phi$ and $\delta\rho$ field fluctuations.  In the covariant formalism, we set these fluctuations as a vector in the field space $\phi^a=(\delta\rho, \phi)$. Then at the linear order, the canonically normalized field fluctuations along and orthogonal to the trajectory now are given by
\be
\phi_T  = \phi^aT_a = \frac{L(\dot\Phi_0+\dot\Phi_1)}{\dot\Phi_{\rm t}} \phi 
 +\frac{\dot\rho_1}{\dot\Phi_{\rm t}} \delta\rho~,~~~~~~\sigma = \phi^a N_a = \frac{\sqrt{L}}{\dot\Phi_{\rm t}} \(-\dot\rho_1 \phi + (\dot\Phi_0+\dot\Phi_1) \delta\rho \) .
\ee
The first order background solution is needed to rotate the field basis along the trajectory.
Here $\phi_T$ corresponds to the adiabatic perturbation that leads to the curvature modes, while $\s$ is the isocurvature ones. 
Beyond the linear order, their definition becomes more complicated in a curved field manifold. 
Certainly we need to address these subtleties before re-examining the effects of heavy moduli and in particular identifying the leading interactions between adiabatic and isocurvature perturbations.

\paragraph{The EFT of inflation with multiple fields}
To simplify the analysis, it turns out the most convenient approach is to consider the multi-field extension of the EFT of inflation \cite{Cheung:2007st}.
We start with the unitary gauge where the field fluctuations along the inflaton direction vanish and the adiabatic perturbations are directly given by the curvature perturbation $\zeta$. 
Then one point in the field space close to the inflaton trajectory $\Phi^a(t)$ can be expressed as
\be
\Phi^a (t,{
\bf x})= \Phi^a (t) + \sigma(t,{
\bf x}) N^a(t)~.
\ee
To identify the leading couplings, we also take the decoupling limit $\epsilon\rightarrow 0$ where the gravitational interactions are negligible.
This is justified as we are mainly interested in the resonance effects that happen deep inside the horizon. 
Then the perturbed Lagrangian takes a simpler form, and 
the mixing between adiabatic and isocurvature perturbations comes from the kinetic term of the two-field system 
\bea \label{g00s}
-\frac{1}{2}G_{ab} \partial_\mu \Phi^a \partial^\nu \Phi^b &\rightarrow&
-\frac{1}{2} g^{\mu\nu} G_{ab} \[ \delta_\mu^0 \dot\Phi^a_B +  \partial_\mu\( \sigma N^a \)   \] \[ \delta_\nu^0 \dot\Phi^b_B +  \partial_\nu\( \sigma N^b \) \] \nn\\
&\subset & - \dot\Phi_{\rm t}  G_{ab} g^{0\mu} T^a  \partial_\mu (\sigma N^b)  \nn\\
&\subset &
 \lambda(t) \delta g^{00} \sigma ~,
\eea
where we have used $G_{ab}T^aN^b=0$ and \eqref{turn} in the last step.
Note that the coupling $\lambda(t)$ is highly oscillatory 
\bea
\lambda \equiv -\dot\Phi_t\Omega  & \simeq & \frac{L'(\rho_0)\dot\Phi_0^2}{2}
+ {L'(\rho_0)\dot\Phi_0 } \dot\Phi_1 - \frac{1}{2} \ddot\rho_1 +... \simeq -\frac{\dot\Phi_0^2}{2\Lambda} \[1+ b_*\cos\( \frac{\Phi_0(t)}{f} + \delta \) \]~,
\eea
with a strong time-dependence $\dot\lambda\sim \omega\lambda$.
In this analysis we have neglected the interactions related to the field space curvature, which only appear in vertices with $\sigma^2$ or higher powers of $\sigma$, and their couplings are suppressed by $1/\Lambda^2$ at least. 

The next procedure of the EFT approach is to perform the gauge transformation to the flat gauge and introduce the Goldstone $\pi$ of the time diffeomorphism breaking. In the decoupling limit, the 00-component of the metric transforms as $\delta g^{00} \rightarrow -2\dot\pi +(\partial_\mu\pi)^2$.
Then the EFT operator \eqref{g00s} leads to the quadratic and cubic interacting Lagrangian in terms of $\pi$ and $\s$ as
\be \label{mix}
\mathcal{L}_{\rm mix} = \lambda(t+\pi) \[-2\dot\pi +(\partial_\mu\pi)^2\] \s = -2\lambda(t)\dot\pi\s -2\dot\lambda(t) \pi\dot\pi\s + \lambda(t) (\partial_\mu\pi)^2 \s +...
\ee
Note that since the coupling $\lambda(t+\pi)$ is oscillating with time, its strong time dependence generates a new cubic interaction $\pi\dot\pi\s$.
As its coupling has $\dot\lambda\gg \lambda H$  for very high-frequency oscillations, this cubic term becomes more important than $\lambda(\partial_\mu\pi)^2 \s$.
It is remarkable to note that compared to \eqref{int23}, great simplification has been achieved in the EFT language. For highly oscillatory multi-field trajectories, we can use one single EFT operator \eqref{g00s} to capture the leading interactions between two types of perturbations.
This simple result is in agreement with the full computation of the cubic action of multi-field inflation with a generic curved field manifold \cite{Garcia-Saenz:2019njm}.

\paragraph{The validity of the single field effective theory}
Next let's re-examine the regime of validity for the single field EFT. We repeat the process of integrating out the heavy fields at the beginning of this subsection, but change our focus to $\pi$ and $\s$ instead. At the linear level, the mixing between the two is given by $-2\lambda\dot\pi\s$, and the $\s$ equation of motion becomes
\be
\(\Box  - m_{\rm eff}^2\) \s = -2 \lambda \dot\pi~,
\ee
where $m_{\rm eff}^2\simeq m^2$ in our consideration.\footnote{
As well known in multi-field literature, the full effective isocurvature mass has two corrections $m^2_{\rm eff} = m^2 + (\dot\Phi_t^2/2)\mathbb{R} -\Omega^2$
with $\mathbb{R}=-2/\Lambda^2$ being the curvature of the hyperbolic field space. In this work we are interested in the regime with $\Omega\sim \dot\Phi_0/\Lambda \ll H$, and thus both the field space curvature and turning corrections are negligible.}
The major difference with the conventional analysis around \eqref{integrateout} is that the coupling of the linear mixing becomes highly oscillatory. If we integrate out the heavy field $\s$ using the approximate solution $\s_0\simeq (2\lambda/m^2)\dot\pi$ as before, we need to make sure that the $\Box\s$ term is still subleading. However, from the strong time dependence of the coupling $\lambda$ we find $\ddot\s_0 \simeq \omega^2\s_0$. Therefore, with a new scale $\omega$ in the system (see Figure \ref{fig:mass}), we are only allowed to integrate out $\s$ when $\omega\ll m$.
Notably, this is known as the adiabaticity condition $\dot\Omega /\Omega \ll m $ that is commonly used as a criterion for the validity of the single field EFT, especially for sharp-turn trajectories \cite{Cespedes:2012hu,Achucarro:2012yr}.

Meanwhile, in stringy embeddings if the lightest modulus field is lighter or comparable to the oscillating frequency of the axion, which is relatively common for UV-complete theories of axion monodromy, then this heavy field will get continuously excited by the oscillating background. In that case, the single field description breaks down. We need a full multi-field treatment that manifests the resonance between the oscillatory coupling and the heavy modulus, as we demonstrated in the bootstrap computation. 
In the next subsection, we shall investigate the new phenomenologies in the two-field regime of axion monodromy inflation with $\omega\gtrsim m$.

\subsection{Phenomenology}

Now we make the explicit connection between the two-field axion monodromy and the bootstrap analysis of resonant cosmological colliders in the previous section.
We shall consider the primordial power spectrum and bispectrum as the main observables, and analyze the sizes of the new signals from the resonance effects between the heavy field and oscillatory couplings.

In the previous subsection, we have identified the mixings between adiabatic and isocurvature perturbations in our two-field extension of axion monodromy inflation.
In terms of the primordial curvature perturbation $\zeta\equiv -H\pi$, the leading interactions in \eqref{mix} now become
\begin{equation}
    \mathcal{L}_{\mathrm{mix}}=(\bar{g}+g_2)\dot{\zeta}\sigma+g_3\zeta\dot{\zeta}\sigma+(\tilde{g}+\tilde{g}_3)(\partial_{\mu}\zeta)^2\sigma,
    \label{5.4E1}
\end{equation}
where the couplings are explicitly given by
\begin{align}
    &~~~~~~~~~~~~~~~~~~\bar{g}=-\frac{\dot{\Phi}_0^2}{H\Lambda}, ~~~~~~~~~~~~~~g_2=\bar{g}b_*\cos\left(\alpha Ht+\delta\right)
    \label{5.4E2}, \\ 
   & g_3=\bar{g}\alpha b_*\sin(\alpha Ht+\delta),~~~~~~~
    \tilde{g}=\frac{\bar{g}}{2H}, ~~~~~~~\tilde{g}_3=\tilde{g}b_*\cos\left(\alpha Ht+\delta\right) .
    \label{5.4E3}
\end{align}
With these interactions, we can make use of the analytical results in Section \ref{sec:collider}, and compute the dominant power spectrum corrections and scalar bispectrum associated to the massive field exchanges.

For corrections to the power spectrum,
we notice that the oscillatory quadratic coupling $g_2$ is suppressed by a factor $b_*$ compared to the constant quadratic coupling $\bar{g}$, so the leading-order oscillatory correction to the power spectrum comes from the Feynman diagram with one oscillatory linear mixing and one constant vertex. Using the analytical result (\ref{4.1E22}), we find
\begin{align}
    &\frac{\delta P_{\zeta}(k)}{P_{\zeta,0}(k)}=\delta n^{\mathrm{col.}}\cos\left(\alpha\log k + \theta\right);
    ~~~~~~~~\delta n^{\mathrm{col.}} = -2\epsilon_0\frac{M_{\mathrm{Pl}}^2}{\Lambda^2}|E_1^P(\mu,\alpha)|b_*~,
    \label{5.4E4.2}
\end{align}
where $E_1^P(\mu,\alpha)$ given in \eqref{4.1E18} is of $\mathcal{O}(0.01)$ as shown in Figure \ref{4.1F1.1}.
The contribution of the exchange diagram containing two oscillatory couplings, which has an oscillation of $\cos(2\alpha\log k)$, is suppressed by $b_*^2$. 
From observation, we have constraints imposed by the Planck data on this type of correction $\delta n \lesssim 0.05$ \cite{Planck:2018jri}, which can be easily satisfied for natural choices of the model parameters $\Lambda$ and $b_*$. 

\begin{figure}
    \centering
    \begin{subfigure}{0.485\textwidth}
        \includegraphics[width=\columnwidth]{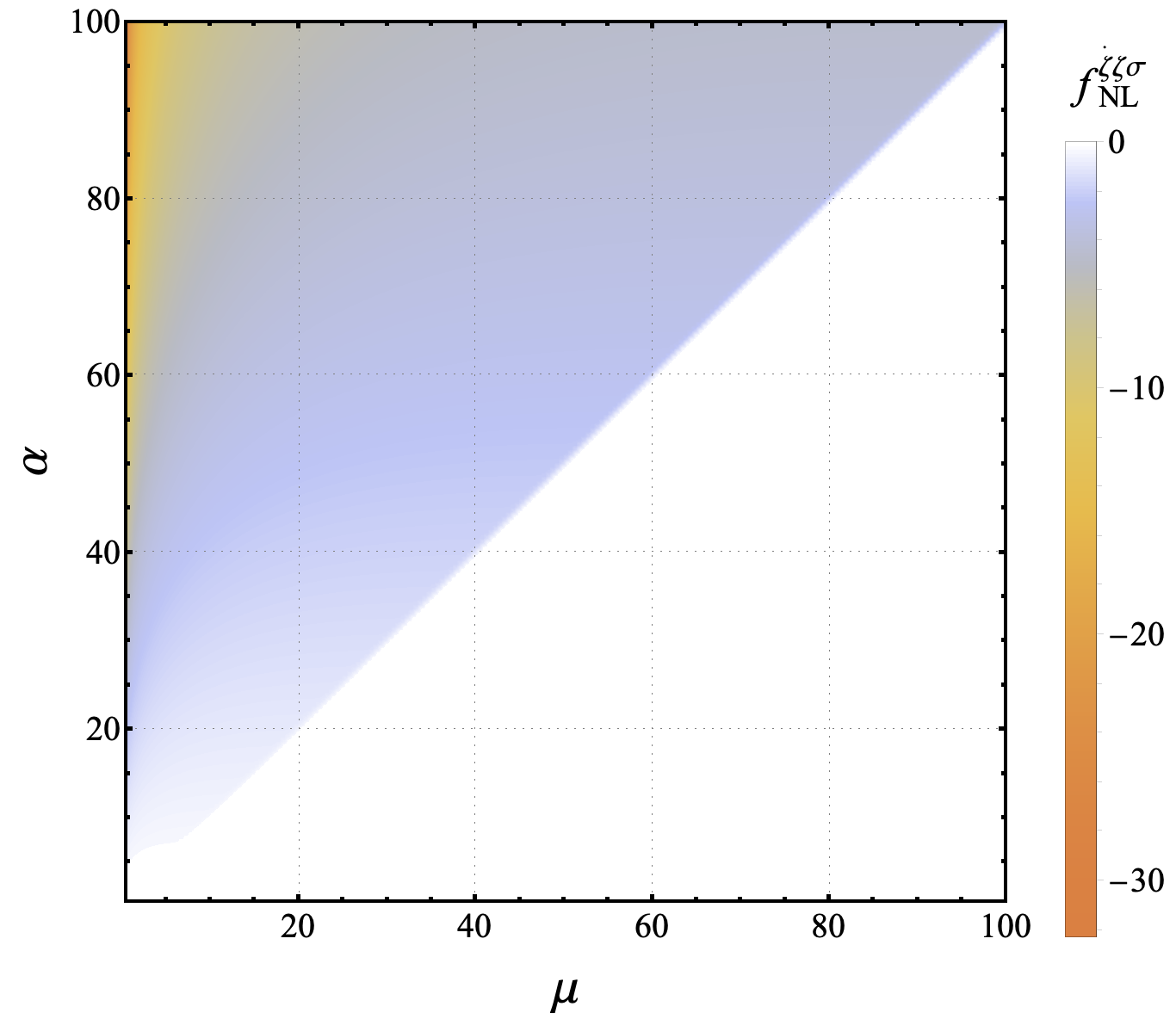}
        \label{5.4F2.1}
    \end{subfigure}
    \hspace{0.1cm}
    \begin{subfigure}{0.485\textwidth}
        \includegraphics[width=\columnwidth]{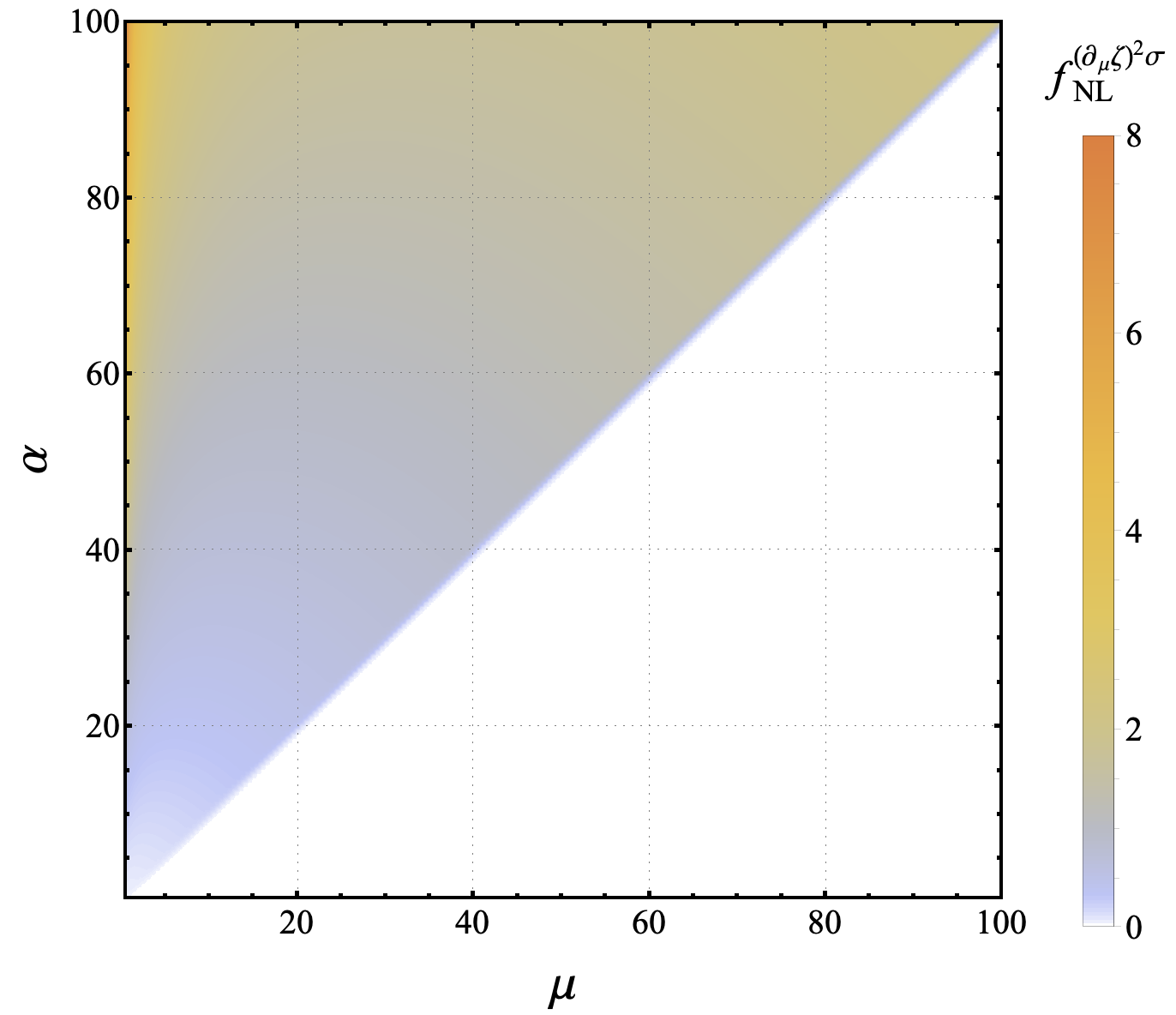}
        \label{5.4F2.2}
    \end{subfigure}   
    \caption{The size of the bispectrum: $f_{\mathrm{NL}}^{\dot{\zeta}\zeta\sigma}$ (left) and $f_{\mathrm{NL}}^{(\partial_{\mu}\zeta)^2\sigma}$ (right) as functions of $\alpha$ and $\mu$. Here we set $\epsilon_0 {M_{\mathrm{Pl}}^2}/{2\Lambda^2}=1$ and $b_*=0.1$.}
    \label{5.4F2}
\end{figure}

For the bispectrum, the dominant contribution arises from the single-exchange diagram involving the quadratic vertex $g_2\dot{\zeta}\sigma$ and the cubic vertex $g_3\zeta\dot{\zeta}\sigma$. This is due to the fact that $g_3$ is enhanced by $\alpha$ compared to the couplings $\tilde{g}$ and $\tilde{g}_3$ for the $(\partial_\mu\zeta)^2\s$ interaction. 
The full shape function is given by \eqref{3.3E5}, which contains both the scale-dependent resonant oscillations and the enhanced shape-dependent collider signal, as we have shown in Section \ref{pheno_ng}. 
Here we mainly focus on the size of the bispectrum.
Matching \eqref{5.4E2} and \eqref{5.4E3} with the convention introduced in Section \ref{pheno_ng}, we obtain $\lambda_{\mathrm{quad}}^{\dot{\zeta}\sigma} = \bar{g}b_*$ and $\lambda_{\mathrm{cub}}^{\dot{\zeta}\zeta\sigma}=\bar{g}\alpha b_*$. 
Substituting these into \eqref{4.2E2} and imposing the hierarchy $\alpha \geq \mu \gg 1$ (with $\alpha\equiv \alpha_1 = \alpha_2$), we derive the resonant cosmological collider signal in squeezed limit
\begin{equation}
    \lim_{k_3\ll k_1=k_2}\langle\zeta\zeta\zeta\rangle'_{\dot{\zeta}\zeta\sigma}=f_{\mathrm{NL}}^{\dot{\zeta}\zeta\sigma}P_{\zeta}(k_1)P_{\zeta}(k_3)\tilde{S}^{\dot{\phi}\phi\sigma}(k_1=k_2, k_3),
    \label{phi_dot_shape}
\end{equation}
with amplitude
\begin{align}
f_{\mathrm{NL}}^{\mathrm{\dot{\zeta}\zeta\sigma}}=-\frac{1}{4}\epsilon_0\frac{M_{\mathrm{Pl}}^2}{\Lambda^2}|E_1^B(\mu, \alpha)||F^{\dot{\phi}\phi\sigma}(\mu,\alpha)|\alpha b_*^2~~\overset{\alpha > \mu \gg 0}{\approx}~~ -\frac{1}{2}\epsilon_0\frac{M_{\mathrm{Pl}}^2}{\Lambda^2}|E_1^B(\mu,\alpha)|\alpha^2 b_*^2
    \label{5.4E6}.
\end{align}
The conventional Boltzmann factor $e^{-\pi \mu}$ is now replaced by $|E_1^B(\mu, \alpha)|$, which as shown in Figure \ref{4.2F1} is of $\mathcal{O}(0.1)$ and no longer exponentially suppressed when $\alpha \geq \mu$. This resonance effect is a universal phenomenon in systems involving massive fields and background oscillations, independent of the specific details of the model. 
Nevertheless, the size of the bispectrum is sensitive to the strength of the couplings, which depends on the model details. 
We notice that it is parametrically easy to achieve large collider signals within the consistent consideration.
For instance, let's consider $b_*=0.1$, and lower $\Lambda$ such that $\epsilon_0{M_{\mathrm{Pl}}^2}/{2\Lambda^2} \sim 1$. 
As shown in the left panel of Figure \ref{5.4F2}, $f_{\mathrm{NL}}^{\dot{\zeta}\zeta\sigma}\sim \mathcal{O}(10)$ can be achieved for various choices of $\alpha$ and $\mu$. 

Another possible contribution to the bispectrum arises from the Feynman diagram associated with the cubic interaction $\tilde{g}_3(\partial_{\mu}\zeta)^2\sigma$. While the coupling constant is not enhanced by the oscillation frequency $\alpha$, we observe that the operator-dependent factor $|F^{(\partial_{\mu}\phi)^2\sigma}|$ in \eqref{F_factor}  increases more rapidly than $|F^{\dot{\phi}\phi\sigma}|$ in the large $\alpha$ regime. 
In this case, we identify $\lambda_{\mathrm{quad}}^{\dot{\zeta}\sigma} = \bar{g}b$ and $\lambda_{\mathrm{cub}}^{(\partial_{\mu}\zeta)^2\sigma}=\tilde{g}b_*$. Subsituting these into \eqref{bisp_dS} with $\alpha_1 = \alpha_2 \equiv \alpha$, we find the squeezed bispectrum
\begin{equation}
    \lim_{k_3\ll k_1=k_2}\langle\zeta\zeta\zeta\rangle'_{(\partial_{\mu}\phi)^2\sigma}=f_{\mathrm{NL}}^{(\partial_{\mu}\phi)^2\sigma}P_{\zeta}(k_1)P_{\zeta}(k_3)\tilde{S}^{\mathrm{dS}}(k_1=k_2, k_3),
    \label{dS_shape}
\end{equation}
where
\begin{equation}
    f_{\mathrm{NL}}^{(\partial_{\mu}\zeta)^2\sigma} = \frac{1}{32}\epsilon_0\frac{M_{\mathrm{Pl}}^2}{\Lambda^2}|E_1^B(\mu,\alpha)||F^{(\partial_{\mu}\phi)^2\sigma}|b_*^2~~\overset{\alpha > \mu \gg 0}{\approx}~ \frac{1}{8}\epsilon_0\frac{M_{\mathrm{Pl}}^2}{\Lambda^2}|E_1^B(\mu,\alpha)|\alpha^2 b_*^2.
\end{equation}
The behaviour of $f_{\mathrm{NL}}^{(\partial_{\mu}\zeta)^2\sigma}$ as a function of $\mu$ and $\alpha$ is illustrated in the right panel of Figure \ref{5.4F2}. As expected from their analytical expressions, $f_{\mathrm{NL}}^{\dot{\zeta}\zeta\sigma}$ is approximately 4 times larger than $f_{\mathrm{NL}}^{(\partial_{\mu}\zeta)^2\sigma}$ for $\alpha>\mu>0$. Therefore, for qualitative discussion, we adopt the shorthand notation $f_{\mathrm{NL}}^{\mathrm{col.}}\equiv f_{\mathrm{NL}}^{\dot{\zeta}\zeta\sigma}$ to represent the size of the resonance-enhanced cosmological collider signal.

Now let's estimate how large this new collider signal could be. First, it is worth mentioning that a large bispectrum can be achieved without inducing significant correction to the power spectrum. 
If we compare $f_{\mathrm{NL}}^{\mathrm{col.}}$ and $\delta n^{\mathrm{col.}} $, there is a typical hierarchy for the enhancement factors $\left|{E_1^B}/{E_1^P}\right|\sim 10$ and $f_{\mathrm{NL}}^{\mathrm{col.}}$ is more enhanced by large $\alpha$. 
To be concrete, we set $b_*=0.1$ and $b_*|E_1^B/E_2^P|=1$ as an example. Then we have $\Lambda$ as the model-dependent parameter, and the sizes of the power spectrum correction and the bispectrum are related by $f_{\mathrm{NL}}^{\mathrm{col.}} = \delta n^{\mathrm{col.}}\alpha^2/4$. If we lower $\Lambda$ to saturate the observational bound $ \delta n^{\mathrm{col.}} \lesssim 0.05$, then for large $\alpha$ we get highly enhanced collider signals. Meanwhile, there is an upper bound on the frequency of the oscillation from the consistency of the EFT $\alpha\ll 400$ \cite{Behbahani:2011it}. Taking this into account, one would be able to have $f_{\mathrm{NL}}^{\mathrm{col.}}\sim \mathcal{O}(100)$ for some particular choices of model parameters.

In addition, even without the coupling to heavy moduli, the axion monodromy model \eqref{model} can also generate the oscillatory correction to the power spectrum and resonant non-Gaussianity.
Thus, it is interesting to compare our results above with the standard predictions from the single-field axion monodromy. For $\delta n$ and $\fnl$, the single-field results are given by \cite{Flauger:2009ab,Flauger:2010ja}
\begin{align}
    &\delta n^{\mathrm{s.f.}} = 3\sqrt{2\pi}\alpha^{-\frac{1}{2}}b_*\label{5.4E7}\\
    &f_{\mathrm{NL}}^{\mathrm{s.f.}} =\frac{3\sqrt{2\pi}}{8}\alpha^{\frac{3}{2}}b_*
    \label{5.4E8}.
\end{align}
For comparison, the resonant collider results in \eqref{5.4E4.2} and \eqref{5.4E6} contain two new ingredients: the enhancement factors and the overall coefficient $\epsilon_0\mpl^2/\Lambda^2$. For the weakly coupled scenario $\Lambda\sim \mpl$, it is more likely that the single-field predictions are still the dominant contribution. 
But for $\Lambda \lesssim 10^{-3/2}M_{\mathrm{Pl}}$ and $b_*\sim 0.1$, it would be more natural to have $f_{\mathrm{NL}}^{\mathrm{col.}}>f_{\mathrm{NL}}^{\mathrm{s.f.}}$.
Consequently, in UV-inspired scenarios of axion monodromy, it is parametrically possible for the enhanced collider signal to dominate over the resonant non-Gaussianity generated by the self-interactions of inflaton. This new phenomenology in the primordial bispectrum shows how sensitive this string inflation model can be to UV physics much heavier than the Hubble scale.


\section{Discussion and Outlook}
\label{sec:concl}


Symmetries play an essential role in the development of the cosmological bootstrap. 
The first implementation of the bootstrap idea exploits the full dS spacetime symmetry \cite{Arkani-Hamed:2018kmz,Baumann:2019oyu,Baumann:2020dch}.
However, in more realistic descriptions of our primordial Universe, some symmetries are broken, either spontaneously or explicitly. 
For instance, the breaking of the dS boost symmetries is commonly seen in $P(X)$-type models or the EFT of inflation with a small sound speed. 
This consideration led to the boostless bootstrap where the invariance of correlators under dS boost symmetries is relaxed \cite{Pajer:2020wxk, Jazayeri:2021fvk, Bonifacio:2021azc, Pimentel:2022fsc, Jazayeri:2022kjy, Wang:2022eop}. 
Meanwhile, in most of the bootstrap analysis so far, scale invariance is still an important assumption. 
Its explicit breaking corresponds to inflation scenarios with scale-dependent features, whose predictions in general depend on specific models. 
Then the vast range of possibilities lacks a unifying
theme, which challenges the bootstrap analysis.
One way out is to search for another guiding principle, such as the discrete shift symmetry \cite{DuasoPueyo:2023viy}. This allows us to systematically re-examine inflation models with globally oscillatory features, which are usually known as resonant non-Gaussianity. 

\vskip4pt
In this work, we extend the bootstrap analysis to inflation with discrete shift symmetry and also intermediate massive scalars. This setup, which can be seen as the cosmological collider with oscillatory couplings, is phenomenologically interesting because of the resonant enhancements for the collider signals of heavy particles \cite{Chen:2022vzh}. 
We start with the three-point scalar seed functions of massive exchange with two conformally coupled scalars and one mixed propagator. Due to the oscillatory couplings, the boundary differential equations for these scalar seeds are modified. Consequently, their solutions demonstrate both the scale-invariant collider signal and the resonant features. Then we apply the weight-shifting operators to derive inflationary bispectra for any kinematic configurations, which leads to new shapes of primordial non-Gaussianity with detectably large signals. 

\vskip4pt
Furthermore, we explore the explicit connection between the bootstrap computation and one particular string-inspired two-field model of axion monodromy. Because of the cosine modulation of the axion potential, the heavy modulus field which was thought to be stabilised can be continuously excited and lead to unsuppressed collider signals. 
This model, which essentially {differs} from the fast/sharp-turn multi-field inflation in literature, can be seen as a minimal extension of the original single field axion monodromy, with couplings to heavy fields under perturbative control and mass corrections negligible. 
The only requirement for the resonance effect is that the axion oscillating frequency should be greater than the moduli mass $\omega\gtrsim m$, which is relatively common in UV-complete theories if we just take the modulus field to be the lightest one. 
Therefore, our finding points to a new type of UV sensitivity in string inflation, that the heavy moduli from flux compactifications may not simply decouple but lead to detectable PNGs.

\vskip4pt
This work intends to build the bridges among several active research areas of primordial cosmology, including cosmological correlators, string inflation, and multi-field EFTs. As the first attempt, it showcases that the bootstrap approach provides a powerful tool of precise computations to re-examine the conventional idea for UV completions of inflation. Therefore, we expect that there would be several exciting directions for future explorations. 
\begin{itemize}
    \item First, for the cosmological bootstrap, it would be interesting to explore extensions to other inflation scenarios with scale-dependent features. In addition to the resonant features with global oscillations, another class of models present localized features, such as the transient changes of the inflaton potential or turning parameters. They generate new types of scale dependence in the resulting cosmological correlators. Equipped with the bootstrap techniques, we expect to have a systematic understanding of these predictions. 
    \item Second, our current discussion on UV embeddings of inflaton relies on several simplified assumptions. For instance, the modulus and axion fields are weakly coupled through a hyperbolic kinetic term. For the 4D EFT, there are certainly other possibilities from string compactifications, like strong couplings and other field space geometries. Would they lead to new types of predictions?
    Meanwhile, we may even start with a 10D picture directly and check if there would be UV-sensitive signatures in cosmological correlators. 
    There have been attempts to study cosmological collider physics with concrete UV embeddings \cite{Kumar:2018jxz,Reece:2022soh,Aoki:2023tjm,Chakraborty:2023eoq,Pimentel:2025rds}. 
    From a fully stringy setup, the UV physics may be linked to information beyond field-theoretic descriptions, such as geometries of compactified dimensions. What would be the characteristic signatures in cosmological correlators? We leave these intriguing questions for future work. 
    \item In the end, the predictions here can already be tested in the current data. Well-motivated in UV theories, they provide cosmological collider signals of heavy masses that can bypass the Boltzmann suppression. The bispectrum templates contain both scale-dependent resonant features and high-frequency collider oscillations, which serve as interesting new targets for the CMB bispectrum data analysis and also future LSS surveys. 
\end{itemize}

\vskip12pt
\paragraph{Acknowledgements} 

We would like to thank Enrico Pajer for the collaboration on a closely related project \cite{Pajer:2024ckd} and comments on a draft.
We are grateful to Xingang Chen for the initial discussion and helpful suggestions.
We acknowledge insightful discussions with Ana Ach\'ucarro, Carlos Duaso Pueyo, Zhehan Qin, Fernando Quevedo, Paul Shellard, Xi Tong, Gonzalo Villa.
DGW is partially supported by a Rubicon Postdoctoral Fellowship from the Netherlands Organisation for Scientific Research (NWO), the EPSRC New Horizon grant EP/V048422/1, and the Stephen Hawking Centre for Theoretical Cosmology.  BZ is supported by the Science and Technology Facilities Council
(STFC) studentship.

\appendix

\section{Useful formulas}
Converting ${}_2F_1\left[\cdots;\frac{u-1}{2u}\right]$ to ${}_2F_1[\cdots;u^2]$ with following identities:
\begin{align}
    {}_2F_1\left[a,b;\frac{a+b+1}{2};z\right]&=\frac{\Gamma\left(\frac{1}{2}\right)\Gamma\left(\frac{a+b+1}{2}\right)}{\Gamma\left(\frac{a+1}{2}\right)\Gamma\left(\frac{b+1}{2}\right)}{}_2F_1\left[\frac{a}{2},\frac{b}{2};\frac{1}{2};(1-2z)^2\right]\nonumber\\
    &\quad + (1-2z)\frac{\Gamma\left(-\frac{1}{2}\right)\Gamma\left(\frac{a+b+1}{2}\right)}{\Gamma\left(\frac{a}{2}\right)\Gamma\left(\frac{b}{2}\right)}{}_2F_1\left[\frac{a+1}{2}, \frac{b+1}{2}; \frac{3}{2};(1-2z)^2\right],
    \label{hyperg1}
\end{align}
and
\begin{align}
    {}_2F_1[a,b;c;z]&=\frac{1}{(-z)^a}\frac{\Gamma(b-a)\Gamma(c)}{\Gamma(b)\Gamma(c-a)}{}_2F_1\left[a,a-c+1;a-b+1;\frac{1}{z}\right]\nonumber\\
    &\quad+\frac{1}{(-z)^b}\frac{\Gamma(a-b)\Gamma(c)}{\Gamma(a)\Gamma(c-b)}{}_2F_1\left[b,b-c+1;-a+b+1;\frac{1}{z}\right],
    \label{hyperg2}
\end{align}
The enhancement factors in the power spectrum correction {can be derived by using the closed-form formulas of scalar seed functions in \cite{Qin:2023ejc}}:
\begin{align}
    &E_1^{P}(\mu, \alpha) \equiv \frac{e^{\pi\alpha/2}}{8}\frac{\Gamma(1+i\alpha)\Gamma\left(\frac{1}{2}+i\mu\right)\Gamma\left(\frac{1}{2}+i\mu+i\alpha\right)}{\Gamma\left(\frac{3}{2}+i\mu\right)\Gamma\left(\frac{3}{2}+i\mu+i\alpha\right)}{}_3F_2\left[\begin{matrix} 1+i\alpha, \frac{1}{2}+ i\mu, 1\\[1ex]\frac{3}{2} + i\mu + i\alpha, \frac{3}{2} + i\mu \end{matrix} \middle| 1\right]~,\label{4.1E18}\\[5pt] 
    &E_2^{P}(\mu, \alpha) \equiv \frac{e^{\pi\alpha}}{16}\frac{\Gamma(1+2i\alpha)\Gamma\left(\frac{1}{2}+i\mu+i\alpha\right)^2}{\Gamma\left(\frac{3}{2}+i\mu+i\alpha\right)^2}{}_3F_2\left[\begin{matrix} 1+2i\alpha, \frac{1}{2}+i\mu, 1\\[1ex]\frac{3}{2} + i\mu + i\alpha, \frac{3}{2} + i\mu + i\alpha \end{matrix} \middle| 1\right]~,\label{4.1E19}\\[5pt] 
    &E_3^{P}(\mu, \alpha) \equiv \frac{1}{8}\frac{\Gamma\left(\frac{1}{2}+i\mu+i\alpha\right)\Gamma\left(\frac{1}{2}+i\mu-i\alpha\right)}{\Gamma\left(\frac{3}{2}+i\mu+i\alpha\right)\Gamma\left(\frac{3}{2}+i\mu-i\alpha\right)}{}_3F_2\left[\begin{matrix} 1, \frac{1}{2}+ i\mu, 1\\[1ex]\frac{3}{2} +i\mu+i\alpha, \frac{3}{2} +i\mu-i\alpha\end{matrix} \middle| 1\right]~,\label{4.1E20}\\[5pt] 
    &E_4^{P}(\mu, \alpha)\equiv \frac{e^{\pi\alpha}}{16}\frac{\Gamma\left(\frac{1}{2}-i\mu+i\alpha\right)\Gamma\left(\frac{1}{2}+i\mu+i\alpha\right)\Gamma\left(\frac{1}{2}-i\mu-i\alpha\right)\Gamma\left(\frac{1}{2}+i\mu-i\alpha\right)}{\Gamma(1-i\alpha)\Gamma(1+i\alpha)}~.
    \label{4.1E20.1}
\end{align}

\section{Derivation of the Boundary Differential Equation}
\label{A:Derive}
In this Appendix, we present technical details of the derivation of the bootstrap equation (\ref{3.2E15}) for the primary scalar seed $\hat{\mathcal{I}}_1^{2\alpha}$  defined by (\ref{3.2E12}) (in the following context, we call it $\hat{\mathcal{I}}$ for simplicity). The main idea, as illustrated in \cite{Pimentel:2022fsc}, is to use the property of the dimensionless mixed propagator, (\ref{3.1E4}) to simplify the integral in the R.H.S. of (\ref{3.2E12}). In this case, we find
\begin{align}
    &\quad\left(\mathcal{O}_{k_3}+\frac{m^2}{H^2}\right)k_3^{1-i\alpha_2}\hat{\mathcal{I}}\nonumber\\ 
    & = \frac{1}{2(-x_2)^{i\alpha_2}}\int_{-\infty}^{0}d\eta (-\eta)^{i\alpha_2-2}\left[e^{ik_{12}\eta}\left(\mathcal{O}_{k_3}+\frac{m^2}{H^2}\right)\hat{\mathcal{K}}_+(k_3\eta, x_1)-c.c.\right]\nonumber\\
    &= \frac{ik_3^{1-i\alpha_2}}{4}\Gamma(1-i\alpha_1+i\alpha_2)\left(-\frac{1}{x_1}\right)^{-i\alpha_1}\left(-\frac{1}{x_2}\right)^{i\alpha_2}\cosh\left[\frac{\pi}{2}(\alpha_1-\alpha_2)\right]\left(\frac{u}{1+u}\right)^{1-i\alpha_1+i\alpha_2}\nonumber\\
    &\quad+\frac{ik_3^{1-i\alpha_2}}{4}\Gamma(1+i\alpha_1+i\alpha_2)\left(-\frac{1}{x_1}\right)^{i\alpha_1}\left(-\frac{1}{x_2}\right)^{i\alpha_2}\cosh\left[\frac{\pi}{2}(\alpha_1+\alpha_2)\right]\left(\frac{u}{1+u}\right)^{1+i\alpha_1+i\alpha_2}
    \label{A1E2}
\end{align}
where $k_t = k_1+k_2+k_3$. We used (\ref{3.1E4}) and then the analytical evaluation of the integral for the last step, which exactly gives the source terms appearing in the R.H.S. of equation (\ref{3.2E15}). 
Now, let us look at the L.H.S. of the equation. Our goal is to rewrite it
in terms of $u\equiv k_3/k_{12}$. Explicitly, we have
\begin{align}
    \left(\mathcal{O}_{k_3}+\frac{m^2}{H^2}\right)k_3^{1-i\alpha_2}\hat{\mathcal{I}}  
    &=\left(k_3^2\partial^2_{k_3}-2k_3\partial_{k_3}-k_3^2\partial_{k_{12}}^2+\frac{m^2}{H^2}\right)k_3^{1-i\alpha_2}\hat{\mathcal{I}}
    \label{A1E4}
\end{align}
where we used $ \eta^2\hat{\mathcal{I}} = -\partial_{k_{12}}^2\hat{\mathcal{I}}$.
Notice that $\hat{\mathcal{I}}$ depends only on the ratio $u\equiv k_3/k_{12}$, using chain rule, we can change $k_3$-derivatives to $k_{12}$-derivatives.
After some algebra, we obtain
\begin{equation}
    \left(\mathcal{O}_{k_3}+\frac{m^2}{H^2}\right)k_3^{1-i\alpha_2}\hat{\mathcal{I}} = k_3^{1-i\alpha}\left[(k_{12}^2-k_3^2)\partial_{k_{12}}^2+2(1+i\alpha_2)k_{12}\partial_{k_{12}}+\frac{m^2}{H^2}-(1-i\alpha_2)(2+i\alpha_2) \right]\hat{\mathcal{I}}.
    \label{A1E8}
\end{equation}
The next step is to convert all $k_{12}$-derivatives into $u$-derivatives using the chain rule.
At the end of the computation, we get
\begin{align}
    \left(\mathcal{O}_{k_3}+\frac{m^2}{H^2}\right)k_3^{1-i\alpha_2}\hat{\mathcal{I}}
    &=k_3^{1-i\alpha_2}\left[u^2(1-u^2)\partial_u^2-2u(u^2+i\alpha_2)\partial_u+\mu^2+\frac{1}{4}+i\alpha_2-\alpha_2^2\right]\hat{\mathcal{I}},
    \label{A1E10}
\end{align}
which is the L.H.S of the bootstrap equation.
Combining (\ref{A1E2}) and (\ref{A1E10}), we eventually obtain equation (\ref{3.2E15}). 

\bibliographystyle{utphys}
\bibliography{references.bib}
\end{document}